\documentclass[twocolumn]{aastex62}

\usepackage{amstext}
\usepackage{amsmath}
\usepackage{amssymb}
\usepackage{graphicx}
\newcommand{\aext}{{\bf a}_{\rm ext}}

\newcommand{\beq}{\begin{equation}}
\newcommand{\eeq}{\end{equation}}
\newcommand{\beqar}{\begin{align}}
\newcommand{\eeqar}{\end{align}}

\shorttitle{}

\shortauthors{}

\begin{document}

\title{Runaway Coalescence at the Onset of Common Envelope Episodes}

\author[0000-0002-1417-8024]{Morgan MacLeod}
\altaffiliation{NASA Einstein Fellow}
\affiliation{Harvard-Smithsonian Center for Astrophysics, 60 Garden Street, Cambridge, MA, 02138, USA}
\affiliation{School of Natural Sciences, Institute for Advanced Study, Princeton, NJ, 08540, USA}
\email{morgan.macleod@cfa.harvard.edu}

\author[0000-0002-0509-9113]{Eve C. Ostriker}
\affiliation{Department of Astrophysical Sciences, Princeton University, Princeton, NJ 08544, USA}

\author[0000-0001-5603-1832]{James M. Stone}
\affiliation{Department of Astrophysical Sciences, Princeton University, Princeton, NJ 08544, USA}

\begin{abstract}
Luminous red nova transients, presumably from stellar coalescence, exhibit long-term precursor emission over hundreds of binary orbits, leading to impulsive outbursts with durations similar to a single orbital period. In an effort to understand these signatures, we present and analyze a hydrodynamic model of unstable mass transfer from a giant-star donor onto a more compact accretor in a binary system. Our simulation begins with mass transfer at the Roche limit separation and traces a phase of runaway decay leading up to the plunge of the accretor within the envelope of the donor. We characterize the fluxes of mass and angular momentum through the system and show that the orbital evolution can be reconstructed from measurements of these quantities. The morphology of outflow from the binary changes significantly as the binary orbit tightens. At wide separations, a thin stream of relatively high-entropy gas trails from the outer Lagrange points.  As the orbit tightens, the orbital motion desynchronizes from the donor's rotation, and low-entropy ejecta trace a broad fan of largely ballistic trajectories. An order-of-magnitude increase in mass ejection rate accompanies the plunge of the accretor with the envelope of the donor. We argue that this transition marks the precursor-to-outburst transition observed in stellar coalescence transients. 
\end{abstract}


\section{Introduction}\label{sec:introduction}

Common envelope episodes occur in binary systems when a giant star engulfs its companion within its gaseous envelope \citep{1976IAUS...73...75P}. Drag forces tighten the orbit of the stellar cores during this phase, depositing orbital energy and momentum into the envelope gas \citep[e.g.][]{1993PASP..105.1373I,2000ARA&A..38..113T,2013A&ARv..21...59I}. The outcome of orbital tightening during this phase can be  a more compact binary and an ejected envelope, or a merged, single object with some retained envelope material \citep{1978ApJ...222..269T}. 

Such a phase of orbital tightening is thought to be crucial in the transformation of binaries into compact systems that can merge within the age of the universe under the influence of gravitational radiation \citep[e.g.][]{2002ApJ...572..407B,2007PhR...442...75K,2008ApJS..174..223B}.  As merging compact binaries are discovered in increasing numbers by gravitational wave interferometers like the LIGO--VIRGO network \citep{2016PhRvD..93l2003A,2016PhRvL.116x1103A,2017PhRvL.118v1101A,2017PhRvL.119p1101A,2017ApJ...851L..35A} there is an increased significance in tracing the evolutionary history that leads to the formation of these objects \citep[see][for a review]{2014LRR....17....3P}. Despite the centrality of the common envelope phase to the formation of compact binaries \citep{1984Natur.309..235V,2002ApJ...572..407B,
2003MNRAS.342.1169V,2013A&A...557A..87T,2017ApJ...846..170T,2018NewA...58...33B}, the details of these encounters, and which binaries map to which outcomes, have remained major sources of theoretical uncertainty for decades \citep{2013A&ARv..21...59I}. 

New, empirical evidence is now offering an avenue to better understand mass ejection in common-envelope-like encounters. It has become clear that the class of transients known as luminous red novae represents emission from cooling, expanding material ejected in either common envelope or stellar-merger encounters \citep{2013Sci...339..433I}. A key object in identifying this connection was the galactic transient V1309 Sco \citep{2010A&A...516A.108M}, which was identified by \citet{2011A&A...528A.114T} in pre-outburst data as an eclipsing binary with a decreasing orbital period. Several extragalactic systems, including M31 LRN 2015 \citep{2015A&A...578L..10K,2015ApJ...805L..18W}, M101OT 2015-1 \citep{2017ApJ...834..107B}, and NGC 4490 OT \citep{2016MNRAS.458..950S}, have added extra evidence and posed new questions. 

These transients exhibit outbursts defined by rapid rise to peak in the optical followed by fading and reddening emission. On timescales of months, dust formation leads to optical extinction and extensive reprocessing into the infrared \citep[e.g.][]{2013MNRAS.431L..33N,2016A&A...592A.134T}. 
Another shared feature among recent red nova  transients is precursor emission in the months to years leading to outbursts themselves \citep[e.g.][]{2011A&A...528A.114T,2017ApJ...834..107B}, strongly suggesting binaries interacting in circular orbits rather than single, impulsive passages \citep[e.g.][]{2011MNRAS.415.2020S}.  The phase of the rise and peak is particularly well sampled in the M31 LRN 2015 transient, in which the rise of the light curve from transient discovery to peak is similar to one orbital period of a test mass on the surface of the observed giant-star progenitor \citep{2017ApJ...835..282M}. 

The occurrence of impulsive transients might seem in tension with hints at circular binary orbits: how can a binary, which is stable over long-term, stellar evolution timescales,  produce a transient with timescale similar to the orbital period? 
Perhaps not surprisingly, the optical photosphere traces rapidly expanding ejecta in these transients, rather than the enveloped binary \citep[e.g.][]{2014ApJ...788...22P,2017ApJ...835..282M}. 
Rather than expanding freely, internal shocks may play a significant role in the ejecta's thermodynamics \citep{2017MNRAS.471.3200M}. 
To decipher how the underlying binary dynamics emerge into various components of the light curve, we need to trace these ejecta with the maximum fidelity possible in our simulation counterparts to these encounters.

In an effort to better understand the evidence about mass ejection in common envelope episodes offered by luminous red nova transients, we develop and present a new simulation method to study the lead-in to common envelope events. We study the phase from the onset of mass transfer from a more massive, evolved primary star onto a lower mass accretor. Because of the mass ratio, mass transfer shrinks the binary separation, leading to an unstable increase in mass-loss rate from the donor.  The secondary star is modeled only gravitationally, in reality, it may be either an unevolved main-sequence star or a compact remnant. We validate our approach through a series of careful convergence studies. We trace this runaway mass transfer and the resulting dynamics and forces that lead to the engulfment of the secondary star within the envelope of the giant-star primary and the onset of a common envelope phase.

We begin with some analytic background for the evolution of binary orbits under the influence of mass exchange in Section \ref{sec:analytic}. We describe our numerical simulation method in Section \ref{sec:method}; a number of tests are presented in Appendix \ref{sec:appendix}.  Using this method, we present and analyze a simulation of the coalescence of a binary with mass ratio 1:0.3 that  begins with mass transfer from a more-massive, giant-star primary to a less massive and more compact accretor in Section \ref{sec:results}.  In Section \ref{sec:lrn}, we argue that the features of runaway orbital inspiral at the onset of common envelope episodes, as seen in our simulations, explain the presence of both precursor emission and impulsive outbursts in luminous red nova transients. In section \ref{sec:conclusion}, we summarize and conclude. 

\section{Analytic Context for Binary Mass Transfer and Orbital Evolution}\label{sec:analytic}

In a binary system that evolves into contact, mass transfer begins from a donor star onto the accretor. The accreting object may accept this mass, in which case the mass transfer is described as conservative in that the system conserves total mass and angular momentum. In this case, there is a simple relationship that describes the rate of change of orbital separation, $a$, in terms of the mass transfer rate and binary mass ratio \citep{1956AJ.....61...49H}.  

If mass loss and the associated  changes in orbit and donor-star structure cause the donor to become less contained within its Roche lobe, the mass-loss rate intensifies. 
\citet{1972AcA....22...73P} estimate the mass-loss rate of a polytropic donor star of index $n$ to be
\beq\label{mdot_analytic}
\dot M \propto -  {M_{\rm d} \over P_{\rm orb} } \left( {R_{\rm d} - R_L \over R_{\rm d}} \right)^{n+{3\over 2}},
\eeq
where $M_{\rm d}$ and $R_{\rm d}$ are the mass and radius of the mass-losing donor star, $R_L$ is the radius of the Roche lobe \citep{1983ApJ...268..368E}, and $P_{\rm orb}$ is the binary orbital period \citep[see][for a derivation]{1994inbi.conf.....S}. Therefore, for $n=3/2$, the exponent is 3. Later, \citet{1987MNRAS.229..383E} validated this proportionality across a small range of mass-loss rates  in two-dimensional hydrodynamic simulations. These scalings indicate that when orbital evolution leads to increased Roche lobe overflow, the mass transfer rate is subject to a runaway to ever-increasing rates. 

In these cases of unstable mass transfer, the mass transfer rate is very likely to increase past the rate at which material may be accepted by the accreting object. This implies that some mass, and with it, some angular momentum, is carried away from the binary. In the extreme limit, the majority of mass transferred is lost from the system, and mass transfer is described as non-conservative. The orbital evolution equation \citep{1963ApJ...138..471H} for the case of fully non-conservative mass loss from a binary  is  
\beq\label{orbit_analytic}
{\dot a \over a}  = -2 {\dot M_{\rm d}  \over M_{\rm d} } \left[ 1  - \left(\gamma_{\rm loss}+{1\over2}\right) {M_{\rm d} \over M_{\rm d} + M_{\rm a}}\right] 
\eeq
where $M_{\rm d}$ is the mass of the donor-star primary, $M_{\rm a}$ is the mass of the accretor, and $\gamma_{\rm loss}$ parameterizes the specific angular momentum of the lost mass as
\beq
\gamma_{\rm loss} = {h_{\rm loss}  \over h_{\rm bin}}
\eeq
which is the ratio of the specific angular momentum of lost material, $h_{\rm loss}$, to the specific angular momentum of the binary, $h_{\rm bin}$, which is its total angular momentum divided by its mass.\footnote{A useful summary is presented in O. Pols' binary-evolution notes, Chapter 7. \url{http://www.astro.ru.nl/~onnop/education/binaries_utrecht_notes/} }

Studying equation \eqref{orbit_analytic} reveals that the evolution of the orbit under non-conservative mass transfer depends critically on the specific angular momentum of the material lost, through the parameter $\gamma_{\rm loss}$. A critical value separates  solutions that imply orbital expansion from those that imply orbital tightening, $\gamma_{\rm loss,crit} = (M_{\rm d}+M_{\rm a})/M_{\rm d} - 1/2$. Because $\dot M_{\rm d}<0$, situations with $\gamma_{\rm loss}<\gamma_{\rm loss,crit}$ lead to widening orbits, while those with $\gamma_{\rm loss}>\gamma_{\rm loss,crit}$ lead to shrinking orbits.

To illustrate the possibilities, consider a scenario where a more massive donor transfers material onto a less massive accretor ($M_{\rm d}>M_{\rm a}$).  We might imagine that material lost from the donor can leave the binary system with \citep{1963ApJ...138..471H}: 
\begin{enumerate}
\item the specific angular momentum of the donor star, as if being blown off in a spherical wind. In this case, $\gamma_{\rm loss}=M_{\rm a}/M_{\rm d}$. 
\item the specific angular momentum of the accretor star, as if being pulled from the donor onto the accretor and re-emitted.  In this case, $\gamma_{\rm loss}=M_{\rm d}/M_{\rm a}$. 
\item the specific angular momentum of the outer Lagrange point near the secondary, $L_2$, which as a saddle point in the potential might allow outflow in the plane of the orbit. In this case, $\gamma_{\rm loss} \approx (1.2)^2 (M_{\rm d}+M_{\rm a})^2 / (M_{\rm d}M_{\rm a})$,  where the numerical factor is related to the lever arm to $L_2$ from the center of mass \citep{1998CoSka..28..101P}. 
\end{enumerate}
These scenarios lead to major differences in orbital evolution. In case (1), $\dot a > 0$ and the binary widens its separation. In case (2), $\dot a < 0$, and the binary orbit tightens. In case (3), again $\dot a < 0$, but with larger magnitude because the $L_2$ point represents a larger lever arm from the system center of mass.

Our knowledge of and intuition for the geometry of mass-transferring binaries \citep[e.g.][]{1975ApJ...198..383L} makes case (1) appear unlikely -- mass is typically taken from the donor to the vicinity of the accretor, which implies an exchange of angular momentum \citep{1963ApJ...138..471H}. Qualitatively, then, the prediction of these analytic models (cases 2 and 3) for unstable mass transfer is  orbital tightening driven by increasingly rapid mass transfer \citep[e.g.][]{1977ApJ...211..881W}. By the time the orbital separation shrinks to be similar to the size of the donor, $a\approx R_1$, we enter a common envelope phase, during which it is difficult to discuss the stars as separate entities.

While these parameterized models provide a framework to understand binary evolution toward coalescence, the do not offer key quantitative details. In particular, while the scaling in \eqref{mdot_analytic} comes from dimensional analysis, the coefficient is unknown. Similarly, the precise magnitude of the order-unity quantity $\gamma_{\rm loss}$ determines the rate and even direction of evolution but is also unknown.  Motivated by this uncertainty, we use hydrodynamical simulations of mass exchange in a binary system to study the runaway coalescence of merging binaries in the remainder of this paper. 

\section{Simulation Models and Method}\label{sec:method}

Simulating the common envelope phase of binary systems has been a decades-long effort in astrophysics, from 1D approximations \citep{1979A&A....78..167M,1978ApJ...222..269T}, to the ``Double-core" series of multi-dimensional efforts which revealed the lack of spherical symmetry in the problem \citep{1984ApJ...280..771B,1989ApJ...337..849T,1991ApJ...373..246T,1994ApJ...422..729T,1994ApJ...431..247T,1995ApJ...445..367T,1995ApJ...451..308Y,1996ApJ...458..692T,1998ApJ...500..909S}, to more-recent three dimensional hydrodynamic \citep[e.g.][]{1996ApJ...471..366R,2008ApJ...672L..41R,2012ApJ...746...74R,2014ApJ...786...39N,2015MNRAS.450L..39N,2017MNRAS.464.4028I,2016ApJ...816L...9O,2018arXiv180503607C} and magnetohydrodynamic \citep{2016MNRAS.462L.121O} calculations with both Eulerian and Lagrangian methods \citep[see a review of methods in][section 4]{2017PASA...34....1D}. The majority of this work has focused on the outcome of the phase and the final separation of the tightened orbit of the stellar cores \citep[see][for a review]{2013A&ARv..21...59I}.  Other efforts have neglected the global evolution while focusing on flow in the immediate vicinity of an embedded object using a ``wind-tunnel''-like formulation \citep{2015ApJ...803...41M,2015ApJ...798L..19M,2017ApJ...838...56M,2017ApJ...845..173M}. 

Here we focus on the lead-in to the common envelope phase itself. 
Early interaction determines the observable properties of common envelope ejecta as well as the initial conditions for the bulk of the common envelope interaction. 
Previous authors of have either neglected the early interaction entirely in hydrodynamic calculations, starting the calculation with the stars already in contact \citep[e.g.][]{2016ApJ...816L...9O}, or focused on other characteristics of the system in their analysis. 
There are several notable exceptions. 
\citet{2017MNRAS.464.4028I} examined the effect of initial separation on simulation progression and final binary separation. 
\citet{2014ApJ...786...39N} tracked ejecta in simulated systems thought to be similar to V1309 Sco, but using a Lagrangian method that, in tracing mass rather than volume, lacks the resolution to follow the low-mass, low-density early interaction. 
Relatedly, \citet{2017ApJS..229...36G} attempted to synthesize observables from  hydrodynamic simulations but found that poor resolution and thermodynamics near the photosphere considerably limited current capabilities. 
\citet{2018MNRAS.477.2584S} and \citet{2017MNRAS.465L..54S} study the interaction of jets with tenuous envelope material during an early, grazing, phase of binary interaction, but these calculations again begin with the binary components in contact and, further, neglect their gravitational interaction. 

Other work of methodological relevance includes simulations of mass transferring and merging binaries. \citet{1995ApJ...438..887R} discuss initial data and the hydrodynamics of merger of polytropic binaries. \citet{2002ApJS..138..121M} and later \citet{2006ApJ...643..381D}, and  \citet{2007ApJ...670.1314M},  developed techniques for the simulation of mass-transferring binaries on Eulerian meshes, and emphasized the importance of frame choice in conserving key quantities \citep[see also related work by][]{2016ApJ...819...94K}. \citet{2017ApJS..229...27M} have recently reviewed this work and compared to particle-based methods.  \citet{2011ApJ...737...89D}, \citet{2011ApJ...737...49L}, and \citet{2012MNRAS.424.2222P} emphasized the importance of relaxed initial conditions (in the binary potential) for the subsequent evolution using smoothed-particle-hydrodynamic approaches.

The approach presented here does not represent a dramatic departure from previous efforts, but we do make a series of numerical choices such that our method is particularly suited to studying this phase of lead-in to engulfment that has become so interesting in light of recently observed transient outbursts. Of particular importance, we adopt an Eulerian method to solve the equations of gas dynamics, which allows us to follow low-density flow in (and outside) stellar envelopes. Because the distribution of this material contributes  to torques in the system, following this  low-mass material is critical to resolving the early interaction. Additionally, we use spherical polar coordinates to simulate the envelope surrounding the giant-star donor, which allows us to preserve the hydrostatic structure of the donor envelope to particularly high fidelity. 

We describe our method in the sections that follow and validate it through a series of convergence studies in Appendix \ref{sec:appendix}. 

\subsection{Simulated System and Reference Frame}\label{sec:system}

\begin{figure}[tbp]
\begin{center}
\includegraphics[width=0.48\textwidth]{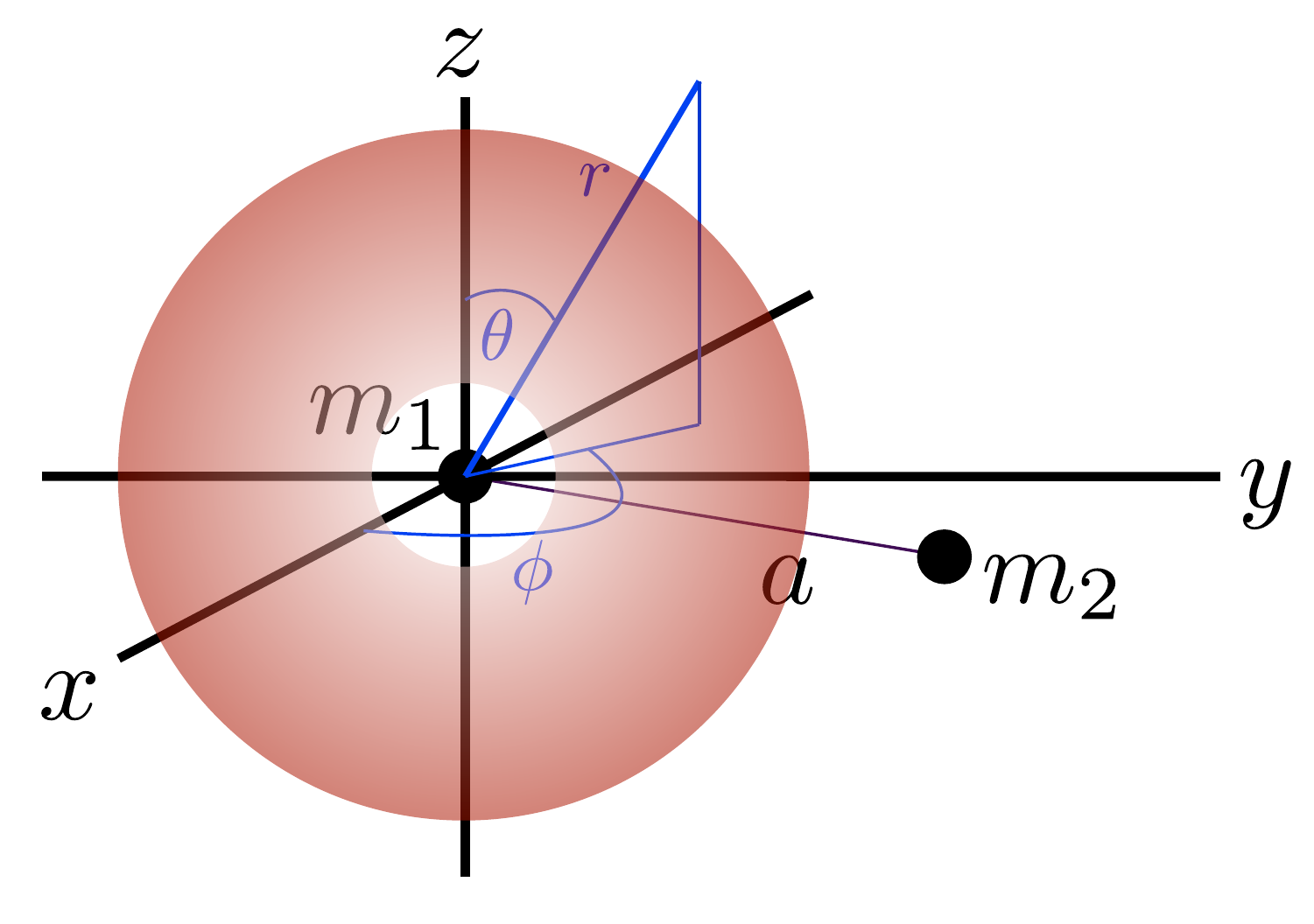}
\caption{Coordinate system and domain of numerical simulations. Two point masses, $m_1$ and $m_2$ interact with the simulated gas. The orbiting, non-rotating simulation frame orbits with $m_1$ and the mesh is allocated in spherical polar coordinates. The secondary object, labeled $m_2$ orbits in the $x-y$ ($\theta=\pi/2$) plane. The simulation domain extends from $-\pi$ to $\pi$ in azimuthal angle $\phi$, from a minimum radius $r_{\rm min}$ to $r_{\rm max}$ in radial coordinate and does not necessarily extend to the poles, with the polar angle, $\theta$, ranging from 0 to $\pi$. }
\label{fig:domaincartoon}
\end{center}
\end{figure}

We consider a binary star system composed of a giant-star donor (primary, object 1) and its accreting companion (secondary, object 2). A cartoon of the simulated system is shown in Figure \ref{fig:domaincartoon}. Our simulations are conducted in spherical polar coordinates in three dimensions. 

Here we consider cases where the companion is a main-sequence star or a compact object and is not resolved on the scale of our simulations. We therefore represent this secondary object only gravitationally with a point mass.  Similarly, the core of the giant star is excised from the computational domain and replaced with a point mass. The simulated system therefore contains two point masses as well as the gaseous envelope of the giant star. 

The total mass of the primary star is $M_1 = m_1 + m_g$, where $m_1$ is the mass of point mass 1, representing the core of the giant star, and $m_g$ is the mass of the gaseous envelope.  The primary star has total radius $R_1$.  The secondary has mass $M_2$, it is made up entirely of a point mass, so $M_2=m_2$. The total system mass $M=M_1 + M_2$, which generates an orbital frequency of $\Omega_{\rm orb} = \left(GM / a^3 \right)^{1/2}$, where $a$ is the binary semi-major axis. Equivalently, the orbital period is given by $P_{\rm orb} = 2\pi \Omega_{\rm orb}^{-1}$. 

We adopt an orbiting, but non-rotating, frame of reference.  The coordinate origin is centered on (and orbiting with) the giant-star donor. The coordinate frame of the calculation is therefore non-inertial, in that it accelerates with the orbital motion. After testing both rotating and non-rotating coordinate frames, we chose the non-rotating frame for the calculations presented here because we found that it more consistently preserved hydrostatic equilibrium of stars of arbitrary rotation. In a rotating frame, when the rotation of the star does not match that of the mesh, hydrostatic equilibrium implies a numerical balance of the centrifugal and Coriolis accelerations in addition to those from gravity and pressure gradients. 

\subsection{Equations Solved}
Here we outline the equations solved in our hydrodynamic evolution, which are the basic expressions of inviscid gas dynamics with source terms related to the binary system and choice of reference frame.  Our simulation is developed within the {\tt Athena++}\footnote{Stone et. al. (in preparation), publicly available at: https://princetonuniversity.github.io/athena} code, which is an Eulerian (magneto)hydrodynamic code descended from {\tt Athena} \citep{2008ApJS..178..137S}. 

\subsubsection{Inviscid Gas Dynamics}
The coupled equations of inviscid gas dynamics that we solve are 
\begin{subequations}\label{gaseq}
\begin{align}
\partial_t \rho  + \nabla \cdot \left( \rho {\bf v} \right) &= 0 , \\
\partial_t  \left( \rho {\bf v} \right) + \nabla \cdot \left( \rho {\bf v} {\bf v} + P {\bf I} \right)  &= - \rho \aext , \\
\partial_t E + \nabla \cdot \left[ \left( E+ P \right) {\bf v} \right] &= - \rho \aext \cdot \bf {v} ,
\end{align}
\end{subequations}
expressing  mass continuity, the evolution of gas momenta, and the evolution of gas energies. In the above expressions, $\rho$ is the mass density, $\rho {\bf v}$ is the momentum density, and $E = \epsilon + \rho {\bf v} \cdot {\bf v} / 2$ is the total energy density and $\epsilon$ is the internal energy density. Additionally, $P$ is the pressure, $\bf I$ is the identity tensor, and $\aext$ is an external acceleration (source term).  The equations above are closed by an ideal gas equation of state, $P=\left(\gamma -1\right) \epsilon$, in which $\gamma$ is the gas adiabatic index or ratio of specific heats. 

\subsubsection{Source Terms}
The source term on the right-hand side of equations \eqref{gaseq}, $\aext$, represents the forces associated with the binary, as well as fictitious accelerations associated with our choice to perform the calculation in a non-inertial frame of reference.

The orbital plane in our simulations is the $x-y$ (or $\theta=\pi/2$) plane, and the angular momentum vector of the orbit points in the $+z$-direction ($\theta=0$). The terms that make up $\aext$ are
\beq\label{source}
\aext = {\bf a}_1 + {\bf a}_2 + {\bf a}_{\rm sg} - {\bf a}_{1,i},
\eeq
representing the acceleration from the gravity of the primary core particle, $m_1$, the secondary, $m_2$; the acceleration from the gradient of the  self-gravitational potential of the gas; and the orbiting-frame correcting acceleration of the primary measured in the inertial frame, respectively. 

Within the code, we derive these accelerations in Cartesian coordinates then convert them to spherical polar coordinates. 
With the $x,y,z$ accelerations defined, we convert back to the local spherical polar coordinates using
\begin{align}
a_r &= \sin(\theta) \cos(\phi) a_x + \sin(\theta) \sin(\phi) a_y + \cos(\theta) a_z  \nonumber \\
a_\theta &= \cos(\theta) \cos(\phi) a_x + \cos(\theta) \sin(\phi) a_y - \sin(\theta) a_z \nonumber \\
a_\phi &= -\sin(\phi) a_x + \cos(\phi) a_y.
\end{align}
Finally, these source accelerations enter the gas momentum and energy equations in \eqref{gaseq}. 

The acceleration from $m_1$ is
\beq
{\bf a}_1 = - {G m_1  \over  |{\bf r}|^3} {\bf r},
\eeq
where $r$ is the vector displacement from the coordinate origin, where the donor-core particle is located. 
The acceleration from the secondary particle is 
\begin{align}
{\bf a}_2 &=  - {G m_2 \over |{\bf r}- {\bf r}_2|^3} ({\bf r}- {\bf r}_2)  \nonumber \\
    &\approx -Gm_2({\bf r}- {\bf r}_2) f_{\rm spline}(|{\bf r}- {\bf r}_2|, r_{\rm soft}) 
\end{align}
where ${\bf r}_2$ is the  vector position of the secondary particle. The approximate version is applied numerically and includes a softening spline at small distances $r\lesssim2 r_{\rm soft}$ from the particle, 
\begin{align}
&f_{\rm spline}(r, r_{\rm soft}) =  \nonumber \\
& \   \begin{cases}
r_{\rm soft}^{-3} \left[  {4 \over 3} - {6 \over 5} u^2 + {1 \over 2} u^3  \right]  & 0 \leq u < 1 \\
r^{-3}\left[ -{1 \over 15} +{8 \over 3}u^3 - 3u^4 + {6 \over 5} u^5 - {1 \over 6}u^6   \right]   & 1 \leq u < 2 \\
r^{-3} &u\geq 2 
\end{cases}
\end{align}
where $u=r/r_{\rm soft}$  \citep[][equation A2]{1989ApJS...70..419H}. Compared, for example, to a Plummer softening, this spline softening has the advantages of converging much more quickly to the Newtonian force when $1 < u < 2$ and (more importantly) being exact outside two softening lengths, $u>2$. 

We treat the self-gravitational interaction of the gas only approximately in our simulation. We adopt
\beq\label{asg}
{\bf a}_{\rm sg} \approx - {G \left[ M_{\rm enc,1}(r) - m_1 \right]  \over  |{\bf r}|^3} {\bf r},
\eeq
where $M_{\rm enc,1}(r)$ is the enclosed mass of the original primary star's (initial) spherical profile at distance $r$. That is, we apply an approximate, ``static self-gravity",  to the mass $m_g$ of envelope gas. This choice most-affects our solution when significant mass is lost from the donor star, or under dramatic departures from spherical symmetry. We discuss the impacts of this choice for total angular momentum conservation further at the end of Appendix \ref{sec:appendix}. 

The acceleration of the donor star relative to the inertial frame is 
\beq\label{a1i}
{\bf a}_{1,i} = {Gm_2 \over |{\bf r}_2|^3}{\bf r}_2 + {\bf a}_{1,{\rm gas}} 
\eeq
where the first and second terms represent the acceleration due to the accretor (particle 2) and the acceleration due to the gas is
\beq\label{agas1}
{\bf a}_{\rm 1,gas} = \int_{\rm vol} \frac{G \rho dV }{| {\bf r} |^3 }  {\bf r}.     
\eeq
We compute the integral above once per hydrodynamic time step.  We apply the inverse of this acceleration to the gas in equation \eqref{source} to correct for the non-inertial frame of the simulation.

\subsection{Secondary Point-Mass Particle}
As described earlier, to simulate the effect of a secondary in the binary system, we add a perturbing particle with mass $m_2$ to the system. We integrate the motion of this particle along an arbitrary orbit, which responds to the primary particle, $m_1$, and gas-distribution accelerations. We assume that the mass of the secondary is constant in time -- which implies that it neither gains nor loses mass due to interaction with the surrounding gas. This is likely an oversimplification, but without much higher spatial resolution it is difficult to even determine the qualitative outcome of accretion versus mass loss. 

\subsubsection{Initial Condition}
The secondary is initialized based on a choice of semi-major axis and eccentricity. It is initialized in the $\hat x$-direction, at apocenter, such that the initial $x$-coordinate (and separation) is $a(1+e)$; the initial $y$- and $z$-coordinates are set to zero. 

The initial velocity is set along the $\hat y$ direction (such that the $x-y$ plane becomes the orbital plane) and with a positive velocity such that the angular momentum is oriented in the $+z$-direction. Together, we have
\begin{subequations}
\begin{align}
x_0 &= a(1+e), \\
v_{y,0} &= \sqrt{{GM\over a}\left({1-e \over {1+e}} \right)}.
\end{align}
\end{subequations}
Finally, on the first call to the particle integrator, we kick the particle velocities backward a half step to achieve staggered position and velocity for the leapfrog integrator described below. 

\subsubsection{Integration}
We integrate the motion of the secondary's particle in response to the gravitational force from the primary-star particle and from the gas distribution. Because of our choice of a non-inertial frame, the primary-star particle is fixed at the origin and the equation of motion for the secondary particle includes extra terms. 

The particle step is currently synchronized with the hydrodynamic time step and is controlled by an integer number of substeps per hydrodynamic step. We use a leapfrog algorithm, with the velocities offset a half step behind the positions in time, in order to be centered across the time step. An evolution step therefore consists of a kick and a drift. 
The kick is
\begin{subequations}
\beq
v_{i+1/2} = v_{i-1/2} + a_i  \Delta t,
\eeq
and the drift is 
\beq
x_{i+1} = x_{i} + v_{i+1/2}  \Delta t,
\eeq
\end{subequations}
which advances the particle positions to the next step. 

\subsubsection{Accelerations on Particle 2}

The acceleration acting on the secondary particle in the orbiting and rotating frame is
\beq
{\bf a}_2 = {\bf a}_{2,1} + {\bf a}_{\rm 2,gas} -  {\bf a}_{1,i},
\eeq
where ${\bf a}_{1,i} $ is given by \eqref{a1i}. The acceleration of point mass 2 from point mass 1 is   
\beq
{\bf a}_{2,1}  =  - {Gm_1 \over | {\bf r}_2|^3} {\bf r_2} .
\eeq
The contribution from the gas distribution within the simulation domain to the secondary particle's acceleration is
\beq\label{agas2}
{\bf a}_{\rm 2,gas} = \int_{\rm vol} \frac{G \rho  dV  }{|{\bf r} - {\bf r_2}|^3 } ({\bf r} - {\bf r_2}) ,     
\eeq
which is summed once per time step. If substeps are used in integrating the particle motions, the gas contributions are not re-integrated each substep, only each hydrodynamic time step.

\subsection{Domain and Mesh}
As described above, we solve the equations of gas dynamics on a spherical polar mesh centered on the giant-star core particle $m_1$.  We employ zones which are uniformly spaced in $\theta$ and $\phi$ coordinates but logarithmically spaced in $r$ so that zone shapes remain roughly cubic across a wide range of radial scales. 

For parallelization, the mesh is broken into mesh blocks. In our fiducial cases, we employ 12 ($r$) x 8 ($\theta$) x 16 ($\phi$) mesh blocks of $24^3$ zones each at our base level. 
We additionally employ two nested levels of static mesh refinement opening about the binary orbital plane. Mesh refinement subdivides each zone by a factor of two in each direction, creating 8 smaller zones.
We achieve these nested levels with three refinement patches. One level above the base, we refine within $0.3\leq r \leq 0.538$ for one mesh block above and below the orbital plane, $7\pi/16 \leq \theta \leq 9\pi/16$ at all azimuthal angles, $\phi$. We also refine the region $0.538 \leq r \leq 6$ and $5\pi/16 \leq \theta \leq 11\pi/16$ at all $\phi$ to one level above the base level. Finally, we refine the region $0.6\leq r \leq 2.5$ and $7\pi/16 \leq \theta \leq 9\pi/16$ to two levels above the base.

The spherical polar coordinate system has singularities at $\theta=0$ and $\theta=\pi$. Zones in these regions can become very elongated and narrow in the $\phi$ direction, unless the resolution is very coarse.  This carries accompanying, and often severe, restrictions on the simulation time step. One way to deal with this problem is to have fewer zones in the $\phi$ direction near the poles than at the equator.  Mesh blocks in {\tt Athena++} are fixed in dimension, but we create effectively larger zones by averaging conserved quantities across neighboring zones near the poles and modifying the time step criteria accordingly. In the zones nearest the pole, we average all of the zones within a mesh block, creating one effective zone per mesh block in the $\phi$ direction. As we move farther from the pole, the number of zones averaged over drops linearly with $\theta$, rounding to the next power of two, such that the zone shape is approximately constant.

\subsection{Units, Initial, and Boundary Conditions}
\subsubsection{Units}

\begin{deluxetable*}{lccccc}\label{units}
\tablecaption{Description of Units}
\tablehead{\colhead{Quantity} & \colhead{Code} & \colhead{Example System A} & \colhead{Example System B} & \colhead{Example System C} } 
\startdata
Mass &  $M_1$ &  $1M_\odot$ &  $4M_\odot$ &  $18M_\odot$ \\
Length &  $R_1$ &  $10R_\odot$ &  $35R_\odot$ &  $220R_\odot$ \\
Grav. constant &  $1$ &  $G$ &  $G$ & $G$     \\
\hline
Time $(\Omega^{-1})$& $\left(G^{-1}M_1^{-1}R_1^{3} \right)^{1/2}$  & 0.583~d  & 1.91~d & 14.2~d \\
Velocity & $\left(GM_1 R_1^{-1} \right)^{1/2}$ & 138~km~s$^{-1}$ & 148~km~s$^{-1}$ & 125~km~s$^{-1}$ \\
Density & $M_1 R_1^{-3}$ & $5.91\times10^{-3}$~g~cm$^{-3}$  & $5.52\times10^{-4}$~g~cm$^{-3}$  & $9.99\times10^{-6}$~g~cm$^{-3}$  \\
Pressure & $GM_1^2 R_1^{-4}$ & $1.13\times10^{12}$~dyn & $1.20\times10^{11}$~dyn  & $1.56\times10^{9}$~dyn\\ 
Mass transfer rate & $\left(GM_1^3R_1^{-3} \right)^{1/2}$ & 627~$M_\odot$~yr$^{-1}$ & 766~$M_\odot$~yr$^{-1}$ & 464~$M_\odot$~yr$^{-1}$  \\
Angular momentum & $\left(GM_1^3 R_1 \right)^{1/2}$ & $1.91\times10^{52}$~g~cm$^{2}$~s$^{-1}$ & $2.86\times10^{53}$~g~cm$^{2}$~s$^{-1}$ & $6.85\times10^{54}$~g~cm$^{2}$~s$^{-1}$\\
Torque & $ GM_1^2 R_1^{-1} $  & $3.80\times10^{47}$~g~cm$^{2}$~s$^{-2}$ & $1.73\times10^{48}$~g~cm$^{2}$~s$^{-2}$ & $5.59\times10^{48}$~g~cm$^{2}$~s$^{-2}$ \\
\hline
Secondary mass & $0.3 M_1$ & $0.3 M_\odot$ & $1.2 M_\odot$ & $5.4M_\odot$ \\
Initial orbital period & 16.3 $\left(G^{-1}M_1^{-1}R_1^{3} \right)^{1/2}$  &  9.49~d & 31.1~d &   231~d \\
Initial orbital frequency & 0.386 $\left(GM_1 R_1^{-3} \right)^{1/2}$ & 0.662~d$^{-1}$ & 0.202~d$^{-1}$  & $2.72\times10^{-2}$~d$^{-1}$ \\
\enddata
\tablecomments{
Together, the unit mass, length, and gravitational constant define the dimensionless unit system we adopt in our simulations, labeled ``Code" above. Derived unit values for various quantities are shown in the middle section. Finally, the lower section lists some properties of the binary system simulated -- the mass of the secondary and the initial orbital period, $P_{\rm orb,0}$, and frequency, $\Omega_{\rm orb,0}$, at the Roche limit separation, $a_0 = 2.06R_1$. We convert our code units above to three example systems.   Example System A is a $1M_\odot$, $10R_\odot$ donor with a $0.3M_\odot$ companion, which has an orbital period at the Roche limit of 9.49~days. Example System B is a $4M_\odot$, $35R_\odot$ donor with a $1.2M_\odot$ companion, which has an orbital period at the Roche limit of 31.1~days (these are roughly the inferred donor-star properties for M31 LRN 2015, though the secondary-star mass is unkown \citep{2017ApJ...835..282M}). Example System C is an $18M_\odot$, $220R_\odot$ donor with a $5.4M_\odot$ companion, which has an orbital period at the Roche limit of 231~days (these are the approximate donor-star properties for M101 OT2015-1, where again the secondary-star properties are unconstrained \citep{2017ApJ...834..107B}). }
\end{deluxetable*}

The simulation is performed in units in which $G = M_1 = R_1 = 1$. This implies that the unit velocity is $\sqrt{GM_1 / R_1}=1$ and the unit time is the dynamical timescale of the primary star $\sqrt{R_1^3 / GM_1}=1$. Quantities can be rescaled based on these units to physical systems. In Table \ref{units} we list the units of primary and derived quantities in the dimensionless simulation and illustrate their rescaling to several physical systems. 

\subsubsection{Initial Condition}
The gas initial condition is a polytropic envelope in hydrostatic equilibrium. The equations governing its structure are
\begin{subequations}
\begin{align}
{d P \over dr}  &= - g \rho, \\
{d \rho \over dr}  &= - { g \rho^2 \over \Gamma_{\rm s} P}, \\
{d m \over dr}  &= 4 \pi r^2 \rho,
\end{align}
\end{subequations}
where $g = G m / r^2$ and $\Gamma_{\rm s}=1+1/n$ is the structural polytropic index of the envelope. 
In general, one must iterate to find a solution to these equations with a given total mass and radius. 
Outside the stellar radius, we continue in hydrostatic equilibrium, but create a higher sound-speed, low-density ``corona" with constant sound speed of $c_{\rm s,c}$ using
\beq
{d \rho \over dr}  = - { g \rho \gamma \over c_{\rm s,c}^2}.
\eeq 
We compute and tabulate a profile separately and interpolate onto the mesh at runtime. We allocate the giant-star core point mass, $m_1 = m(r_{\rm in})$, based on this profile and the mesh inner radius.

The initial $r$ and $\theta$ components of the gas velocity are set to zero. In some cases we want the envelope to rotate about the axis of the orbital angular momentum. We specify a rotational frequency, $\Omega_{\rm env,0} = f_{\rm corot} \Omega_{\rm orb,0}$. Then,
\begin{subequations}
\beq
v_\phi = \Omega_{\rm env,0} R_{\rm cyl}  = \Omega_{\rm env,0} r \sin (\theta) 
\eeq
within the envelope $(r<1)$. Outside the envelope $(r>1)$, 
\beq
v_\phi = { \Omega_{\rm env,0}  \sin^2 (\theta) \over R_{\rm cyl}  } 
\eeq 
\end{subequations}
implying constant specific angular momentum (of the surface of the star at that polar angle to the material).

\subsubsection{Relaxation Scheme}
Once the initial conditions are allocated on the mesh, we relax toward the fully dynamical portion of the simulation in two ways, by damping gas velocities at early times to allow the structure to settle into a numerically-hydrostatic equilibrium, and by turning on the force of the secondary object slowly following this relaxation period.

In the case of a rotating envelope, the hydrostatic structure is different (as a function of polar angle $\theta$) from the non-rotating initial profile.  Injecting non-equilibrium conditions quickly leads to shock formation and the destruction of the envelope without some sort of damping and relaxation scheme. 

We adopt a damping source term, which is applied (included in $\aext$ of \eqref{source}) for a time $t_{\rm relax}$ at the beginning of the simulation,
\beq
{\bf a}_{{\rm damp}} = - {\bf{v} \over \tau_{\rm damp}}
\eeq
where $ \tau_{\rm damp}$ is a damping timescale, and the damping is applied in the $r$- and $\theta$-directions only. 
The damping timescale is set to $\tau_{\rm damp}=\tau_{\rm damp,0}=1$ for $t<0.2 t_{\rm relax}$ and \beq
\tau_{\rm damp}=\tau_{\rm damp,0} 10^{2(t-0.2t_{\rm relax}) / 0.8t_{\rm relax}}
\eeq
so that the damping time increases to 100 by $t=t_{\rm relax}$. 

During $t<t_{\rm relax}$ the secondary mass is not active. We then turn on the source term from the secondary mass slowly after the relaxation of the initial hydrostatic profile. For times $t_{\rm relax} < t \leq t_{\rm relax} + t_{\rm mass}$, we apply a fractional factor to the source terms associated with the gravity of the secondary. This factor increases linearly from 0 at $t=t_{\rm relax}$ to 1 at $t=t_{\rm relax} + t_{\rm mass}$. By turning on this source term over $t_{\rm mass}$ significantly greater than the envelope dynamical time, we reduce the effects of any initial transient due to impulsively applying the gravitational force of the secondary to the envelope. We typically adopt $t_{\rm relax}=t_{\rm mass}=15$ such that the simulation is fully active after 30 donor-star dynamical times.

\subsubsection{Boundaries}

 In the $r$-direction, we impose a ``diode'' boundary condition on the outer boundary that is an outflow-only boundary that does not allow inflow onto the grid $(v_r\geq 0)$.   The inner $r$ boundary condition is reflecting. 
 In the $\theta$-direction, we apply ``polar'' boundary conditions that wrap the mesh appropriately to allow free passage of flow through the pole.  In the $\phi$-direction, we apply periodic boundaries so that flow wraps fully around the $-\pi$ to $\pi$ domain.

\subsection{Diagnostics}\label{sec:diagnostics}

We record several diagnostic quantities at runtime to facilitate the analysis of the binary system.  We record the position velocity of the secondary particle. We also record the gas accelerations on each of the particles, equations \eqref{agas1} and \eqref{agas2}. We calculate the (Cartesian) position and velocity of the particle and gas center of mass in the simulation frame,
\begin{subequations}
\begin{align}
{\bf r}_{\rm com} &=  \frac{ m_1 {\bf r_1} + m_2 {\bf r_2} + \int_{\rm vol}  \rho dV   {\bf r}  } { m_1 + m_2 +  \int_{\rm vol}  \rho dV  }  , \\
{\bf v}_{\rm com} &=  \frac{ m_1 {\bf v_1} + m_2 {\bf v_2} + \int_{\rm vol}  \rho dV   {\bf v}  } { m_1 + m_2 +  \int_{\rm vol}  \rho dV  }  ,
\end{align}
\end{subequations}
where all quantities are in the simulation frame (which implies ${\bf r_1}={\bf v_1}=0$). These quantities  allow the correction of simulation quantities to the inertial frame in the analysis. 

With the center-of-mass position and velocity computed, we also compute quantities related to the exchange of angular momentum between the simulated components of gas and particles. 
The particles' net angular momentum relative to the center of mass is 
\begin{multline}\label{Lparticle}
{\bf L}_p = ({\bf r_1} - {\bf r}_{\rm com}) \times m_1 ({\bf v_1} - {\bf v}_{\rm com})   \\
+ ({\bf r_2} - {\bf r}_{\rm com}) \times m_2 ({\bf v_2} - {\bf v}_{\rm com}),
\end{multline}
where all of the initial system orbital angular is in the $\hat z$ component. The net gas angular momentum is 
\beq\label{Lgas}
{\bf L}_g = \int_{\rm vol} ({\bf r} - {\bf r}_{\rm com}) \times ({\bf v} - {\bf v}_{\rm com}) \rho dV .
\eeq
Finally, gas can leave the simulation domain through the outer $r$ boundary, carrying with it angular momentum. We record the instantaneous rate of angular momentum loss through the cell faces at the outer boundary,
\beq\label{Lgasout}
{\bf \dot L}_{g,\rm out} = \int_{\rm area}  ({\bf r} - {\bf r}_{\rm com}) \times ({\bf v} - {\bf v}_{\rm com}) \rho v_r dA ,
\eeq
where $\rho v_r dA$ is the mass flux through the surface element of area $dA$.  
The total system angular momentum at time $t$ is, therefore,
\beq\label{Ltot}
{\bf L}_{\rm tot}(t) = {\bf L}_p(t) + {\bf L}_g(t)  + \int_0^t {\bf \dot L}_{g,\rm out} (t') dt'. 
\eeq

The total angular momentum is a conserved quantity in a closed physical system. There are, however, reasons this may not be precisely the case numerically. First, we do not treat the backreaction of the inner-$r$ boundary condition on particle-1's motion (loss of angular momentum from the outer-$r$ boundary of the domain is recorded through \eqref{Lgasout}). Second,  we treat gas-particle gravitational interaction directly, but gas self-gravity only approximately with a static potential. Finally, our equations of gas dynamics are written in a form which explicitly conserves linear momenta along the coordinate directions, but not angular momenta, so some departure from perfect conservation is to be expected. The measurement of ${\bf L}_{\rm tot}$  is therefore a useful diagnostic of the degree to which our model conserves system angular momentum as it is exchanged between the particle and gas components.  We typically find that angular momentum is conserved to the percent level as particle and gas angular momenta exchange by roughly a factor of  two (see Appendix \ref{sec:appendix} for further tests and discussion).

%%%%%%%%%%%%%%%%%%%%%%%%%
%   RESULTS
%%%%%%%%%%%%%%%%%%%%%%%%%

\section{Results: Model Merging System}\label{sec:results}

In this section, we follow the properties of gas flow and orbital evolution in a binary system that is initiated with mass transfer from a giant-star donor to a less massive accretor. The orbit first evolves slowly, but enters into a phase of runaway coalescence. We introduce the simulated system in Section \ref{sec:param} and the overall evolution in \ref{sec:orbit}.  We follow the exchange of angular momentum through gravitational torques in Section \ref{sec:masstransferangmom}. Finally, we discuss the evolution of outflow morphology  as the orbital separation shrinks in Section \ref{sec:flow}. 

%%%%%%%%%%%%%%%%%%%%%%%%%
%  Orbit
%%%%%%%%%%%%%%%%%%%%%%%%%
\subsection{Simulation Parameters and Initial Conditions}\label{sec:param}

\begin{figure}[tbp]
\begin{center}
\includegraphics[width=0.4\textwidth]{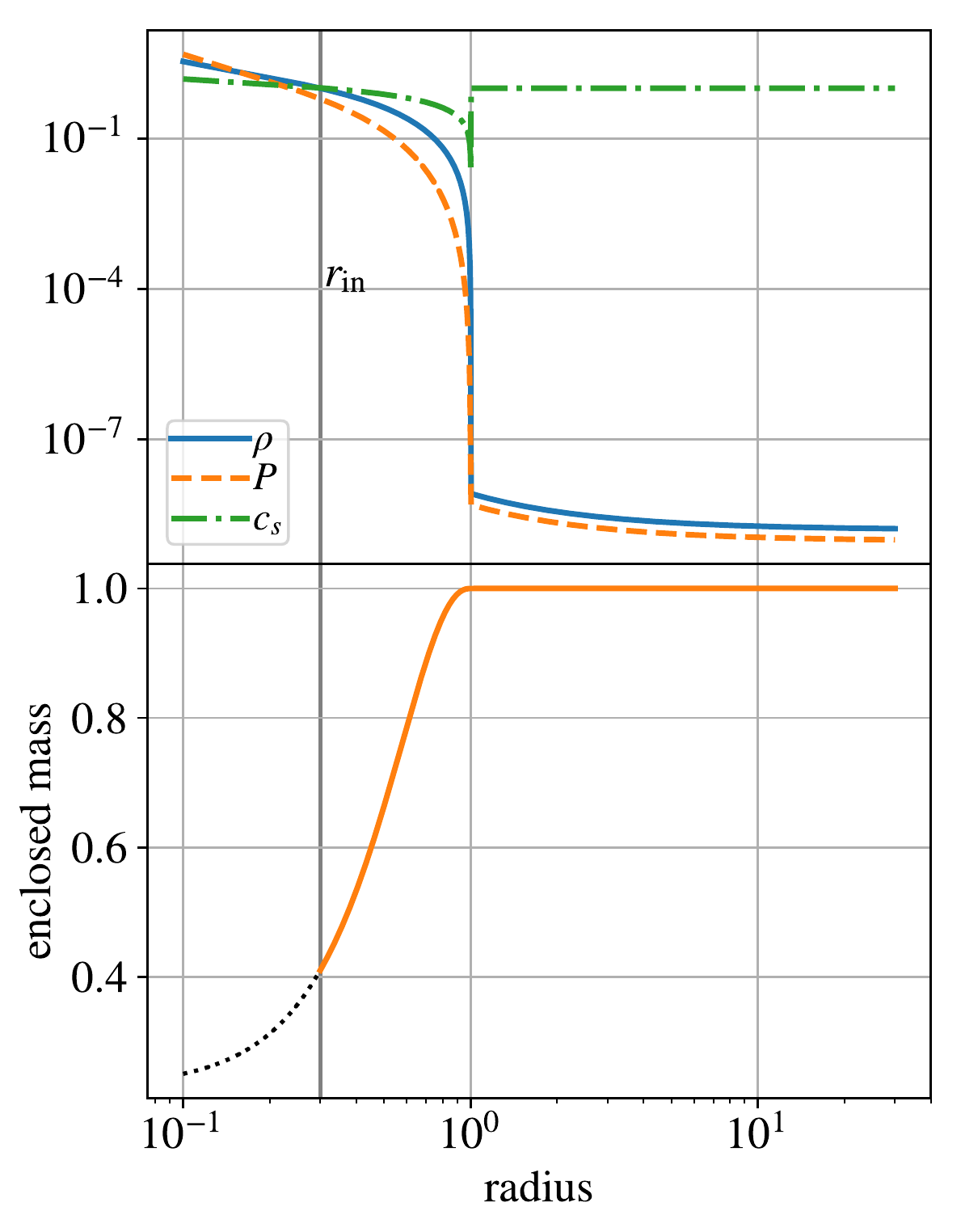}
\caption{1D initial condition for a polytrope with a point-mass core that is mapped onto the simulation domain. The envelope has polytropic structural coefficient $\Gamma_{\rm s}=5/3$. Within $R_1=1$, pressure, density, and sound speed are those of the stellar envelope. Outside, we impose a low-density, constant sound-speed atmosphere in hydrostatic equilibrium. The total mass of this background (the mass outside $R_1$) is roughly $2\times10^{-4}M_1$.   }
\label{fig:ic}
\end{center}
\end{figure}

We simulate a system with a mass ratio $q= M_2/M_1=0.3$, which implies $M_1=1$ and $M_2=0.3$ in our code units (similarly $R_1=1$). The primary-star mass, $M_1$ is made up of the excised core, $m_1$ and the simulated envelope gas. We use a 1D model (shown in Figure \ref{fig:ic}) in which 25\% of the mass is within the inner 10\% of the radius -- similar to a red giant branch star of in which the core mass is roughly one-quarter the total mass. Our model adopts a polytropic coefficient and gas adiabatic exponent identical to one another, $\Gamma_{\rm s}=\gamma=5/3$.  We excise an inner region of $r<r_{\rm in}=0.3R_1$, and include this mass in $m_1$. The outer radius of the domain is $r_{\rm out} = 30R_1$. This implies $m_1\approx0.41M_1$ and an envelope gas mass of approximately $0.59M_1$. In a grid-based hydrodynamic calculation, we cannot embed the stars in vacuum; therefore, in our 1D profile, we apply a transition to a low-density,  high-sound-speed medium at a pressure of $10^{-8}$, implying a total background mass (outside the star) of approximately $2\times10^{-4} M_1$.

Our fiducial model adopts an initial separation of $a_0 = 2.06R_1$, the analytic Roche limit separation for our mass ratio under the \citet{1983ApJ...268..368E} approximation. The initial orbit is circular, $e_0=0$. We set the donor envelope in synchronous rotation with the orbital motion such that $f_{\rm corot}=1$ and $\Omega_{\rm env,0} = \Omega_{\rm orb,0}$. The 1D profile is relaxed onto the computational mesh for 15 code time units (roughly 15 dynamical times of the donor star). Next, the gravity of $M_2$ is linearly turned on for 15 time units such that after 30 dynamical times, the simulation is fully active.

%%%%%%%%%%%%%%%%%%%%%%%%%
%  Model Merging System
%%%%%%%%%%%%%%%%%%%%%%%%%
\subsection{A Model of Runaway Binary Coalescence}\label{sec:orbit}

\begin{figure}[tbp]
\begin{center}
\includegraphics[width=0.23\textwidth]{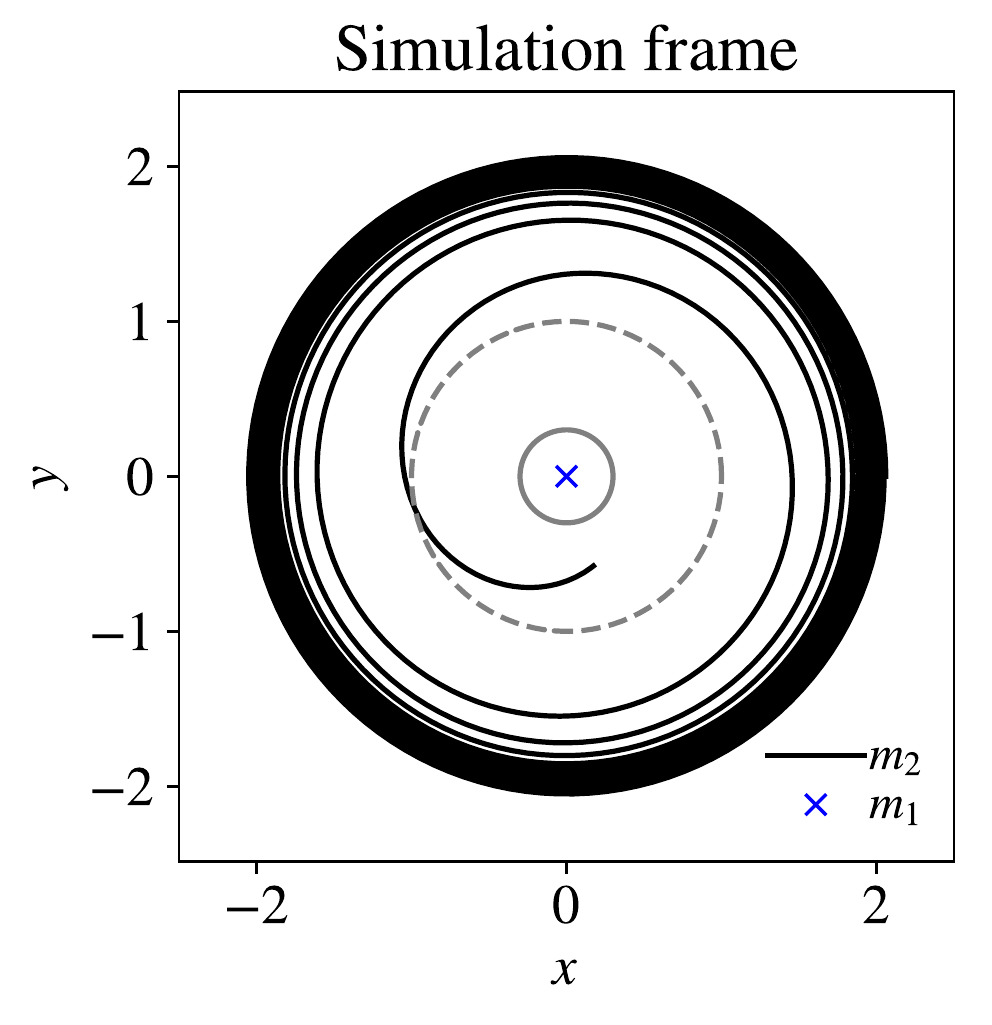}
\includegraphics[width=0.23\textwidth]{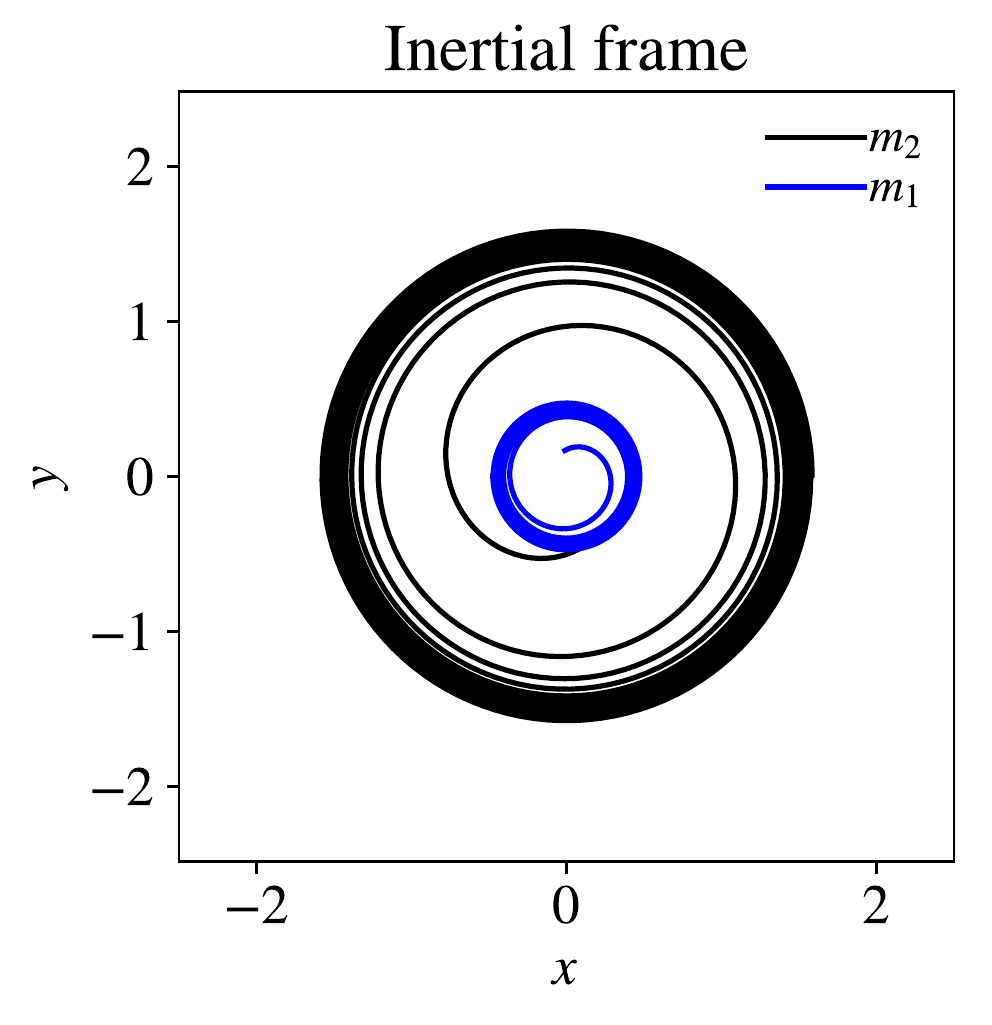}
\includegraphics[width=0.48\textwidth]{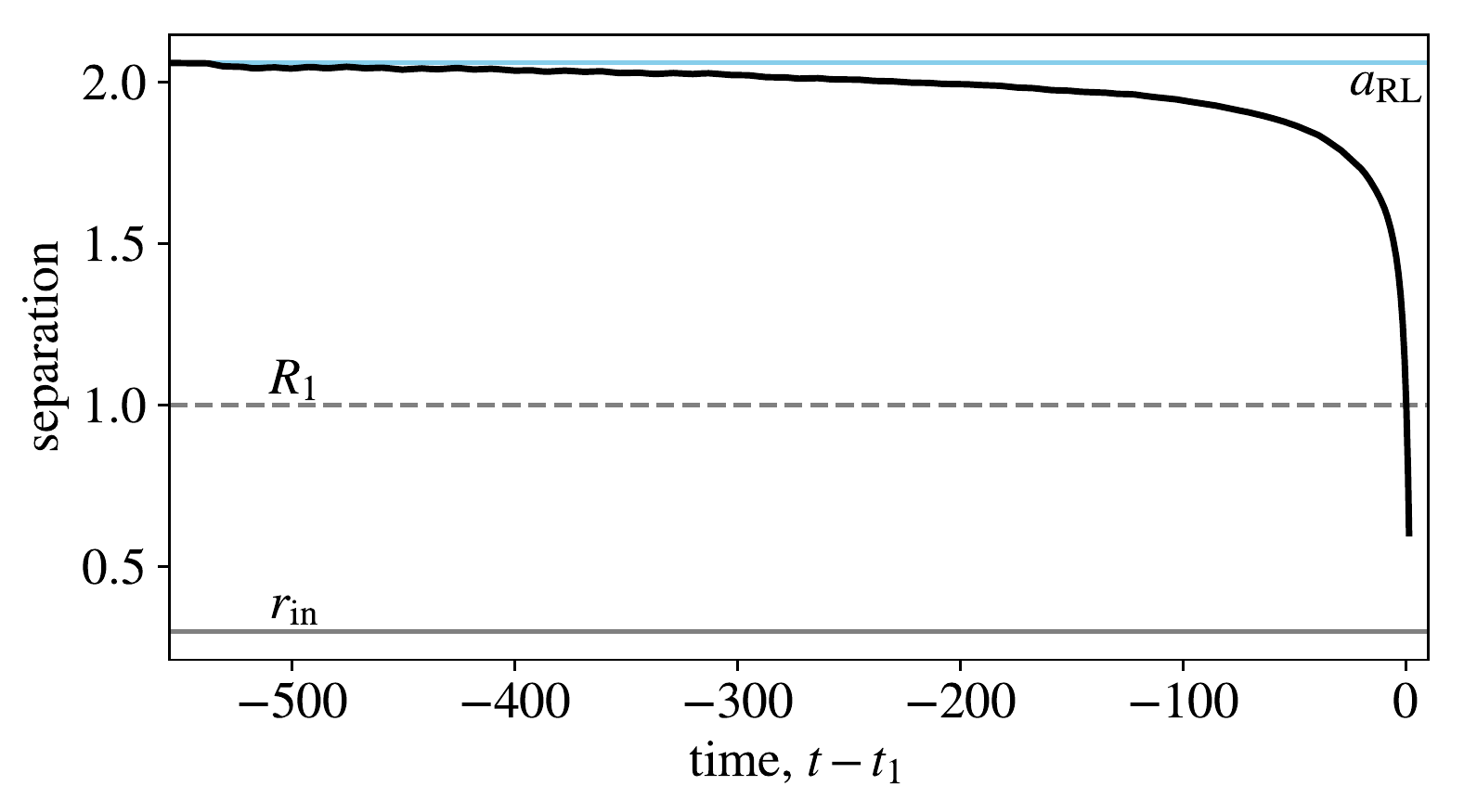}
\caption{Evolution of the orbit of the two stellar core particles, $m_1$ and $m_2$, for the duration of our simulation. The upper panels show the orbital evolution in the simulation and inertial frames, respectively. The lower panel tracks the evolution of the orbital separation as a function of simulation time. In this panel the size of the star is marked with a dashed grey line, and the size of the inner boundary is marked with a solid grey line. The Roche Limit separation is marked with a blue line. Time $t_1$ is when the separation is equal to $R_1$.  }
\label{fig:orbit}
\end{center}
\end{figure}

Our model system's evolution begins at the Roche limit separation $a_{\rm RL} \approx 2.06 R_1$, and terminates at $a=0.6R_1$. Here, we trace the evolution of the orbit and gas flow.

\subsubsection{Orbit}
As we see in Figure \ref{fig:orbit}, the orbital evolution begins with many tightly wound orbits with minimal change in separation, but eventually leads to a phase of dramatic inward spiral toward coalescence.  The upper panels of Figure \ref{fig:orbit} show the path of particles $m_1$ and $m_2$ throughout the simulation.  In the simulation frame, the donor-core particle, $m_1$, is always at rest at the origin, and $m_2$ orbits it, as shown in the black trajectory.  In the inertial frame,  both objects orbit around  their mutual center of mass.

The evolution of orbital separation is most clearly seen in the lower panel of Figure \ref{fig:orbit}. 
While the separation initially evolves slowly in time, it exhibits a runaway decrease to smaller separations and binary coalescense as the orbit shrinks (see similar behavior in simulations of \citet{2014ApJ...786...39N}, their Figure 2, and \citet{2017MNRAS.464.4028I}, Figure 4). At separations $a\approx R_1$, the rate of evolution of the orbit is rapid. We  denote a reference time, $t_1$,  defined by the time when the separation equals the initial radius of the donor,
\beq
a(t_1) \equiv R_1,
\eeq
where we approximate the binary semi-major axis with the separation of the two core particles as $a\approx |{\bf r}_1-{\bf r}_2|$. This is a reasonable approximation because the orbit remains relatively circular.

\begin{figure}[tbp]
\begin{center}
\includegraphics[width=0.48\textwidth]{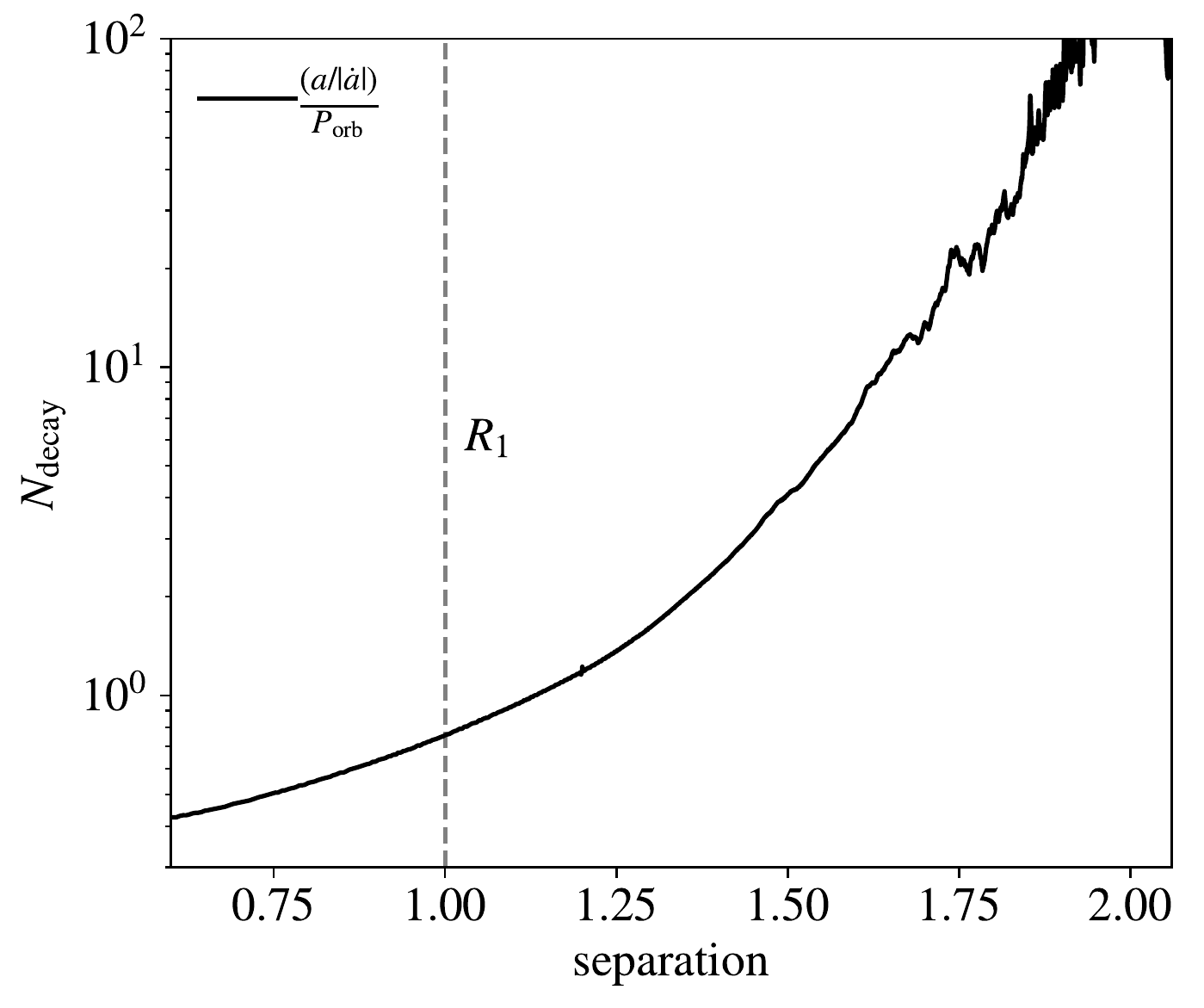}
\caption{Orbital decay rate, expressed as $(a/|\dot a |)/P_{\rm orb}$, the number of orbital periods over which the semi-major axis changes by of order itself. As the separation decreases from near the Roche separation to coalescence, the orbit goes from evolving very slowly $N_{\rm decay}\gg 1$ to evolving rapidly, on the same order as the orbital period, $N_{\rm decay}\approx 1$.  }
\label{fig:ndecay}
\end{center}
\end{figure}

Figure \ref{fig:ndecay} further explores the accelerating rate of orbital decay noted in Figure \ref{fig:orbit}. 
Here, we parameterize this as $N_{\rm decay} = (a / |\dot a|) / P_{\rm orb}$, the orbital decay timescale divided by the instantaneous orbital period. The rate of change of the orbit becomes increasingly rapid as the binary comes closer to coalescence. The orbit first evolves slowly, with $N_{\rm decay}\gtrsim 10^2$, for separations $a\gtrsim 2 R_1$ (90\% of the Roche limit separation). When the separation is $a \approx1.5R_1$, we find $N_{\rm decay}< 10$, indicating that the orbit is rapidly plunging to tighter separations even prior to the engulfment of the secondary within the envelope of the donor. Finally, for separations $a\approx R_1$, the orbital decay timescale is similar to the orbital period of the binary. 

\subsubsection{Gas Flow}
The consequences of this runaway decay of the binary orbit are dramatic. The binary interaction begins quite gently (and, in fact, the realistic system would include a much more gradual lead-in than is possible to simulate here). Even so, the system enters a phase of dynamical coalescence where the orbit is shrinking on timescales similar to the orbital period. This rapid coalescence indicates transition from a very long phase of quasi-hydrostatic evolution for the binary components to one where the evolution timescale is so short that equilibrium cannot be maintained and subsequent interaction is truly hydrodynamic in nature.

\begin{figure*}[tbp]
\begin{center}
\includegraphics[width=0.95\textwidth]{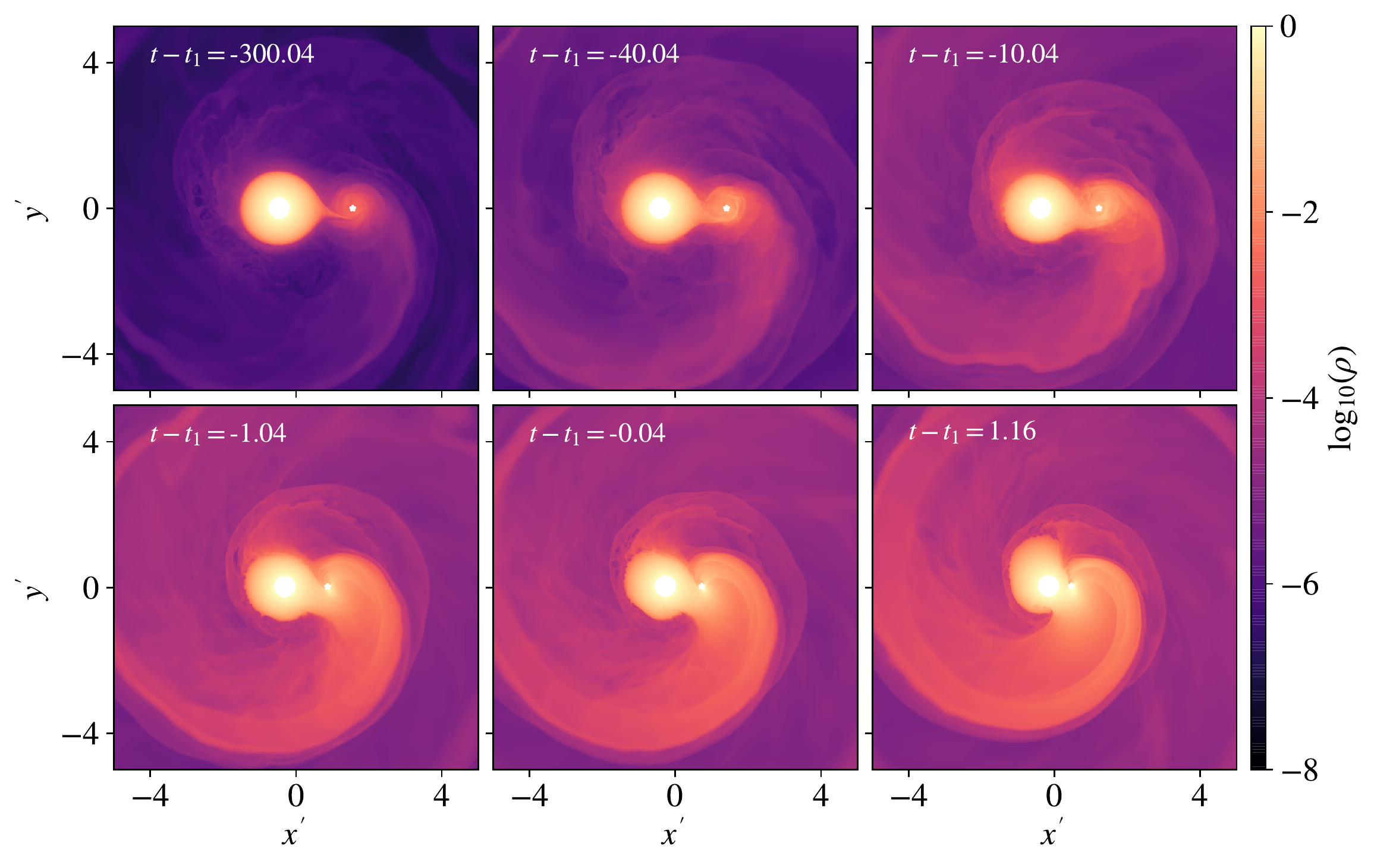}
\includegraphics[width=0.95\textwidth]{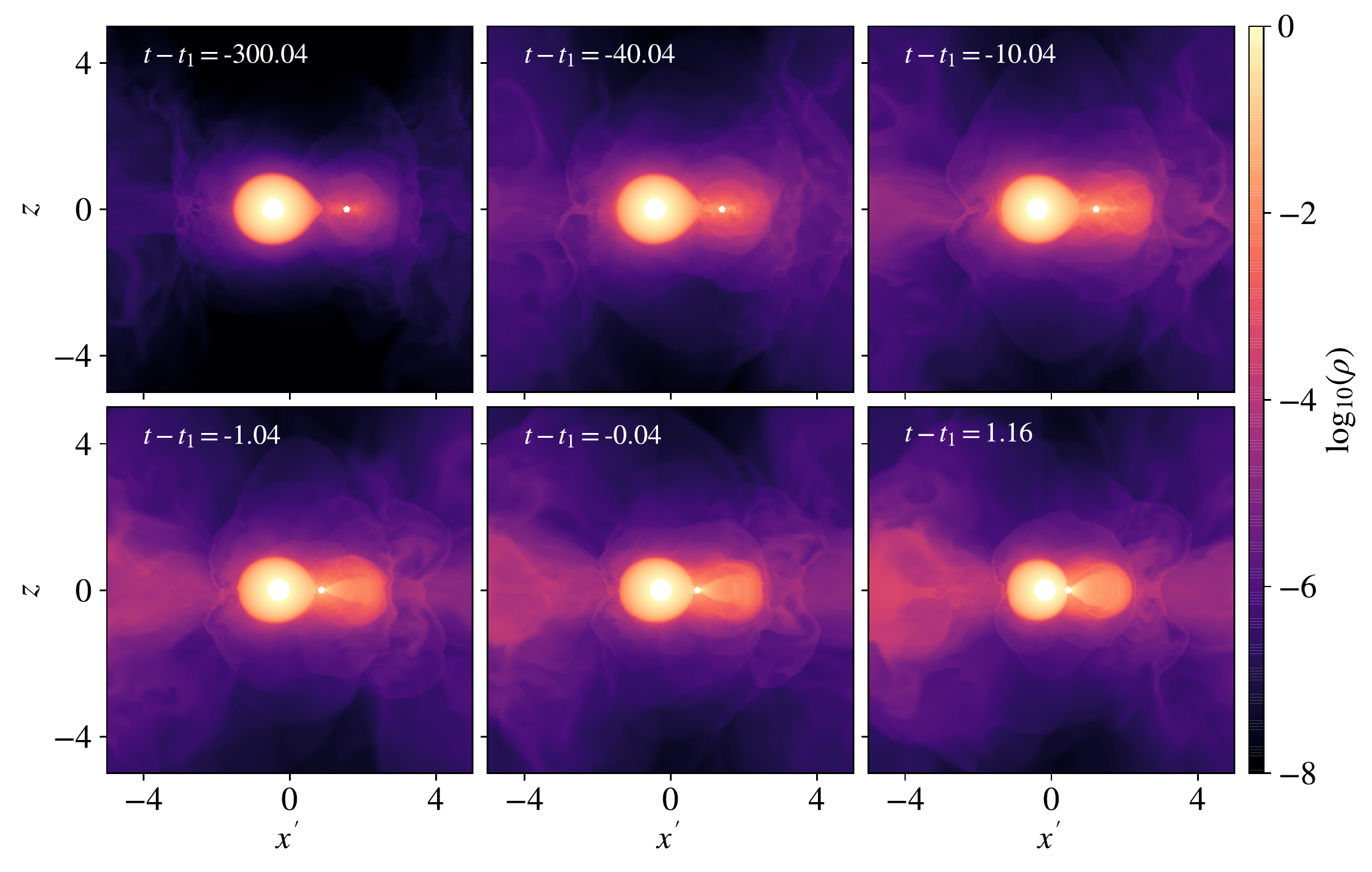}
\caption{Slices of gas density through the orbital plane (top) and perpendicular plane (bottom) for the duration of our simulation. For clarity, we plot along $x'$- and $y'$-axes, which are rotated about the $z$-axis relative to the simulation frame, such that the line of separation between $m_1$ and $m_2$ lies along $x'$. The origin in $x'$ and $y'$ is at the particle-and-gas system's center of mass.  Note that snapshots are much more closely spaced in the time series near the time of binary coalescence, $t_1$.  (An animation of this figure is available in the journal's online version).  }
\label{fig:ts_dens}
\end{center}
\end{figure*}

The evolution to an increasingly dynamic encounter is visualized in Figure \ref{fig:ts_dens} with a time series of slices of gas density in the orbital midplane (upper panel) and perpendicular plane (lower panel).  Slices are rotated relative to the frame in which they are computed (see Figure \ref{fig:domaincartoon} and section \ref{sec:system}) such that the $x'$-axis is parallel to the orbital separation vector, and shifted to have origin located at the particle-and-gas system center of mass.  We note that the snapshots here are not separated equally in time, and we see that evolution in the gas properties evolve largely with binary separation, rather than time, during the system's runaway coalescence. 

Common features in the flow include the exchange of mass from the donor envelope toward the accretor \citep{1975ApJ...198..383L} and then into tails which extend from near the outer $L_2$ (near the accretor) and $L_3$ (near the donor) Lagrange points \citep{1979ApJ...229..223S,2016MNRAS.455.4351P,2016MNRAS.461.2527P}.  This material forms a spiral pattern in the frame that rotates with the binary because, in moving radially outward, it no longer has sufficient angular momentum to maintain corotation with the binary \citep{1979ApJ...229..223S}. The changing pitch angle of this spiral reflects varying radial to orbital velocity ratios. At early times, the morphology of Roche lobe overflow and disk formation near the secondary is reminiscent of simulations of mass transfer by low-velocity winds \citep[e.g.][]{2017MNRAS.468.4465C}. As the system trends toward merger, the outflow morphology changes from a low-density stream originating largely near $L_2$, to a broad, mass-rich fan of ejecta at $t\approx t_1$. 
In slices perpendicular to the orbital plane, we  observe that the outflow is concentrated toward the plane of the orbit. Plumes of trailed-off material from the $L_2$ and $L_3$ points can be seen to self-intersect and interact in this plane \citep{2016MNRAS.455.4351P,2016MNRAS.461.2527P}, with those interactions shaping and broadening the outflow tails. 

A possible origin for dynamical coalescence of a binary is an innermost stable circular orbit, which manifests as a minimum angular momentum and energy in equilibrium sequences \citep{
1993ApJ...406L..63L,2001AIPC..575..176B}. Such a feature, for example, is present in the mergers of double neutron stars \citep{1996PhRvD..54.3958L,1997PhRvL..79.1182B,2010nure.book.....B}. The dynamical importance of this effect is restricted primarily to very stiff equations of state, in which case the innermost circular orbit may arise at larger separations than mass exchange. For a polytrope of index $n=3/2$ interacting with a point mass, dynamical instability occurs at separation similar to the stellar radius with our mass ratio of 0.3 \citep{1993ApJS...88..205L}. However, because mass transfer is initiated at nearly twice this separation, equilibrium configurations are not realized at these small separations (as shown in Figure \ref{fig:ts_dens}). This suggests that the mass exchange shown in Figure \ref{fig:ts_dens}, rather than an innermost stable circular orbit, plays the most significant role in runaway binary coalescence.

Based on the snapshots of Figure \ref{fig:ts_dens}, we take a moment to reconsider the appropriateness of our choice of an adiabatic ($\gamma=5/3$) equation of state for the simulated gas. A first useful property is the optical depth of material, $\tau \approx \rho \kappa L$, for a column of length $L$.  The magnitude of the optical depth depends on the rescaling of our dimensionless system to physical coordinates. If we adopt $L\approx R_1$, then, in terms of donor properties, 
\beq
\tau \approx 10^9 \left( \frac{\rho}{M_1/R_1^3} \right) \left( \frac{R_1}{10R_\odot} \right)^{-2} \left( \frac{M_1}{M_\odot} \right) \left( \frac{\kappa}{1~{\rm cm}^2{\rm g}^{-1}} \right), 
\eeq
where the first term represents density in simulation-scaled units. We note that $\kappa$ varies by orders of magnitude over the density and temperature conditions that may be encountered. Nonetheless, for $\kappa\approx 1~{\rm cm}^2{\rm g}^{-1}$, densities and solar-mass donors, in code units, of $\rho \gtrsim 10^{-9} M_1/R_1^3$ represent optically thick material when $R_1 = 10R_\odot$. This nominal optical depth  scales with stellar properties, if $R_1=10^3 R_\odot$; only densities of $\rho \gtrsim 10^{-5} M_1/R_1^3$ are optically thick.  These scalings indicate, that, for more compact giant-star donors, much of the simulated material is optically thick, suggesting inefficient cooling,  and that the adiabatic treatment of the gas thermodynamics is reasonable as a first approximation. 

%%%%%%%%%%%%%%%%%%%%%%%%%
%  Torques
%%%%%%%%%%%%%%%%%%%%%%%%%
\subsection{Mass Transfer and Angular Momentum Evolution}\label{sec:masstransferangmom}

In order to understand the runaway orbital evolution described in the previous section, we examine the angular momentum budget of our simulated system and its flow between the particles and gas, mediated by gravitational torques. 

\subsubsection{System Angular Momentum}\label{sec:angmom}

\begin{figure}[tbp]
\begin{center}
\includegraphics[width=0.48\textwidth]{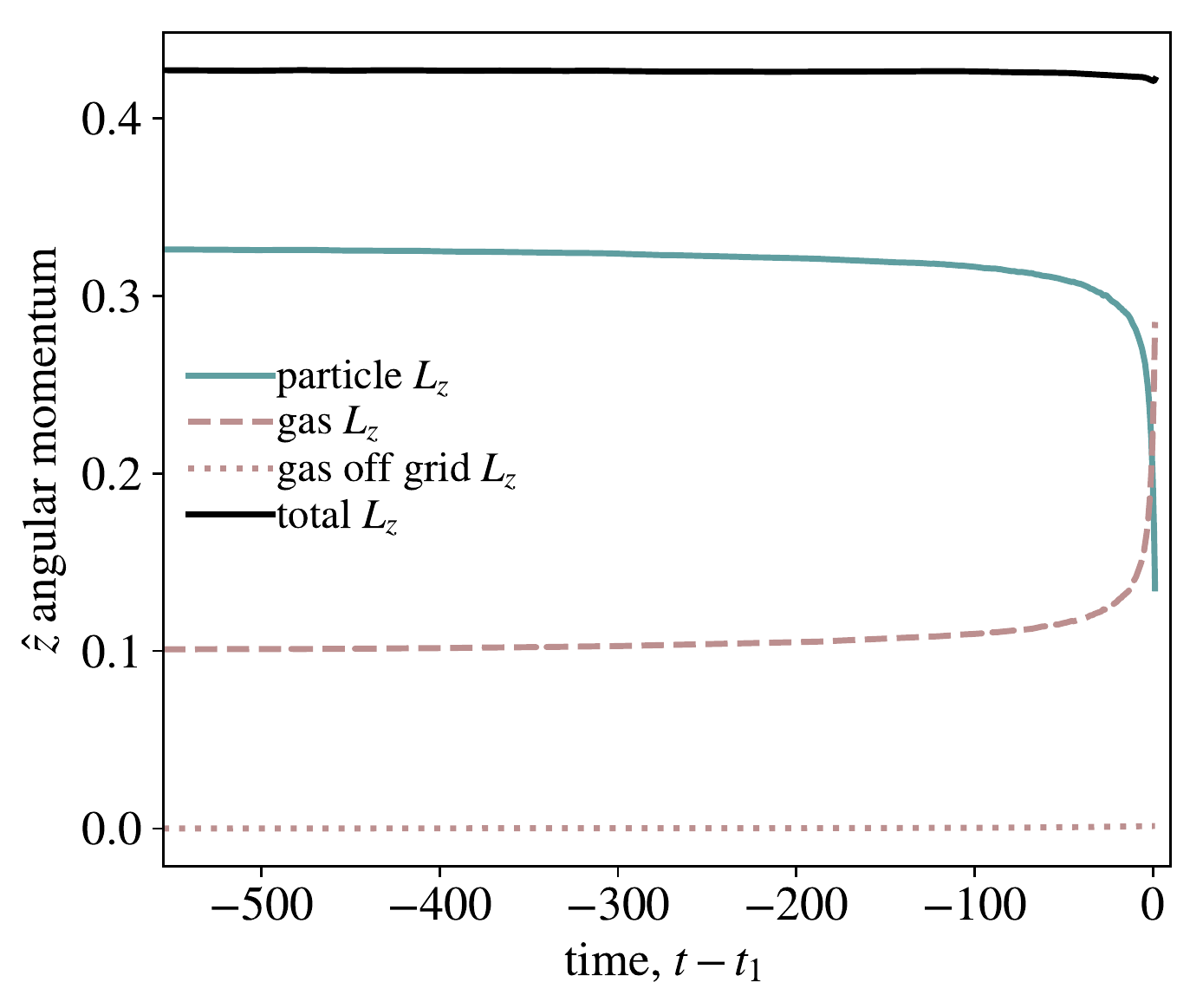}
\caption{Decomposition of angular momentum in the particle-gas system throughout the simulation duration. Particle angular momenta decrease through exchange with the gas. The total angular momentum is conserved to within approximately 2\% as the particles  and gas exchange roughly a factor of two in angular momentum.  }
\label{fig:angmom}
\end{center}
\end{figure}

We begin by examining the total angular momentum of the particle and gas components of our simulated binary, the $z$-component of of which is plotted in Figure \ref{fig:angmom}. 
The lines plotted in this diagram show angular momenta relative to the system's center of mass. The particle angular momentum represents the net angular momenta of the primary and secondary particles relative to the center of mass, as described by equation \eqref{Lparticle}. The gas angular momentum is the total angular momentum of the gas on the computational mesh, equation \eqref{Lgas}, while the ``gas off grid $L_z$'' is the cumulative angular momentum carried by gas leaving the computational domain, which results from a time integral of equation \eqref{Lgasout}. Finally, the total angular momentum of the system is the sum of these components, equation \eqref{Ltot}.

The particle angular momenta are observed to decrease, at first gradually, then rapidly, representing the orbital decay discussed previously. The net gas angular momentum mirrors this behavior -- as angular momentum is lost from the particle pair, it is given to the gas. Some of this gas leaves the domain during the simulation, carrying angular momentum with it, but the majority of the gas remains on the mesh. Importantly, we note that the total system angular momentum is nearly conserved during our simulation (conservation of angular momentum is not guaranteed in our method of solving the evolution equations for the particle-gas system, as discussed in Sections \ref{sec:diagnostics} and  \ref{sec:Lconservation}). We find an approximately 2\% change in total angular momentum as the angular momenta of the particles and gas exchange by a factor of approximately two.

\subsubsection{Mass Flow and Gas Specific Angular Momenta}\label{sec:massflow}

\begin{figure}[tbp]
\begin{center}
\includegraphics[width=0.48\textwidth]{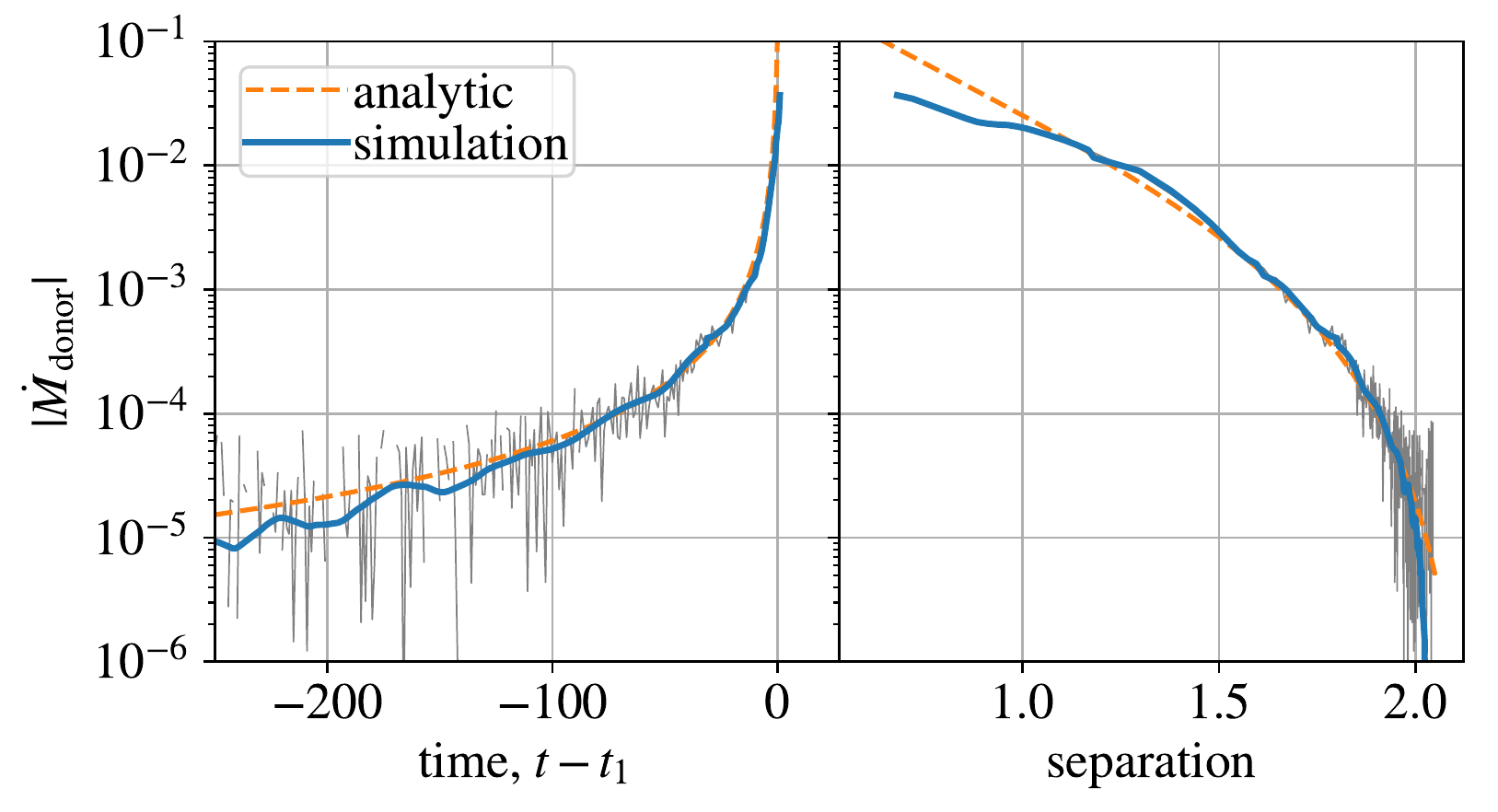}
\caption{Mass loss rate from the donor star increases by orders of magnitude as the system tightens. Mass loss rate is measured from the simulation as the instantaneous flux of mass through a spherical surface with radius $r=1$.  Lines marked ``simulation", are smoothed; the instantaneous values are highly variable, especially at large separation, and are shown in grey in the background.   The dashed line compares to the analytic formula for the expected mass-loss rate, equation \eqref{mdot_analytic} with constant of proportionality equal to one, and an assumed donor radius of $R_{\rm d}=1.05 R_1$, which is the equatorial radius of the rotating stellar profile following relaxation (see Figure \ref{fig:3dhse}). With these choices, the analytic expression produces a remarkable fit for the simulated result.    }
\label{fig:masstransfer}
\end{center}
\end{figure}

As the particle pair's orbital angular momentum decays, angular momentum is clearly exchanged with the gas. It is interesting to begin to examine the details of this flow of angular momentum. Following the analytic description of mass transfer in binary systems outlined in Section \ref{sec:analytic}, 
we break this down into a flux of mass and its distribution of specific angular momenta. 

In Figure \ref{fig:masstransfer}, we show the rate at which material is lost from the donor star. As the encounter proceeds toward a common envelope phase, where the stellar cores are engulfed in a shared envelope, the mass of the individual stars becomes less well defined. Here, we estimate the mass-loss rate by computing the instantaneous flux of mass through a spherical surface at $r= R_1$.  In Figure \ref{fig:masstransfer}, grey lines in the background show the instantaneous mass transfer rate, which is highly variable at large separations. The solid blue line shows a moving average of the instantaneous values (of varying width in time -- widest at early times such that the bin width encompasses roughly equal masses at each separation). 

The most remarkable feature of Figure \ref{fig:masstransfer} is how much the mass transfer rate increases as the binary proceeds toward coalescence. Over the simulation duration, approximately 34 initial orbital periods, the mass transfer rate increases by four orders of magnitude.  The late-time rise is especially rapid; over the several dynamical times it takes for the binary separation to shrink from 1.5$R_1$ to $R_1$, the mass transfer rate increases by an order of magnitude. 

Figure \ref{fig:masstransfer} also shows a comparison to the analytic formula to estimate the mass transfer rate, equation  \eqref{mdot_analytic}. Under several assumptions, we achieve a remarkable fit between this expression and our simulation results. First, we adopt a constant of proportionality of 1 in equation \eqref{mdot_analytic}. Next, we find that the fit is best represented if we use the volume-averaged radius of the oblate, rotating donor-star profile, which is $R_{\rm d}\approx1.04R_1$ as shown in Figure \ref{fig:3dhse}. With these calibrations, we find that the simple expression of equation \eqref{mdot_analytic} provides a very good description of the simulation result, especially at separations larger than $R_1$.

\begin{figure*}[tbp]
\begin{center}
\includegraphics[width=0.95\textwidth]{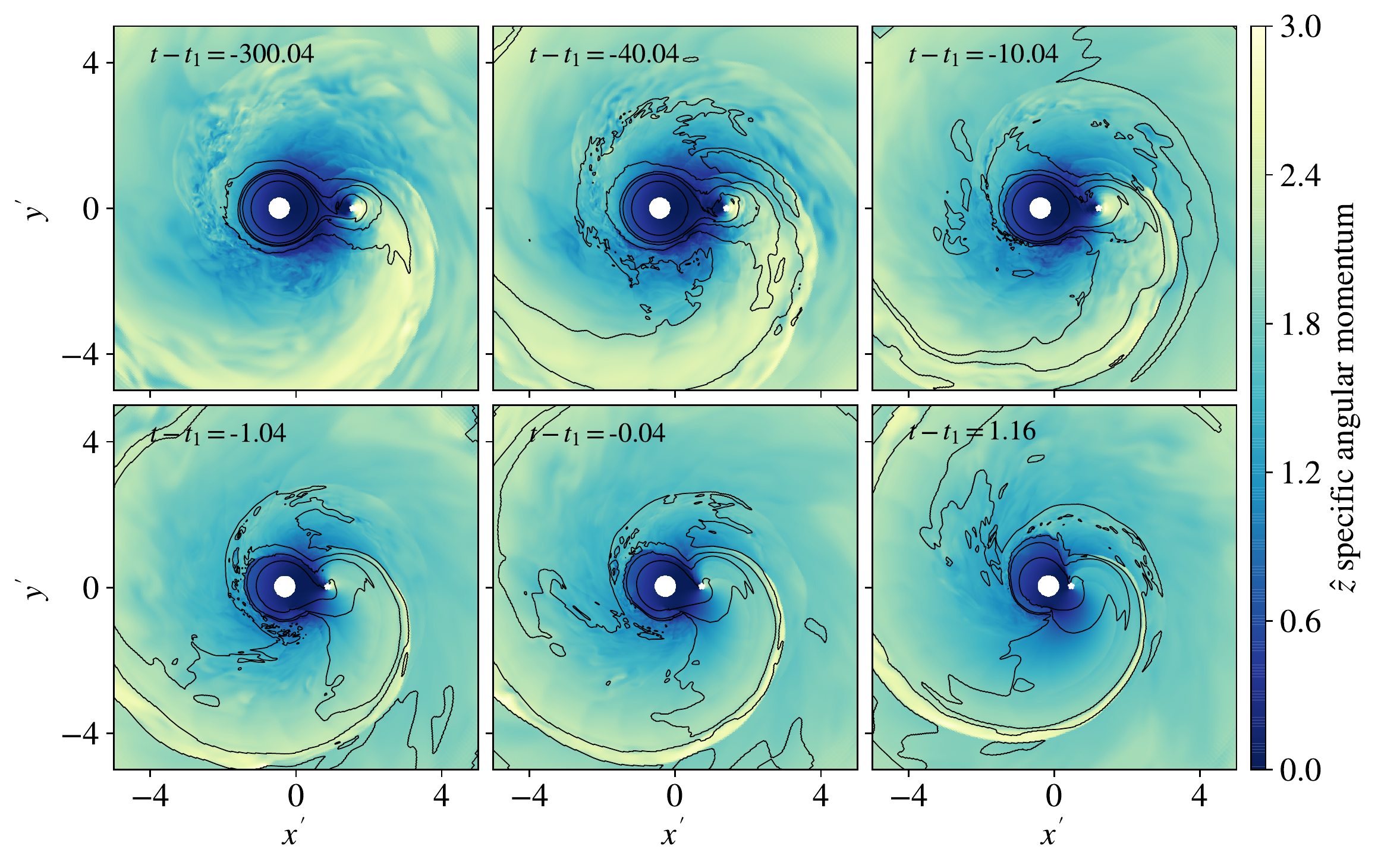}
\caption{Specific angular momentum of material in the orbital midplane, at the same times and rotated $x'$-$y'$ frame as Figure \ref{fig:ts_dens}. Contours show decades of density from $10^{-5}$ to $10^{-1}$~$M_1/R_1^{-3}$. In each frame, material is drawn from low specific angular momentum near the donor to higher specific angular momentum near the accretor. In the earliest snapshot, material is trailed off with relatively high specific angular momentum, roughly from the outer Lagrange point ($L_2$). As the separation tightens slightly in the subsequent snapshots the specific angular momentum of ejecta decreases, and we see an associated broadening of the outflow stream. In the lower panels at times near binary coalescence, even as the orbit shrinks rapidly, the specific angular momentum of ejecta are approximately constant as the evolving outflow morphology approximately compensates for the tightening orbit.    }
\label{fig:ts_angmom}
\end{center}
\end{figure*}

As outlined in Section \ref{sec:analytic}, the determining factor in orbital evolution with mass loss is not just the rate at which mass is lost, but the angular momentum content it carries with it. Here we examine the specific angular momentum of material lost from the donor star's envelope. We begin in Figure \ref{fig:ts_angmom} by mapping the specific angular momentum of material in the orbital plane, in the same series of snapshots as Figure \ref{fig:ts_dens}. A few useful features emerge from this diagram, particularly when compared to the overplotted density contours. Across all the snapshots, we see that material making up the bulk of the donor envelope has relatively low specific angular momentum, typically less than $(G M_1 R_1)^{1/2}$. Material streaming from the donor toward the accretor acquires specific angular momentum through that gravitational interaction. This transition can be observed in the gradient of specific angular momenta leading up to the accretor position. Finally, some material trails off from the the binary, especially near the outer Lagrange point, $L_2$. This outflow contains the highest specific angular momenta in the slices.

\begin{figure}[tbp]
\begin{center}
\includegraphics[width=0.48\textwidth]{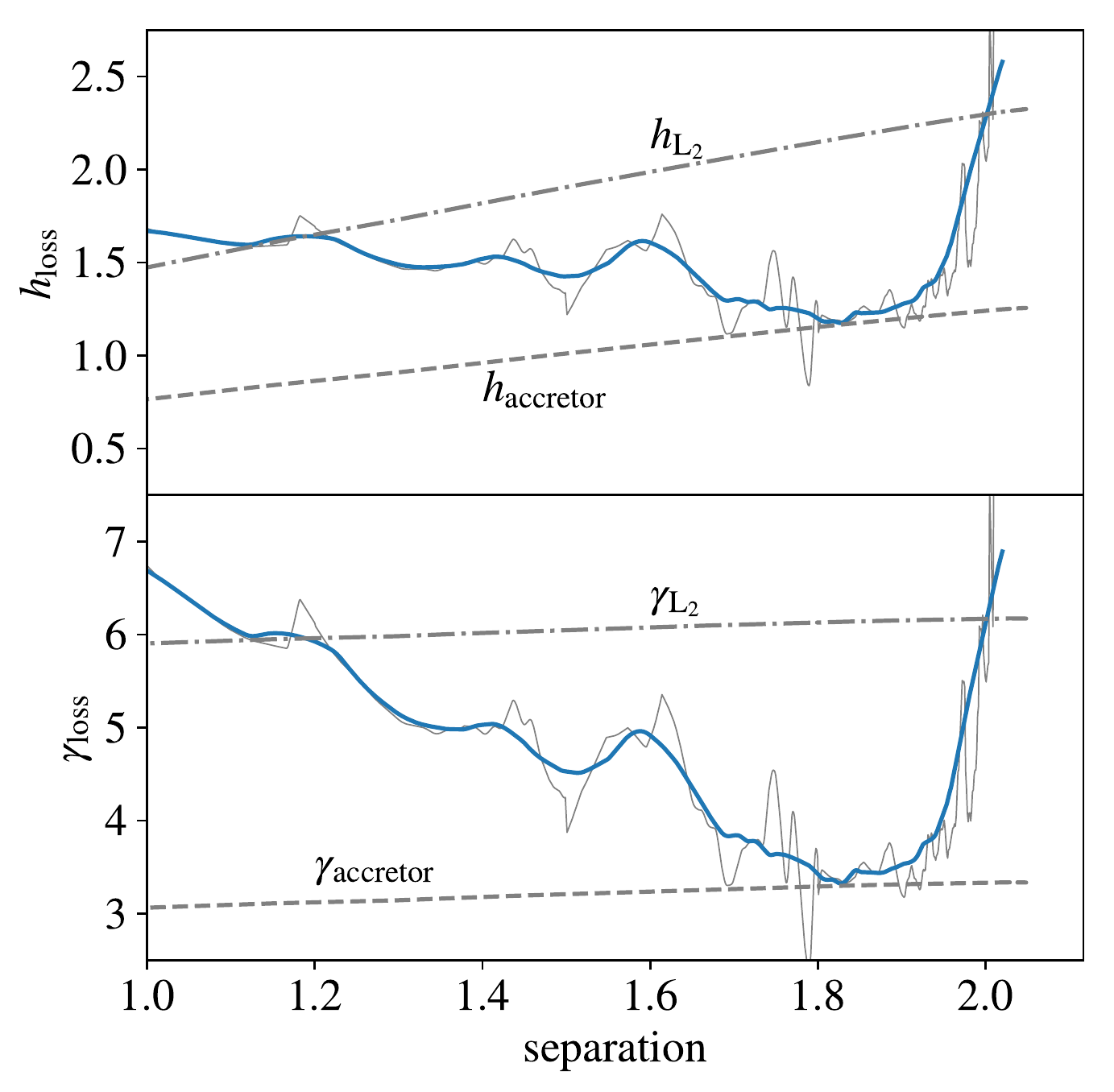}
\caption{Specific angular momentum, $h_{\rm loss}$, and dimensionless specific angular momentum, $\gamma_{\rm loss}$, as a function of binary orbital separation. The solid blue line shows smoothed simulation results, as in Figure \ref{fig:masstransfer}, and the dashed lines show reference quantities for context. Material lost from the binary system carries specific angular momentum similar to the $L_2$ outer Lagrange point at the widest separations. This reduces to near the specific angular momentum of the accretor at a separation of approximately $1.8R_1$, but then increases even as the orbit tightens, exceeding the angular momentum of even the $L_2$ point at a separation of $R_1$. Details on the calculation of these quantities are given in the text.  }
\label{fig:specificmomgamma}
\end{center}
\end{figure}

We quantify the changing specific angular momentum distributions that can be seen in Figure \ref{fig:ts_angmom} further in  Figure \ref{fig:specificmomgamma}, where we estimate the average specific angular momentum of material lost as a function of binary orbital separation.  We also compute the dimensionless quantity $\gamma_{\rm loss}$, which relates the specific angular momentum to the specific angular momentum of the binary; see equation \eqref{orbit_analytic}. 
The average specific angular momentum of material carried away from the binary system is computed as $h_{\rm loss} \approx  - \dot L_{\rm gas}/\dot M_{\rm donor}$, where $\dot L_{\rm gas}$  is the $\hat z$ component of the gas angular momentum about the center of mass, shown in Figure \ref{fig:angmom}. We use smoothed versions of the mass transfer rate and angular momentum transfer rate, as described in Figure \ref{fig:masstransfer}. The specific angular momenta of the accretor and $L_2$ points are also shown for context. 

The lower panel of Figure \ref{fig:specificmomgamma} shows the conversion of the same diagram into the unitless quantity used to describe orbital evolution in equation \eqref{orbit_analytic}, $\gamma_{\rm loss} = h_{\rm loss} M_{\rm bin}/L_{\rm bin}$ (the specific angular momentum of the material lost in units of the specific angular momentum of the binary).  To make this conversion, we must approximate $M_{\rm bin}$ and $L_{\rm bin}$, even when the system is coalescing, and the separation of mass into two stars is increasingly unclear.   In Figure \ref{fig:specificmomgamma}, we adopt the mass within $r=1R_1$ as the instantaneous value $M_{\rm d}$, and we treat the angular momentum as the angular momentum of a circular binary at the current separation. 

Despite the approximate nature described above of the quantities displayed in Figure \ref{fig:specificmomgamma},  useful trends emerge. Near the Roche limit, material is trailed off from the binary with specific angular momentum similar to, or slightly larger than, the angular momentum of the outer Lagrange point \citep[e.g.][]{1979ApJ...229..223S,1998CoSka..28..101P}. As the binary tightens, we see that $h_{\rm loss}$ and $\gamma_{\rm loss}$ drop, reaching minima that are nearly a factor of two lower near a separation of $1.8R_1$ or roughly 87\% of the Roche limit separation. As the binary trends toward tighter separations, we observe that the specific angular momenta of the escaping material actually increases, despite the tightening binary. Relative to the binary's angular momentum, as shown with $\gamma_{\rm loss}$, this implies an even steeper increase.  As might be inferred from Figure \ref{fig:ts_angmom}, the change in specific angular momenta of mass loss from the binary is reflected in the evolving morphology of outflows from the binary pair, a point to which we will return in Section \ref{sec:flow}.

\begin{figure}[tbp]
\begin{center}
\includegraphics[width=0.48\textwidth]{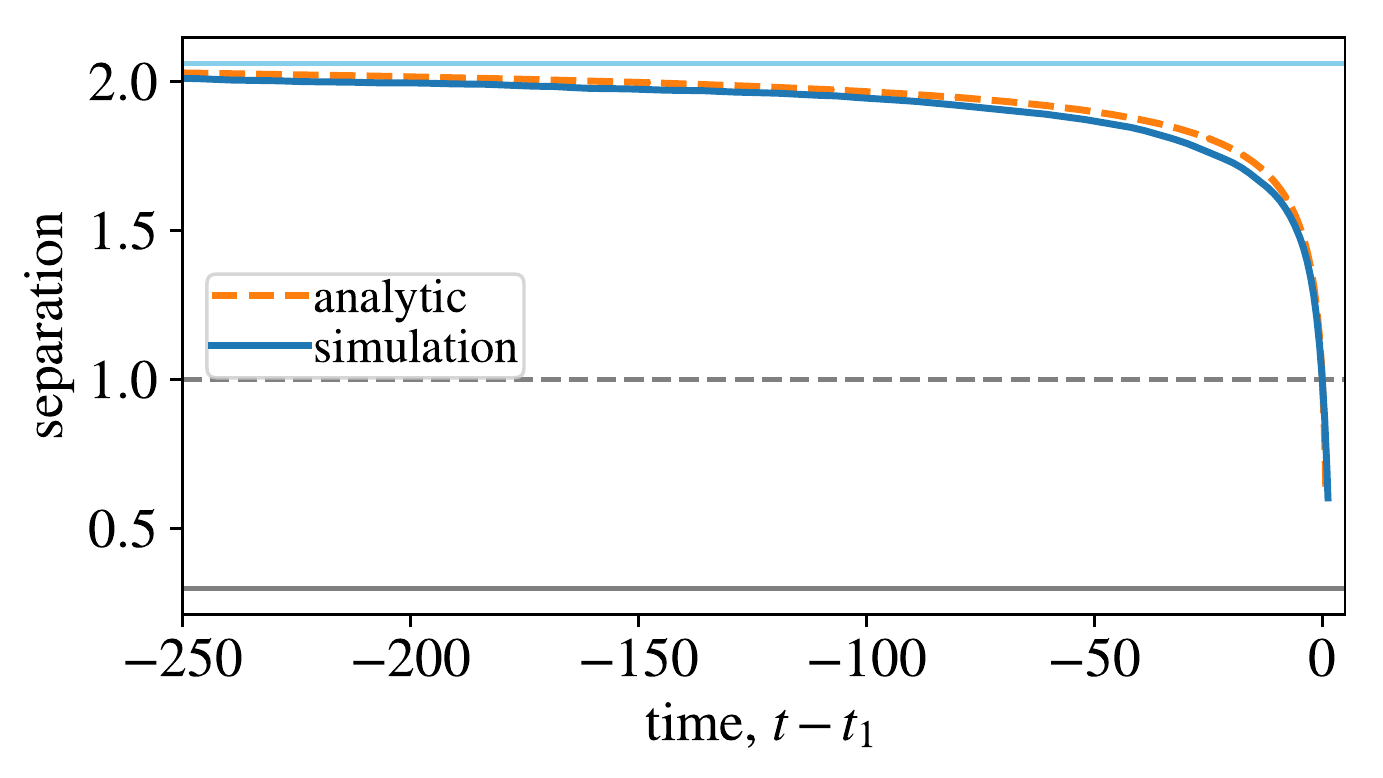}
\caption{Reconstructed orbital evolution from the analytic expression of equation \eqref{orbit_analytic} compared to the simulation result. In the analytic expression, we adopt the parameters described in Figure \ref{fig:masstransfer} for the mass-loss rate and the specific angular momentum of $L_2$ for material lost. With these choices adopted, the analytic description very successfully reproduces the runaway orbital evolution observed in the simulation.   }
\label{fig:sepanalytic}
\end{center}
\end{figure}

In the figures and discussion above, we have measured each of the parameters one needs to assign to integrate the parameterized differential equation of orbital evolution under the influence of mass loss from a binary, equation \eqref{orbit_analytic}. We use these ingredients from the simulation to reconstruct the orbital evolution of the binary and compare to our full simulation result in Figure \ref{fig:sepanalytic}. In particular, we adopt the analytic mass transfer rate and associated constants as plotted in Figure \ref{fig:masstransfer}, equation \eqref{mdot_analytic}. Motivated by the results of Figure \ref{fig:specificmomgamma}, we apply $\gamma_{\rm loss}\approx \gamma_{\rm L_2}$.  The resulting orbital separation evolution is strikingly similar to the simulation result. 

The similarity of these results offers insight into the physical mechanism of runaway coalescence in the binary system. The fact that the mass-loss rate from the donor is an extremely steep function of separation combined with the fact that the typical angular momentum of ejecta is large -- similar to that of the $L_2$ outer Lagrange point -- indicates that runaway mass loss leads to runaway orbital decay.

\subsubsection{Driving Orbital Decay: Torques}\label{sec:torque}

\begin{figure*}[tbp]
\begin{center}
\includegraphics[width=0.95\textwidth]{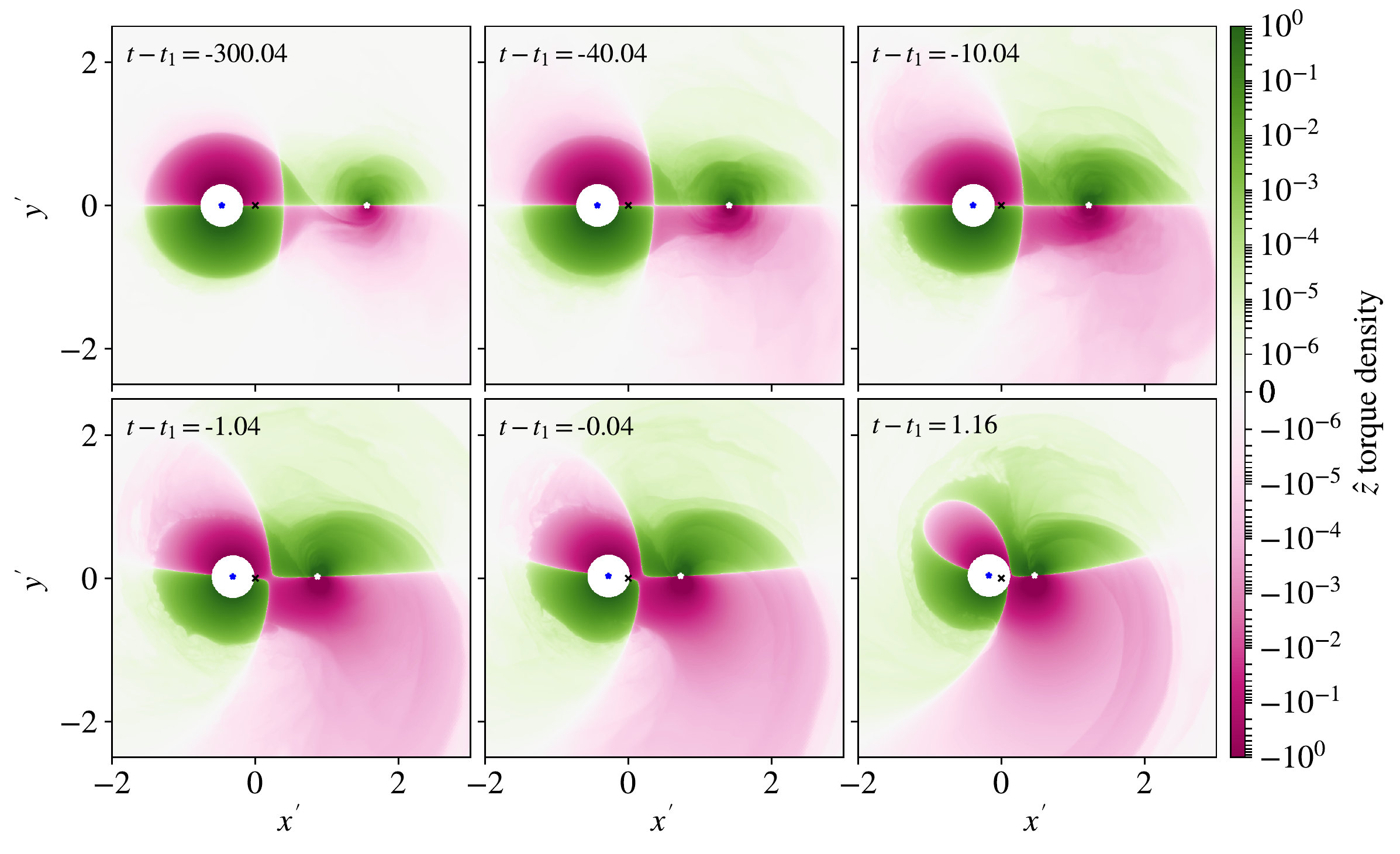}
\caption{Torque volume density in slices through the rotated ($x'$-$y'$) orbital plane (see Figure \ref{fig:ts_dens}). Positive torques indicate material pulling the binary components forward in their orbit; negative torques indicate material pulling backward. An imbalance in positive and negative torques generates a negative net torque on the system, which drives it, increasingly, toward coalescence. In later snapshots, $t\sim t_1$, as the donor's envelope is distorted, the particle-gas system center of mass (black cross) moves away from the line of centers between the two particles, generating curved contours of positive and negative torque. At late times, the gas distribution around the accretor is highly asymmetric, leading to the net negative torque that drives orbital decay.    }
\label{fig:ts_torque}
\end{center}
\end{figure*}

In this section, we turn to the gravitational forces within the binary system in order to understand how  low specific-angular-momentum gas near the donor is pulled away and then expelled with higher specific angular momentum in the vicinity of the accretor. The backreaction of these forces drives the orbital evolution of the pair of stellar cores within their gaseous surroundings. 

A key quantity is the torque on the pair of particles, which describes their rate of change of angular momentum. 
The gas distribution applies a gravitational force to the two particles; the force per unit volume on particle $i=1,2$ from a parcel of fluid with density $\rho$ is
\beq
{{\bf F}_{{\rm grav},i} \over {\rm Vol}} = \frac{G m_i \rho}{|{\bf r}-{\bf r}_i |^3} ({\bf r}-{\bf r}_i ).
\eeq
 The torque on particle $i$ relative to the system's center of mass as a result of this force also depends on its lever arm relative to the center of mass, 
 \beq\label{torquedens}
 {{\bf \tau}_{{\rm grav},i} \over {\rm Vol}} = ({\bf r}_i-{\bf r}_{\rm com}) \times ({\bf r}-{\bf r}_i ) \frac{G m_i \rho}{|{\bf r}-{\bf r}_i |^3} .
 \eeq
Integrating equation \eqref{torquedens} over volume yields the net torque on particle $i$. The net torque on the binary is the sum of the torques on each component, $\tau_{\rm grav} = \tau_{\rm {grav},1} + \tau_{\rm grav,2} = \dot L_{\rm particle}$.

In Figure \ref{fig:ts_torque}, we map out the $z$-component of torque density on particles 1 and 2 relative to the particle-gas center of mass (torque per unit volume)\footnote{See \citet{2017MNRAS.469.4258T} for a similar analysis in the context of disk accretion onto binary supermassive black holes.} We slice in the orbital midplane for the same snapshots as shown in Figures \ref{fig:ts_dens} and \ref{fig:ts_angmom}. 
The color bar in Figure \ref{fig:ts_torque} includes positive and negative values. Gaseous material colored in green is pulling the stellar cores forward in their orbital motion. By contrast, material colored in pink drags the system backward. Because the net torque (and thus the rate of change of particle angular momenta -- as shown in Figure \ref{fig:angmom}) is given by an integral over the full volume, much of the positive and negative torques within the domain cancel out. If the gas distribution (and quantity of pink versus green material) is perfectly symmetric, there is no net torque. 

Figure \ref{fig:ts_torque} shows increasing asymmetry of positively and negatively torquing material as the binary advances to smaller separations. This asymmetry is particularly pronounced about the accretor.  Looking at the magnitude of torque density, we see strongly increasing magnitudes of torque densities near the accretor as the mass flux through the binary system increases. Both increased asymmetry and larger magnitudes, therefore, contribute to the increasing net torque, and accelerating binary coalescence.

\begin{figure}[tbp]
\begin{center}
\includegraphics[width=0.48\textwidth]{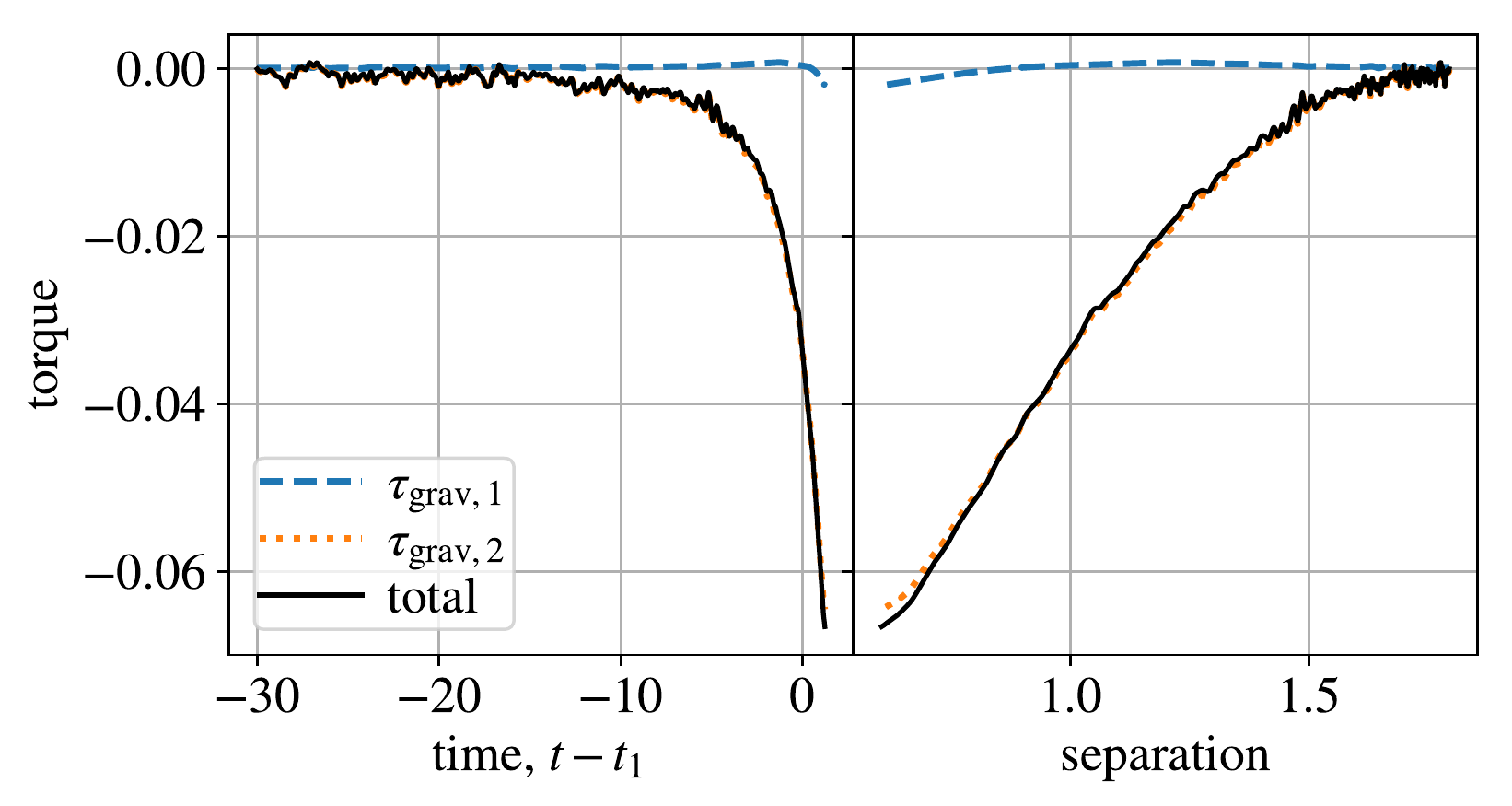}
\caption{Integrated torques from the gas distribution on the two particles as a function of time and separation. The majority of the net torque is applied to the accretor particle. }
\label{fig:torque_time_sep}
\end{center}
\end{figure}

\begin{figure}[tbp]
\begin{center}
\includegraphics[width=0.48\textwidth]{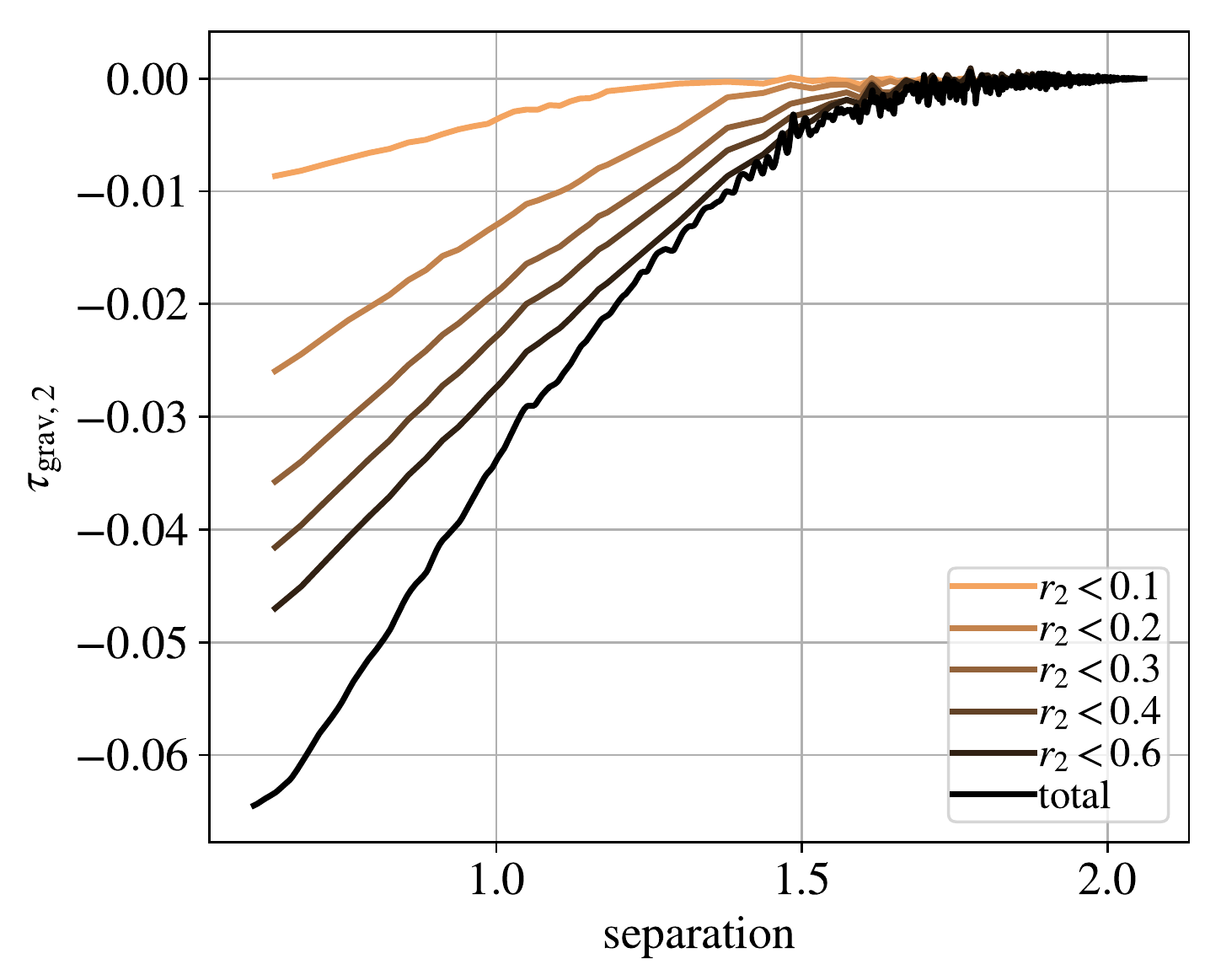}
\caption{Contribution to torque on accretor particle, 2, from within radii from the accretor between 0.1 and 0.6$R_1$. The bulk of the torque on the accretor, and therefore also on the binary system, is generated in the vicinity of the accretor.  }
\label{fig:torque2}
\end{center}
\end{figure}

Inference of the net torque from Figure \ref{fig:ts_torque} requires an ``integral by eye'' and does not include material above and below the orbital midplane. In Figures \ref{fig:torque_time_sep} and \ref{fig:torque2} we integrate the net torques and separate them into their components.  We begin our analysis of net torques with Figure \ref{fig:torque_time_sep}, which shows the $z$-component of the gravitational torque on particles 1 (donor core) and 2 (accretor) from the gas distribution with respect to the full system's center of mass. The panels of this figure show the torque with respect to time (left panel) and orbital separation (right panel).

There are several immediate conclusions to draw from Figure \ref{fig:torque_time_sep}. First, the magnitude of the net torque is much smaller than the maximum torque densities represented in Figure \ref{fig:ts_torque}. This indicates that the cancelation of positively and negatively torquing gas is relatively efficient. Second, the majority of the net torque is on the accretor particle (2). Referencing the midplane maps of torque density in Figure \ref{fig:ts_torque}, we see that this is not because of lack of torquing material near the donor, but because the donor envelope remains relatively symmetric for much of the coalescence and thus generates little net torque. 
Finally, since the net torque represents the rate of change of angular momentum, we see the imprint of the runaway orbital decay here as well. As the binary separation tightens, the torque increases, indicating an ever more rapid runaway of orbital decay.

In Figure \ref{fig:torque2}, we examine the spatial distribution of torques on the accretor particle versus binary separation. In addition to the net torque on particle 2, $\tau_{\rm grav,2}$, we show the torque within spherical regions centered on the accretor particle of radii between 0.1 and 0.6$R_1$. This decomposition indicates that much of the net torque on the accretor particle (and therefore also on the binary as a whole) originates in its immediate vicinity.  By comparison to Figure \ref{fig:ts_torque}, we see that this is the region that encompasses much of the asymmetry in positively and negatively torquing gas. A similar conclusion is reached by \citet{2017MNRAS.469.4258T} in studying disk accretion onto binary black holes -- significant torques are generated in the immediate vicinity of the black holes -- implying that resolving these scales is important.  As the separation decreases to be similar to $R_1$, a larger fraction of the torque is generated outside $0.6R_1$, by the broad fan of outflowing ejecta. Nonetheless, even in this most extreme case, the majority of the torque does not come from the largest-scale flow distribution. 

The trend of the binary system toward runaway merger, studied in the orbital properties in Section \ref{sec:orbit} and in the angular momentum budget in Section \ref{sec:angmom}, is a consequence of the forces acting on the stellar cores due to the distribution of gas originating from the donor envelope. As shown in Section \ref{sec:massflow}, a dramatically increasing flux of mass transfers from the donor to the vicinity of the accretor. Through gravitational torques originating primarily near the accretor particle, this material acquires specific angular momentum similar to (or somewhat larger than) that of the accretor particle (Section \ref{sec:massflow}), and is expelled from the binary. This accelerating loss of angular momentum leads to the runaway plunge of the accretor within the envelope of the donor and the onset of the common envelope phase of the binary system.

%%%%%%%%%%%%%%%%%%%%%%%%%
%  Flow
%%%%%%%%%%%%%%%%%%%%%%%%%
\subsection{Changing Flow Morphology during Coalescence}\label{sec:flow}

We have noted above a changing outflow distribution as the binary orbit tightens, seen particularly clearly in Figure \ref{fig:ts_dens}. Here we examine this changing morphology in more detail and trace its origin to the desynchronization of the orbital motion with donor-star rotation.

\subsubsection{Evolving Flow Morphology}

\begin{figure*}[tbp]
\begin{center}
\includegraphics[width=0.95\textwidth]{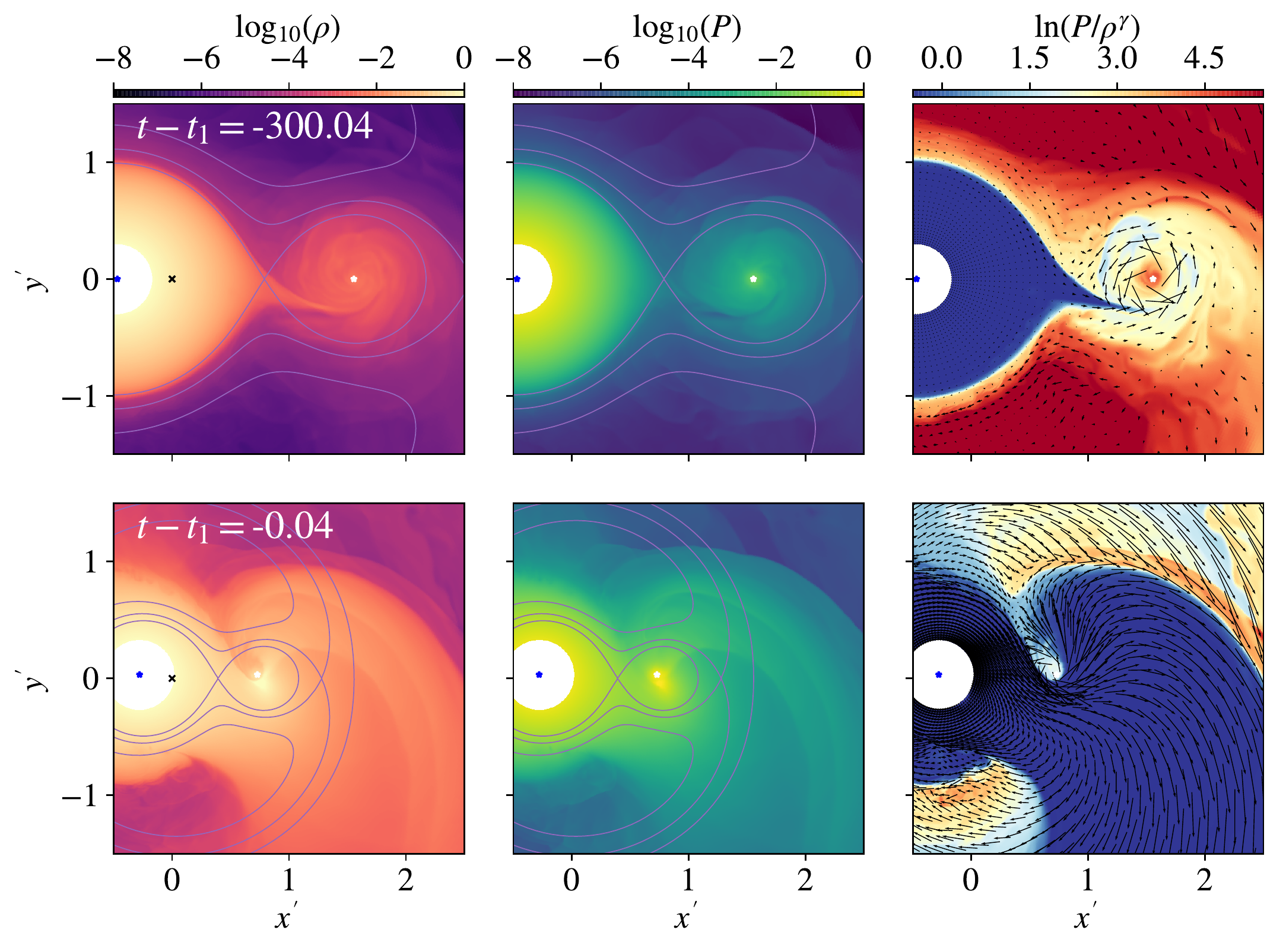}
\caption{A comparison of flow properties at early and late stages in the binary coalescence in the $x'$-$y'$ rotated orbital plane. The left panels show density, the center pressure, and the right specific entropy. The left and center panels overplot the instantaneous Roche lobes, while the right adds velocity vectors in the instantaneously corotating frame. A dramatic transformation in the morphology of flow through and outward from the binary can be seen in this comparison. At early times, material in the donor's envelope is at rest in the corotating frame, and transfers to the accretor in a thin stream. Later, material in the donor's envelope is in motion in the corotating frame, and fans outward following its interaction with the accretor.  }
\label{fig:earlylate}
\end{center}
\end{figure*}

We examine the transformation of flow properties in Figure \ref{fig:earlylate}.  The upper panels show an early snapshot, approximately 300 dynamical times prior to coalescence; the lower panels show a snapshot near $t_1$, when the accretor is plunging within the envelope of the donor. The left panels show density, the center panels show pressure, and the right panels show gas specific entropy. In the left and center panels we overlay the instantaneous Roche lobe. The Roche lobes are approximated by assuming the full donor mass $M_1=1$ and are shown centered on the system's center of mass, with contours corresponding to the potential of the $L_1$, $L_2$, and $L_3$ Lagrange points, respectively -- clearly, these are a more meaningful guide in the early snapshot in which the binary is still separated.  In the right panels, we plot gas velocities in the instantaneously corotating frame. 

The biggest distinction in flow morphology between early and late snapshots comes in the nature of gas flow between and outward from the two binary components. In the early frame (upper panels of Figure \ref{fig:earlylate}), the highest density gas is largely confined to the Roche lobes of the two stars, and gas density and pressure fall off roughly tracing contours of equipotential. A relatively thin stream of stellar material transfers from the donor to the accretor. This material assembles into a rotating flow with a high-pressure and high-specific-entropy core within the accretor-particle softening radius.  Surface layers of higher-specific entropy gas surround both stars, similar to the description of \citet{1976ApJ...209..536S}. Material outflowing from the binary has a relatively high specific entropy and exits near the $L_2$ point. 

In the later snapshot  (shown in the lower panels of Figure \ref{fig:earlylate}), material is flung outward in a broad fan of ejecta \citep[see also Figure 1 of][]{1996ApJ...471..366R}. The width of this fan is similar to its size. Comparison with Figure \ref{fig:ts_dens} shows that that the ejecta also fan out in opening angle in the perpendicular plane, and that the leading edge of this material quickly catches the trailing edge. Material is unconfined by the instantaneous Roche lobes of the pair, and its specific entropy shows that it is largely unshocked or mildly shocked in its ejection, by contrast to the much higher entropies observed near $L_2$ in the early snapshot. 

\subsubsection{Desynchronization}

A crucial clue  to understanding the transformation of flow morphology lies in the velocity vectors plotted in Figure \ref{fig:earlylate}. Early in the simulation, material within the donor envelope is nearly at rest in the instantaneously corotating frame, and it acquires velocity in falling toward the accretor. The later snapshot shows that the donor envelope is in motion in the corotating frame. Although the donor star started out in synchronous, solid-body rotation with the orbit, orbital decay has desynchronized the angular frequencies of the orbit and the donor's envelope. In the corotating frame, donor envelope material appears to counterrotate. It is ejected on a range of trajectories that seem to depend primarily on impact parameter relative to the accretor.

\begin{figure}[tbp]
\begin{center}
\includegraphics[width=0.49\textwidth]{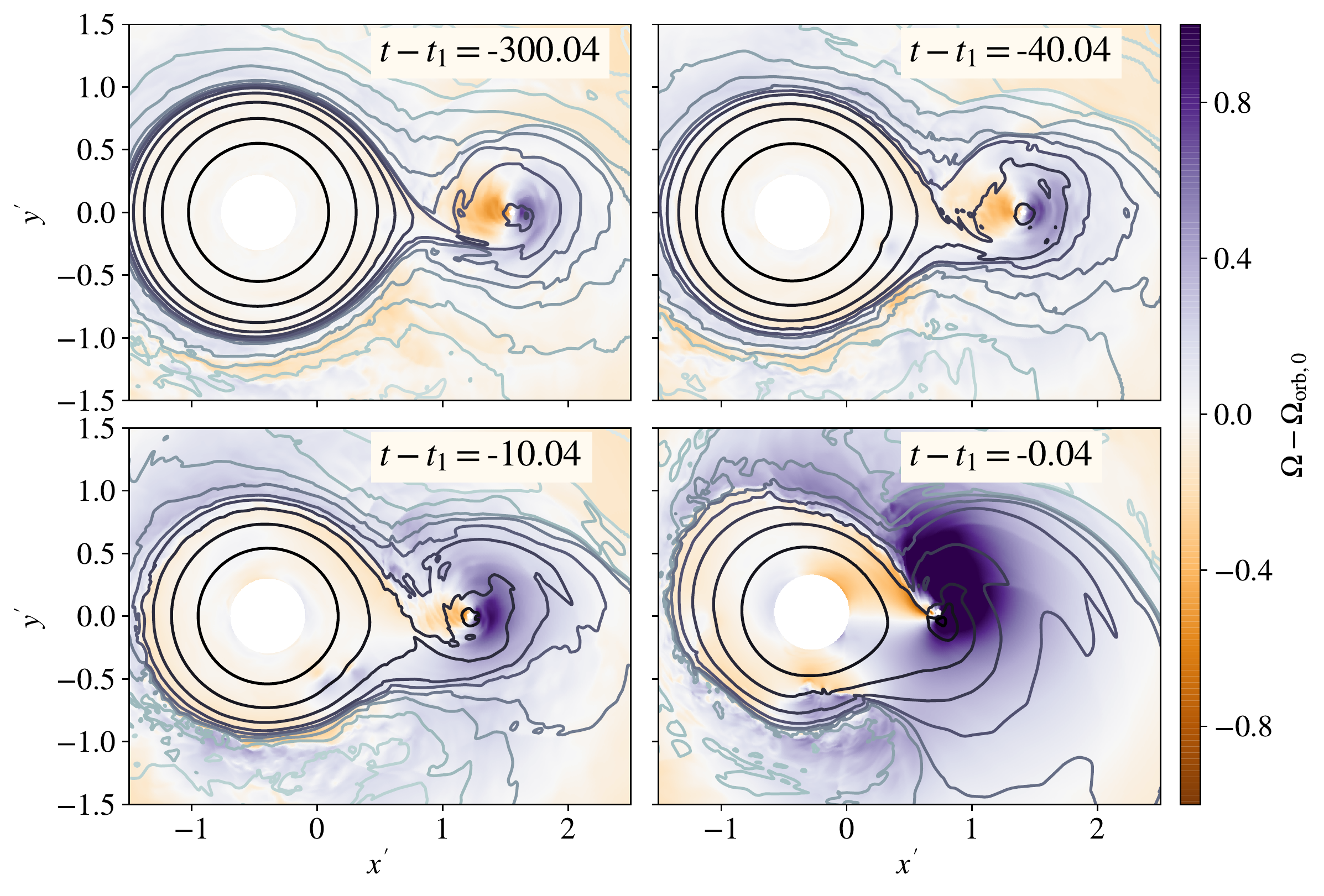}
\includegraphics[width=0.49\textwidth]{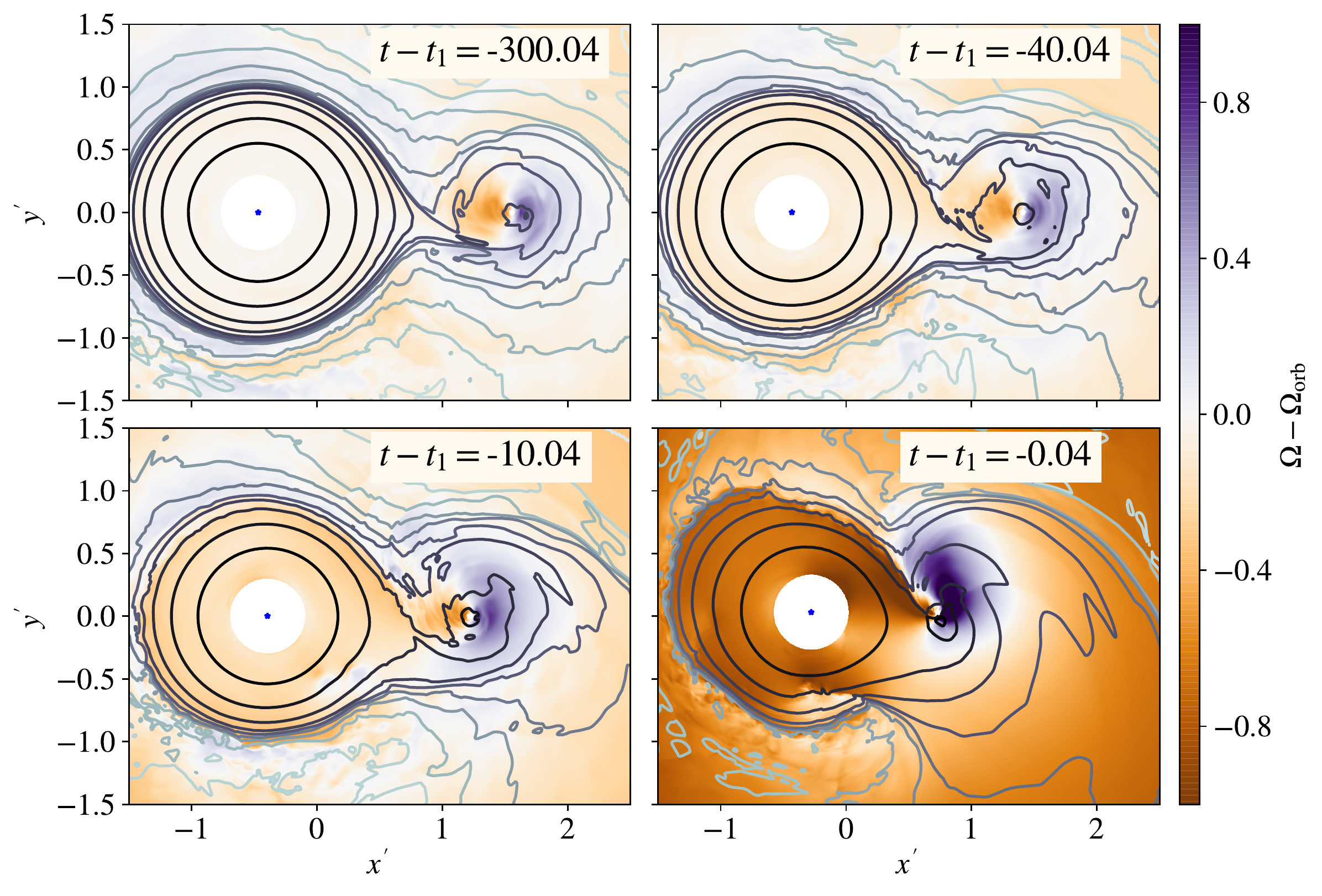}
\caption{ Angular frequency of material about the donor-star origin, $\Omega = v_\phi / (r\sin \theta)$, relative to the initial, $\Omega_{\rm orb,0}$, or instantaneous, $\Omega_{\rm orb}$, orbital velocity in the rotated orbital midplane. Contours show density in half-dex intervals. The gas is initialized in synchronous rotation with the orbit, but desynchronizes with the orbit over time: the gas retains close to its initial rotation even as the orbit shrinks and the orbital frequency increases. Thus, relative to the instantaneous orbital frequency (lower panel), material in the donor envelope appears increasingly counter-rotating.  }
\label{fig:omegamap}
\end{center}
\end{figure}

Figure \ref{fig:omegamap} explores desynchronization of orbit and donor envelope further through mapping the angular velocity of material in the orbital plane. The color scale shows the material's angular velocity about the donor, $\Omega = v_\phi/(r \sin \theta)$, relative to the initial ($\Omega_{\rm orb,0}$, upper panel), or instantaneous, ($\Omega_{\rm orb}$, lower panel) orbital angular velocity.  In the earliest snapshots, we observe that the donor envelope starts in synchronous rotation with the binary orbit. The upper panels show that the angular velocity of material in the orbital plane is largely unchanged, even as the orbit decays. With the exception of strong flow near the accretor, the primary feature in angular velocity maps at the time of coalescence is a quadrupolar perturbation of the velocity field, due to the gravitational pull of the donor. 
When we  shift our reference frequency to that of the instantaneous orbital frequency (lower panels), Figure \ref{fig:omegamap} shows donor envelope material counterrotating in the instantaneous orbital frame. 

\begin{figure}[tbp]
\begin{center}
\includegraphics[width=0.48\textwidth]{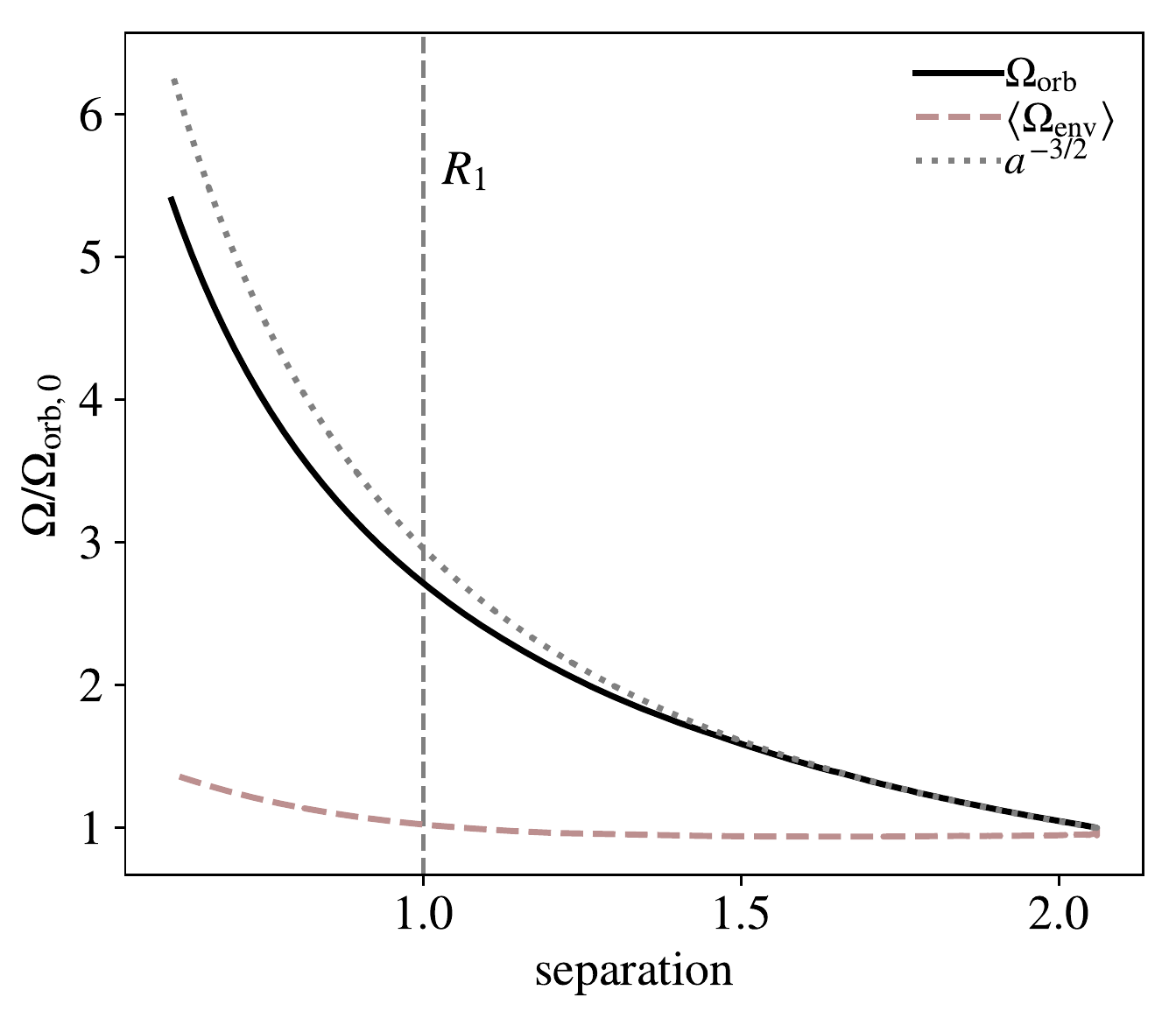}
\caption{Angular frequency of orbital motion and donor star gaseous envelope rotation. The envelope gas initial rotates at the orbital frequency at the Roche limit. As the separation decreases, the orbital frequency goes up, roughly as $a^{-3/2}$, but the envelope still rotates at approximately its original frequency, increasing slightly only when $a<R_1$. Orbital angular momentum lost during orbital decay does not go into spinning up the envelope prior to coalescence. This implies increasing desynchronization of the accretor and the neighboring envelope gas, such that by the time $a\approx R_1$, the secondary skims across the surface of the donor star's envelope.}
\label{fig:omegasep}
\end{center}
\end{figure}

By remaining close to its original rotation rate, the donor envelope loses its initial synchronization with the shrinking orbit. Such behavior is qualitatively described as a ``loss of corotation" by \citet{2001ASPC..229..239P} and \citet{2013A&ARv..21...59I}. Figure \ref{fig:omegasep} shows this desynchronization quantitatively. We compare the orbital angular frequency to the donor envelope angular frequency as a function of orbital separation. To do so, we compute the mass-averaged angular velocity of envelope material, defined crudely as material with $\rho>10^{-3} M_1/R_1^3$. We observe that the orbital angular frequency increases as the separation shrinks over the course of the simulation. This behavior is close to the $a^{-3/2}$ scaling of two point masses, but is slightly shallower because of the decreasing enclosed mass within the orbital separation (both due to mass ejection and as $a<R_1$).  This increasing orbital frequency is not matched by the angular frequency of the envelope, which is relatively constant.  Therefore, while the system begins mass transfer in synchronous rotation, this state is rapidly lost as the orbital separation shrinks toward $a\approx R_1$.

\subsubsection{Outflow Morphology in Desynchronized System: Comparison to Ballistic Trajectories}

\begin{figure*}[tbp]
\begin{center}
\includegraphics[width=0.9\textwidth]{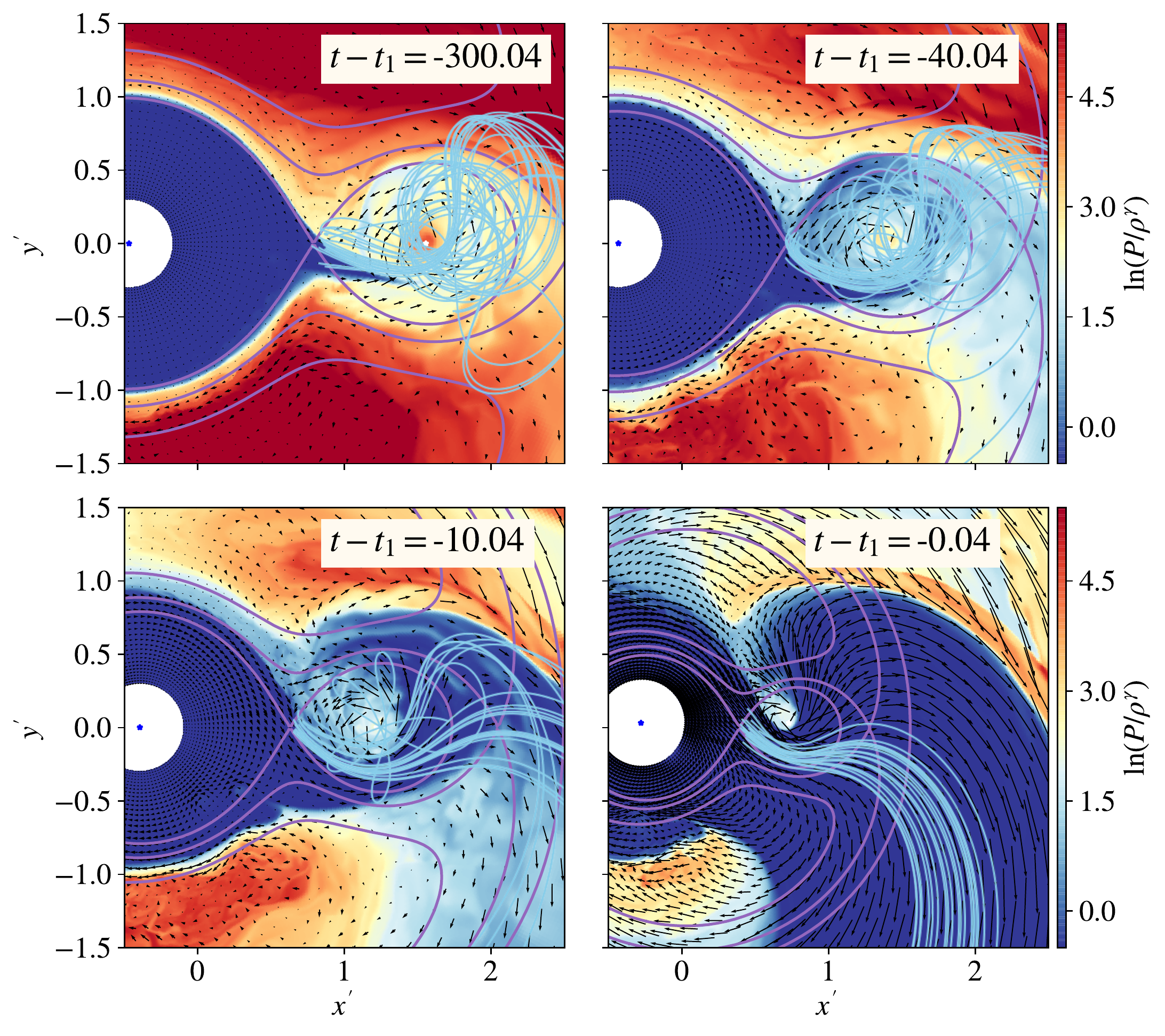}
\caption{Comparison of ballistic trajectories in binary potential to the gas flow in the rotated, $x'$-$y'$, orbital midplane. We plot randomly sampled free-particle trajectories in the corotating frame on top of gas specific entropy (color) and velocity in the corotating frame (vectors). Free trajectories are initialized from the flow near the $L_1$ point. Free trajectories mirror the evolving flow morphology in the binary. Early on, ballistic trajectories are confined to the accretor's Roche lobe. In later snapshots, desynchronization of the orbital motion relative to the donor's rotation implies that material crossing from donor to accretor is flung out in a broad fan that closely mirrors free trajectories.  }
\label{fig:traj}
\end{center}
\end{figure*}

The loss of synchronization between donor rotation and orbit carries implications for gas crossing from the donor toward the accretor, and as a consequence, for the morphology of outflow from the binary. The manner in which material overflows from the binary is particularly affected. Figure \ref{fig:traj} compares ballistic trajectories integrated in the potential of point masses of mass $M_1=1$ and $M_2=0.3$ in circular orbit at the orbital separation with the fluid flow (showing specific entropy in color scale and velocity with vectors).  Ballistic trajectories are initialized at randomly sampled positions near the $L_1$ point, with velocities interpolated from the simulation mesh. 

In early snapshots, Roche lobe overflow leads to the formation of rotating structures around the accretor. Free trajectories self-intersect within the Roche lobe. In the gas dynamic calculation, gas streams collide and circularize, increasing the specific entropy of the gas through shocks. In either case, as discussed in the context of Figure \ref{fig:earlylate}, material is largely confined by the Roche contours of equipotential. Material that is lost is trailed off from the $L_2$ point after being shock-heated to higher  entropy and pushed over the potential barrier by the pressure gradient away from the accretor.  

At times approaching $t_1$, the stream of gaseous ejecta becomes broader and also progressively lower entropy.  Relatedly, an increasing number of free trajectories exit the Roche lobes. By $t\approx t_1$ all of the free trajectories are lost and carried away in the tail, and the free particle streamlines nearly mirror the gas velocities.  A similar flow morphology can be observed in ``wind tunnel'' simulations that assume complete desynchronization and model the supersonic flow of material with a steep density gradient past the accretor \citep{2015ApJ...803...41M,2017ApJ...838...56M,2017ApJ...845..173M}. Of particular relevance here is the role of the steep density gradient near the donor-star limb -- as dense gas is focused and redirected outward along ballistic trajectories, it encounters little impediment to its expansion \citep{2015ApJ...803...41M,2017ApJ...838...56M}. 

These features indicate that the mode of mass shedding from the binary  transforms as the binary desynchronizes: from being pushed out by pressure gradients to being flung out along ballistic trajectories. 
This changing mode of mass loss additionally leads to the change in outflow morphology from a narrow stream to a broad fan.

\subsection{Non-synchronous Initial Rotation}

\begin{figure}[tbp]
\begin{center}
\includegraphics[width=0.48\textwidth]{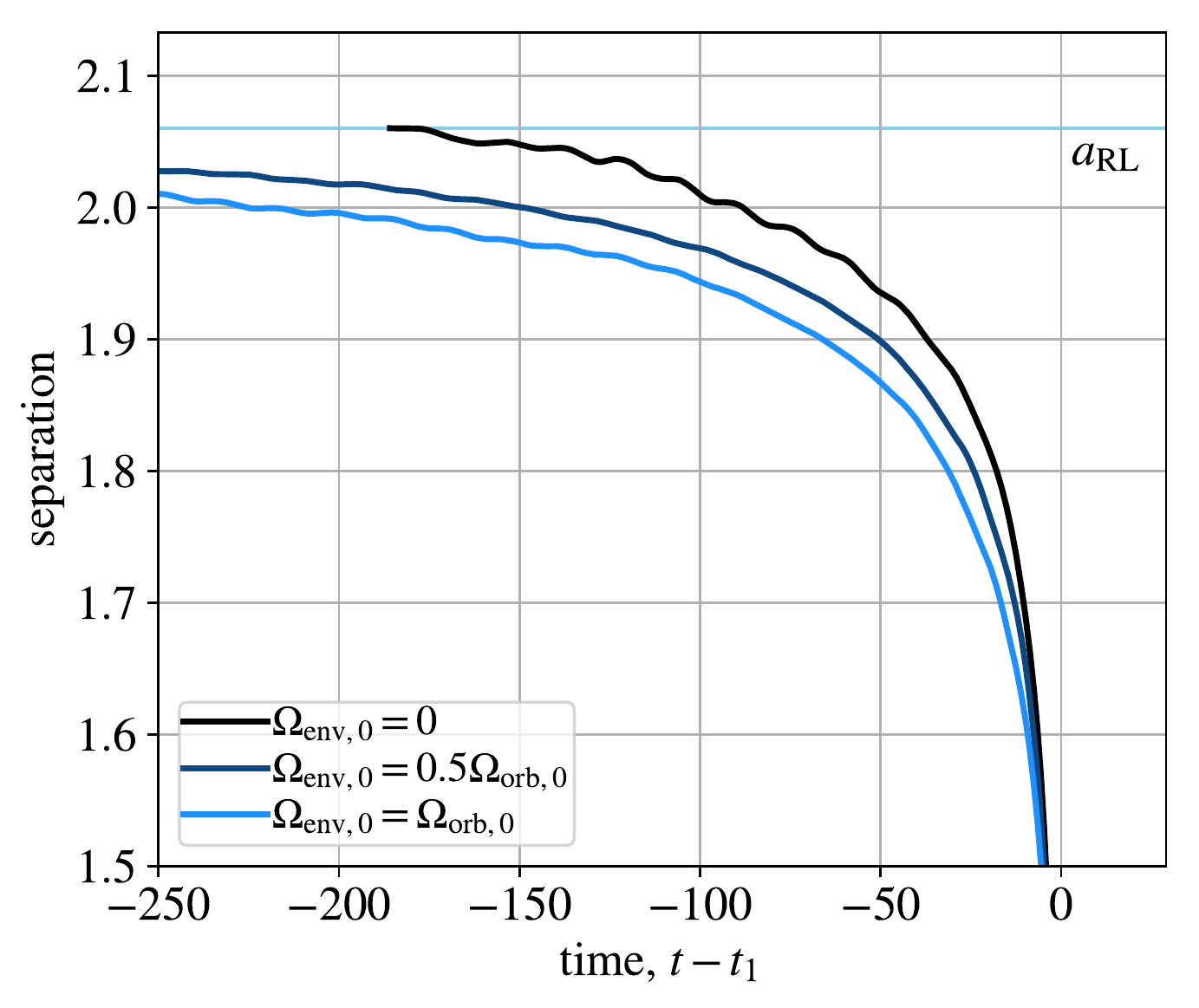}
\includegraphics[width=0.48\textwidth]{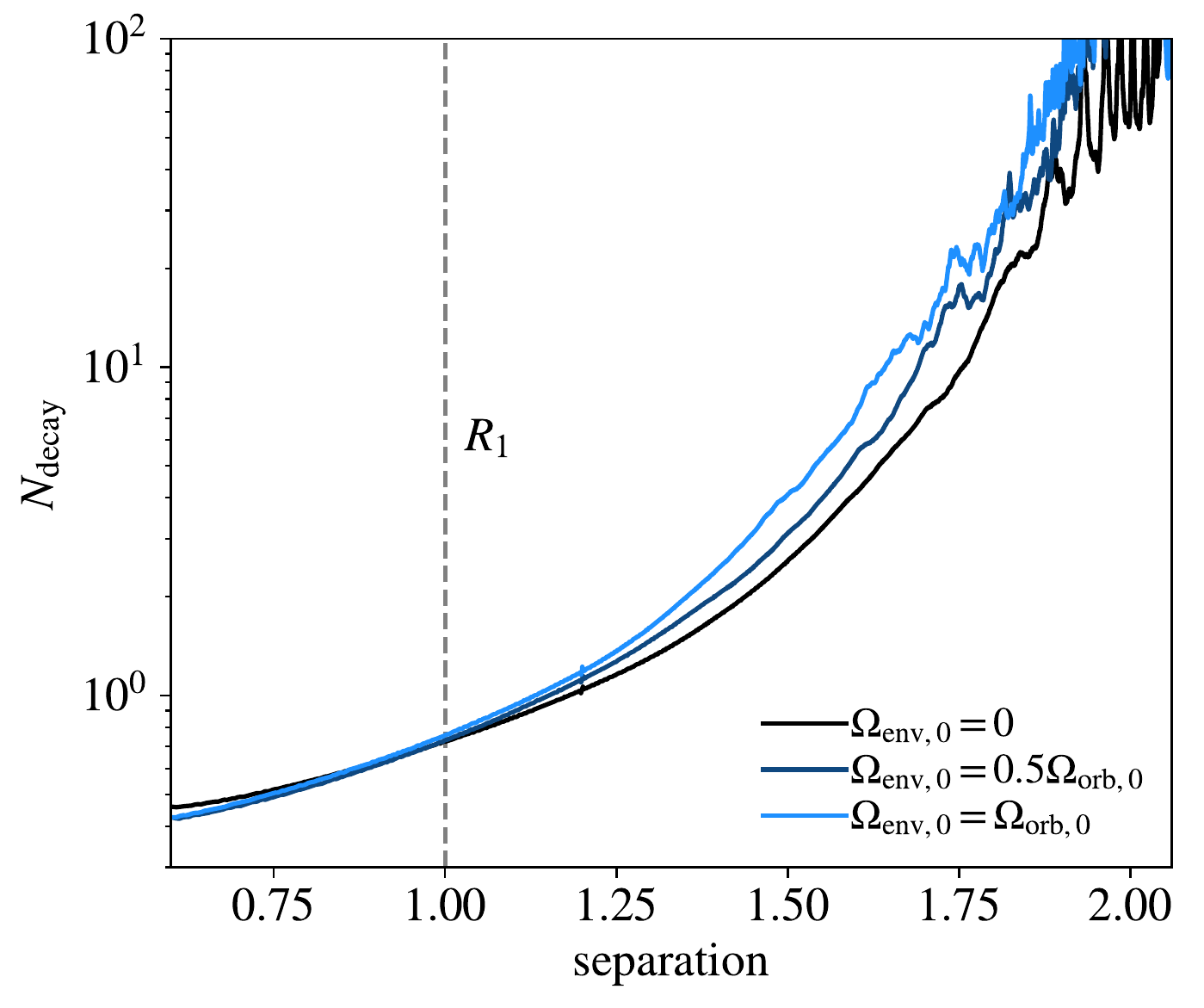}
\caption{Comparison of orbital evolution under different cases of initial donor rotation. We initialize simulations with $\Omega_{\rm env,0}=0$ and $\Omega_{\rm env,0}=0.5\Omega_{\rm orb,0}$, in addition to our fiducial case of $\Omega_{\rm env,0}=\Omega_{\rm orb,0}$. This comparison shows that non-synchronous initial rotation at the onset of Roche lobe overflow leads to somewhat more rapid orbital decay. The difference is most pronounced at large separations where the initially-synchronous cases remain synchronized and vanishes as the orbits tighten and all cases naturally desynchronize.  }
\label{fig:noncorot}
\end{center}
\end{figure}

We have discussed the role of desynchronization of the orbital motion with donor rotation in giving rise to a highly asymmetric flow around the accretor, which enhances the net torque on the binary and drives it toward merger.  We might hypothesize that initial stellar rotation that is non-synchronous would enhance this effect and lead to more rapid orbital evolution. 
To study the magnitude of this effect, we ran simulations otherwise identical to our fiducial case with $\Omega_{\rm env,0}=\Omega_{\rm orb,0}$, but which had envelopes with initial angular velocities of $\Omega_{\rm env,0}=0$ and $\Omega_{\rm env,0}=0.5\Omega_{\rm orb,0}$. 

As we show in Figure \ref{fig:noncorot}, initial rotation does have an effect on the rate of orbital decay, with more rapid coalescence in the non-synchronous cases. The effect is most pronounced at large separations -- at which the synchronous case has not yet fallen far from co-rotation. By the time $a\approx R_1$, all cases exhibit similar behavior because even the synchronous case has desynchronized. 

Overall, the magnitude of the effect of the initial spin of the star is relatively mild. This indicates that  the changing flow morphology due to synchronous or non-synchronous rotation is a secondary aspect of runaway orbital coalescence. 
Instead, the  the dominant  aspect of the runway coalescence is the increasingly mass-rich outflow as the secondary carves deeper into the envelope of the primary star (Figures  \ref{fig:masstransfer} and \ref{fig:ts_torque}).

%%%%%%%%%%%%%%%%%%%%%
% DISCUSSION
%%%%%%%%%%%%%%%%%%%%%
\section{Connection to Astronomical Transients}\label{sec:lrn}

The key question that motivated this study is whether impulsive, luminous transients can originate from binary systems at the onset of common envelope episodes. We have shown that runaway orbital decay leads to increasing mass ejection rate from the binary during the final plunge toward coalescence. Here we argue that this runaway orbital decay and mass ejection are the mechanisms that lead to the production of ``luminous red nova'' transients at the onset of common envelope phases.

\subsection{Impulsive Outbursts in Stellar Coalescence Transients}

Luminous red nova transients are thought to originate in the coalescence of binary systems. The strongest evidence for this identification is the galactic transient V1309 Sco, which was an eclipsing binary with decreasing orbital period \citep{2011A&A...528A.114T} that exhibited a 10 magnitude outburst \citep{2010A&A...516A.108M}, after which no periodicity remained \citep{2011A&A...528A.114T}.  Much of the rest of the luminous red novae class of transients are assembled based on spectroscopic and photometric similarity to this transient. 

Several recent extragalactic transients have offered new insights through particularly well-sampled, multicolor light curves. M31 LRN 2015,  which reached peak brightness in January 2015 \citep{2015A&A...578L..10K}, has high-cadence multicolor photometry during the rise to peak brightness \citep{2015ApJ...805L..18W,2015A&A...578L..10K}.  A progenitor source is detected in {\it Hubble Space Telescope} data a decade prior to outburst \citep{2015ATel.7173....1D,2015ApJ...805L..18W} and in other archival datasets \citep{2015ATel.7173....1D}.  Another transient, M101OT 2015-1 \citep{2017ApJ...834..107B} was detected in outburst just a month later, in February 2015. The rise to peak occurred while the object was behind the sun, but there is a decade of exquisite multicolor photometry of the luminous pre-outburst source \citep{2017ApJ...834..107B}

A surprising lesson from the M31 LRN 2015 transient was that the rise to peak brightness occurs over nearly the same timescale as the orbital period around the surface of the giant star observed in pre-outburst data \citep{2017ApJ...835..282M}. Another universal feature of these transients is precursor emission and a gradual brightening over tens to hundreds of orbital periods prior to coalescence, which was first commented on in the light curve of V1309 Sco \citep{2011A&A...528A.114T}. This precursor brightening is seen to be accompanied by slightly declining photosphere temperatures in the multicolor photometry of M101OT 2015-1\citep{2017ApJ...834..107B}.  

Long-term precursor emission suggests an interaction that occurs steadily over a number of orbits prior to the outburst. The progressive brightening indicates that the interaction intensifies over this time. These precursor signatures rule out a stellar collision or very eccentric encounter as an explanation for the impulsive nature of the outbursts. A relatively-circular binary system is the simplest explanation -- and yet, how can such a system be stable over a stellar evolution timescale, perhaps billions of orbits, and then power an outburst over the course of a single orbit?

\subsection{Connection to Simulated Systems}

Here we argue that precursor emission and outburst share a common origin in the runaway coalescence of a circular binary. 

{\it Precursor emission:} 
Mass transfer in our simulated system soon leads to outflows from the inner binary pair as material is unable to pile up in the vicinity of the accretor. Pressure gradients divert flow outward from the binary, and it is trailed off from the inner pair near the outer, $L_2$ and $L_3$, Lagrange points (as described by  \citet{1979ApJ...229..223S} and seen, for example, in the earliest snapshot of Figure \ref{fig:ts_dens}). 

We have extensively discussed the role of outflows in carrying angular momentum from the binary, but this outflow also leads to an expanding photosphere radius. \citet{2016MNRAS.455.4351P, 2016MNRAS.461.2527P} have modeled the gas dynamics and thermodynamics of outflows from the $L_2$ point of a binary system, and have applied this modeling to very convincingly explain the pre-outburst evolution of the V1309 Sco transient \citep{2014ApJ...788...22P, 2017ApJ...850...59P}. In particular, \citet{2017ApJ...850...59P} showed that the evolving eclipse morphology in the decade prior to outburst can be traced to increasing (orbital-phase-dependent) obscuration due to increasing outflow. \citet{2014ApJ...788...22P} and \citet{2017ApJ...850...59P} show that periodicity vanishes as V1309 Sco brightens because the photosphere exceeds the radius of the binary.

Here we have shown that these outflows arise naturally as a consequence of mass exchange, that they intensify significantly as the binary separation shrinks, and that the torques associated with generating the outflow drive the evolution of the binary orbit. We therefore argue that heated ejecta and an expanding photosphere power the observed precursor emission.

{\it Impulsive outburst:}
Two features of our simulation explain the ability of coalescing binaries to produce impulsive outbursts. We observe a runaway decay of binary orbital separation (Figure \ref{fig:orbit}) that becomes so rapid that the decay timescale is similar to the orbital period, $\dot a / a \approx P_{\rm orb}$, when $a \approx R_1$ (Figure \ref{fig:ndecay}). Associated with (and directly responsible for) the runaway orbital decay, mass ejection from the binary intensifies rapidly in the final moments of the system's coalescence. 

\begin{figure}[tbp]
\begin{center}
\includegraphics[width=0.48\textwidth]{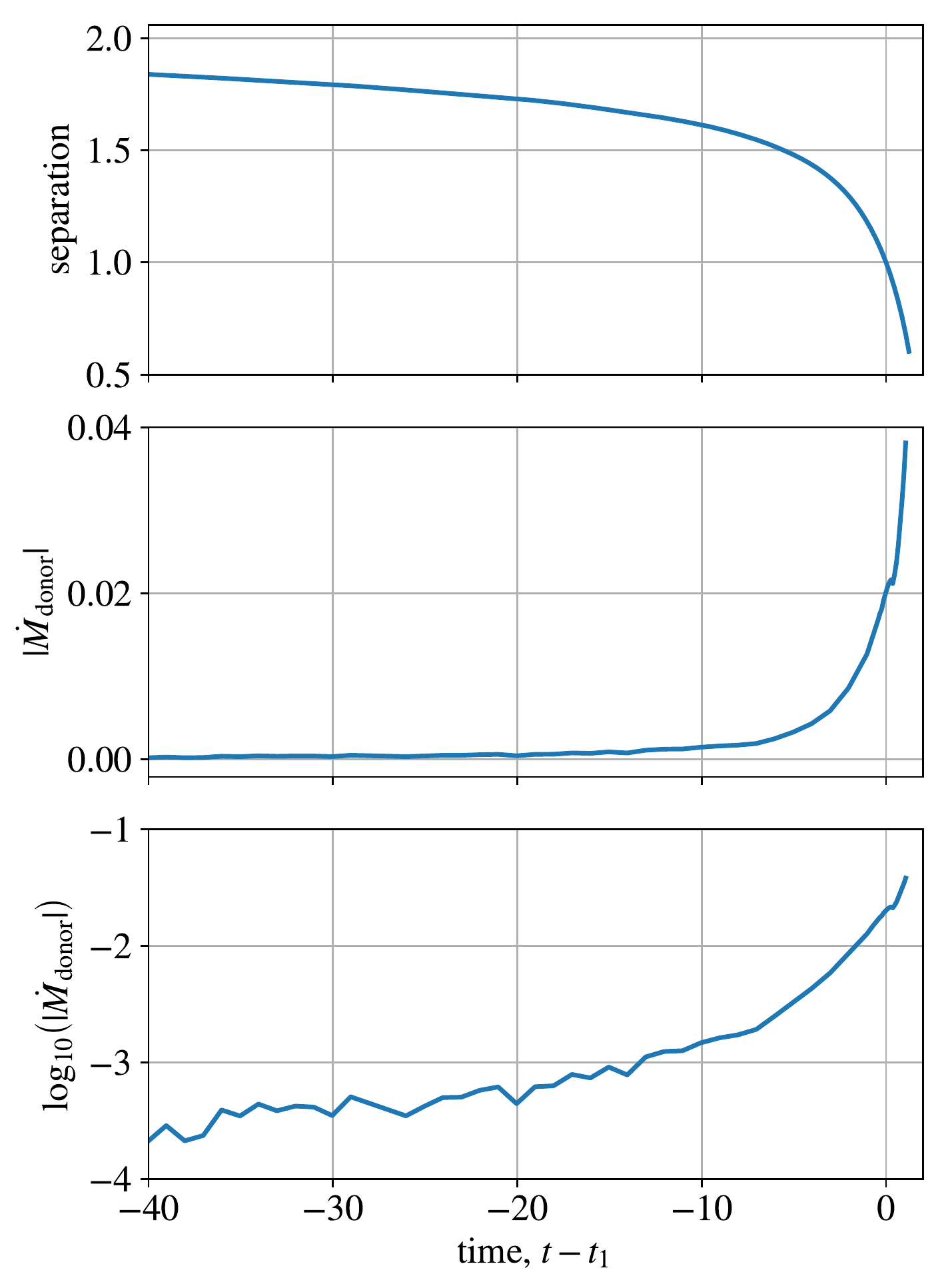}
\caption{ Zoom-in on orbital decay and donor mass-loss rate in the final dynamical times prior to binary coalescence. As the orbit plunges to smaller separations within a few dynamical times of $t_1$, the mass ejection rate from the donor increases by an order of magnitude. We suggest that the origin of impulsive outbursts in luminous red nova transients is the rapid escalation of the mass ejection rate when the binary separation is similar to $R_1$.   }
\label{fig:outburst}
\end{center}
\end{figure}

We summarize the simultaneous orbital decay and mass loss with a zoom-in to the final dynamical times pre-coalescence in Figure \ref{fig:outburst}. Over several donor-star dynamical times, the orbital separation goes from $1.5R_1$ to less than $R_1$ and the mass-loss rate from the donor star increases by an order of magnitude. Impulsive mass ejection accompanies impulsive orbital decay as the accretor star plunges within the envelope of the donor. We argue that the outburst phase of luminous red nova transients originates from this phase of runaway coalescence leading to the engulfment of the accretor and the onset of the common envelope phase. 

Finally, we note a changing outflow morphology as the binary coalesces -- from a thin stream to a broadening fan of essentially ballistic ejecta (Figure \ref{fig:traj}).  The detailed dynamics and thermodynamics of this outflow surely imprint themselves on outbursts of luminous red nova transients.

\section{Conclusion}\label{sec:conclusion}

Many open questions remain -- we have considered just a single binary configuration -- and many physical processes have been neglected in our model, which only treats the hydrodynamics of gravitational interaction between donor-star gas and the stellar cores.  Nonetheless, from the examination of this simplified system, we draw some broad conclusions that carry implications for the coalescence of binary systems at the onset of stellar mergers and common envelope episodes. 

\begin{enumerate}
\item Our simulation of unstable mass transfer in a binary system  from a primary-star donor of mass $M_1=1$ onto an accretor of mass $M_2=0.3$ begins at the Roche limit separation. The binary separation decreases, at first slowly, then increasingly rapidly. The orbit eventually enters into a phase of runaway decay, plunging from separation of $1.5R_1$ to less than $R_1$ in roughly an orbital period (Figures \ref{fig:orbit} and \ref{fig:ndecay}).  

\item Angular momentum lost from stellar cores transfers to the gas pulled from the donor's envelope (Figure \ref{fig:angmom}) and is ejected with an average specific angular momentum between that of the accretor and the $L_2$ point (Figure \ref{fig:specificmomgamma}). 

\item The mass-loss rate (Figure \ref{fig:masstransfer}) and specific angular momentum of material lost (Figure \ref{fig:specificmomgamma}) allow one to analytically reconstruct the orbital evolution, using equation \eqref{orbit_analytic}, which produces a remarkable fit to the simulated result (Figure \ref{fig:sepanalytic}). This reconstruction indicates that escalating mass loss drives much of the runaway orbital decay. 

\item The gravitational torques (Figure \ref{fig:ts_torque}) that exchange angular momentum between the particles and gas are mostly applied by the accretor particle (Figure \ref{fig:torque_time_sep}) in its immediate surroundings (Figure \ref{fig:torque2}).

\item The morphology of outflow from the binary changes as the orbital separation tightens, from a thin, high-entropy stream from the $L_2$ and $L_3$  outer Lagrange points to a broad fan of low-entropy material on largely ballistic trajectories (Figures \ref{fig:ts_dens}, \ref{fig:earlylate}, and \ref{fig:traj}). Desynchronization of the orbital motion with donor-star rotation  (Figure \ref{fig:omegasep}) is the origin of this transition in outflow properties (Figure \ref{fig:traj}). 

\item Characteristic features of luminous red nova transients -- precursor emission and impulsive outbursts -- find common origin in runaway orbital decay and mass ejection leading to a common envelope phase. The orbit shrinks from well-separated to engulfed, and the mass ejection rate increases by an order of magnitude in a single orbital period at separations similar to the radius of the donor (Figure \ref{fig:outburst}). 

\end{enumerate}

 An alternative path to binary coalescence, most relevant for systems  with mass ratio close to unity, involves the formation of a contact binary through stable, rather than unstable, mass transfer \citep{1994A&A...290..119P}. Such a system eventually becomes dynamically unstable \citep[e.g.][]{1993ApJS...88..205L} due to subsequent stellar evolution and wind-carried angular momentum loss. While we might expect the late stages of merger to proceed similarly \citep[see, for example, the simulations of][]{2011ApJ...737...49L,2015ApJ...806..135H}, much of the early dynamics -- and additionally the observational features -- will be different from the unstable mass transfer scenario modeled here. 

Our study of the dynamics of coalescing binaries was motivated by surprises in new time-domain observational data of luminous red nova transients,  which appear to originate in dynamically unstable mass transfer in binary systems. Future efforts to expand these samples are underway in the form of searches for low-temperature extragalactic optical/infrared transients \citep[e.g.][]{2017ApJ...839...88K,2018PASP..130c4202A}. 
Future simulation work can tighten the connection to the observable properties of luminous red nova transients through the progressive inclusion of currently unmodelled physical effects like magnetohydrodynamics, more realistic equations of state with partial ionization, radiative transfer, as well as a broader range of binary system properties. 
More broadly, our conclusions suggest the powerful role new empirical constraints are playing in opening a window into long-uncertain physics of common envelope events, especially through comparison to simulated counterparts of these systems.

\acknowledgements

We gratefully acknowledge many colleagues for helpful conversations that shaped this work, especially Andrea Antoni, Eric Blackman, Thomas Baumgarte, Nadia Blagorodnova, Matteo Cantiello, Orsola De Marco, Rosanne Di Stefano,  Paul Duffell, Adam Frank, Jonathan Grindlay, Tomasz Kaminski, Mansi Kasliwal, Abraham Loeb, Phillip Macias, Ariadna Murguia-Berthier, Priya Natarajan, Enrico Ramirez-Ruiz, Melinda Soares-Furtado, Jennifer Sokoloski, Thomas Tauris, Rich Townsend, Scott Tremaine, Matias Zaldarriaga, and the ``Evolved stars and binaries'' discussion group at the CfA.

This research made use of {\tt astropy}, a community developed core Python package for Astronomy \citep{2013A&A...558A..33A}. 
M.M. is grateful for support for this work provided by NASA through Einstein Postdoctoral Fellowship grant number PF6-170169 awarded by the Chandra X-ray Center, which is operated by the Smithsonian Astrophysical Observatory for NASA under contract NAS8-03060. 
Support for program \#14574 was provided by NASA through a grant from the Space Telescope Science Institute, which is operated by the Association of Universities for Research in Astronomy, Inc., under NASA contract NAS 5-26555.
The work of E.C.O. was supported by a grant from the Simons Foundation (grant no. 510940). 
Resources supporting this work were provided by the NASA High-End Computing (HEC) Program through the NASA Advanced Supercomputing (NAS) Division at Ames Research Center.

\bibliographystyle{aasjournal}
%\bibliography{ce}

\appendix
\section{Numerical Convergence Studies}\label{sec:appendix}

In this section, we present a  series of tests validating the convergence and robustness our simulation results with respect to numerical parameters.

\subsection{Donor Hydrostatic Equilibrium}\label{sec:hse}

\begin{figure}[tbp]
\begin{center}
\includegraphics[width=0.75\textwidth]{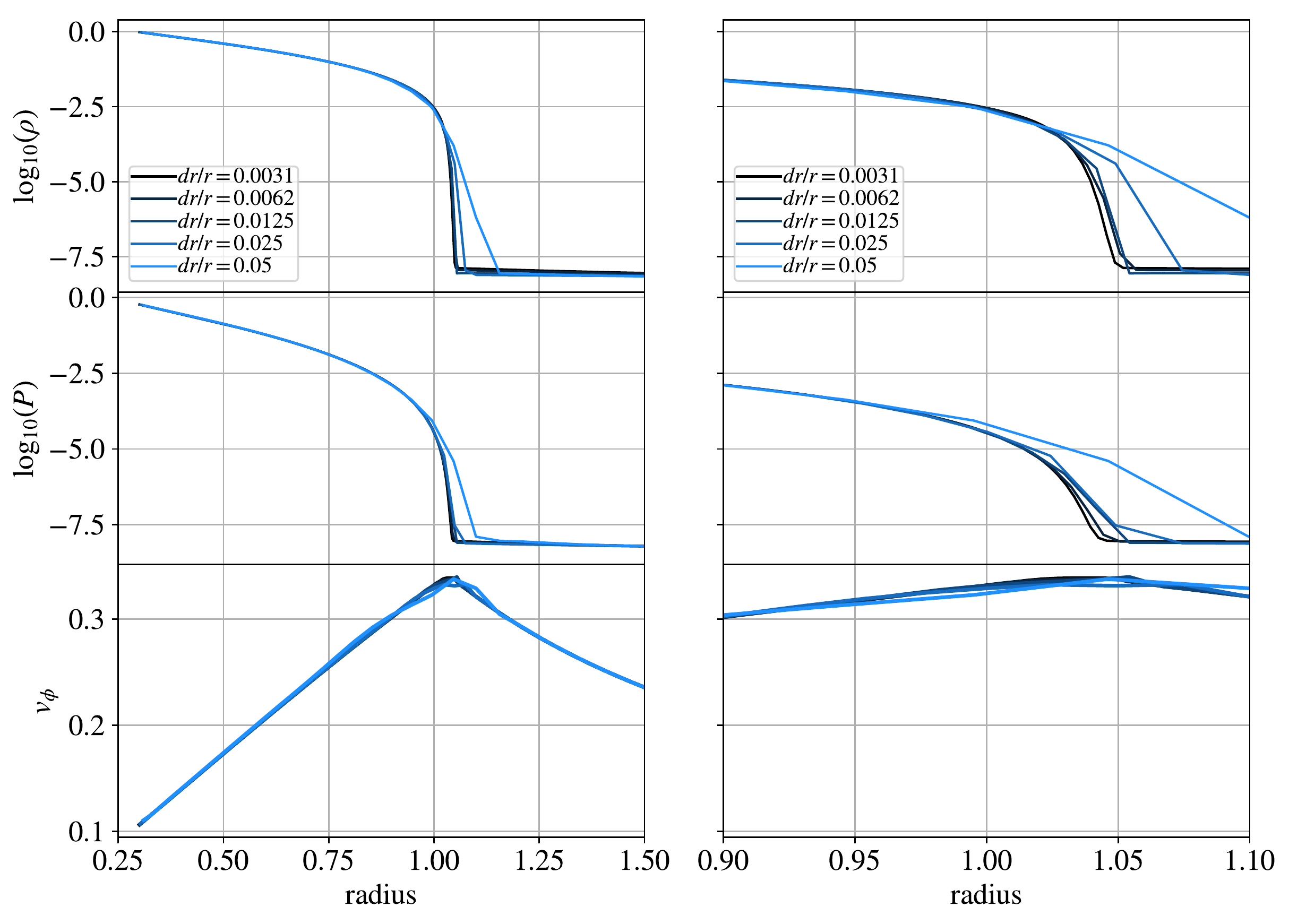}
\caption{ One-dimensional tests of hydrostatic equilibrium at varying spatial resolution.  Zones are logarithmically spaced in radius, and therefore have constant $dr/r$.    }
\label{fig:1dhse}
\end{center}
\end{figure}

\begin{figure}[tbp]
\begin{center}
\includegraphics[width=0.4\textwidth]{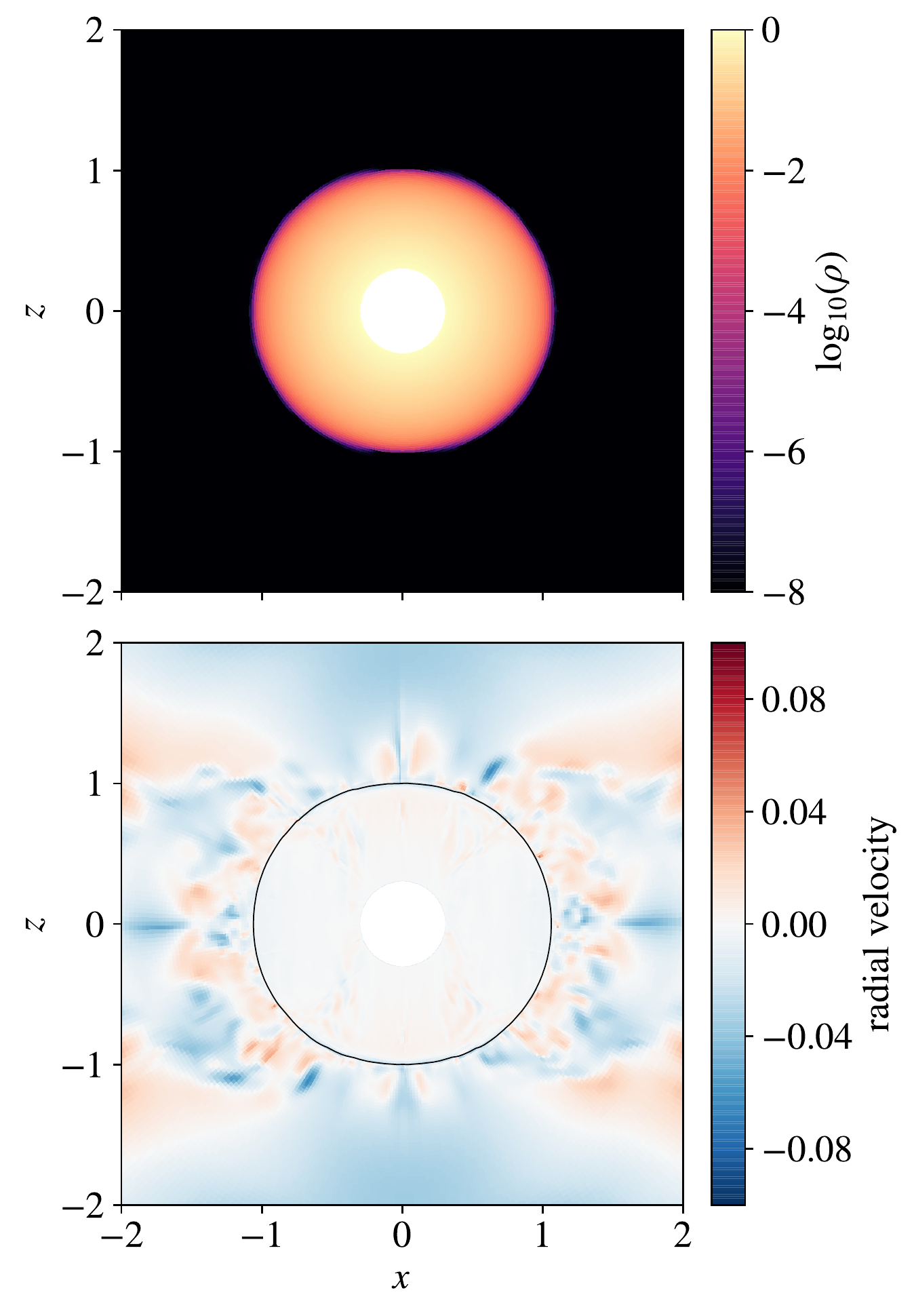}
\hspace{0.5cm}
\includegraphics[width=0.48\textwidth]{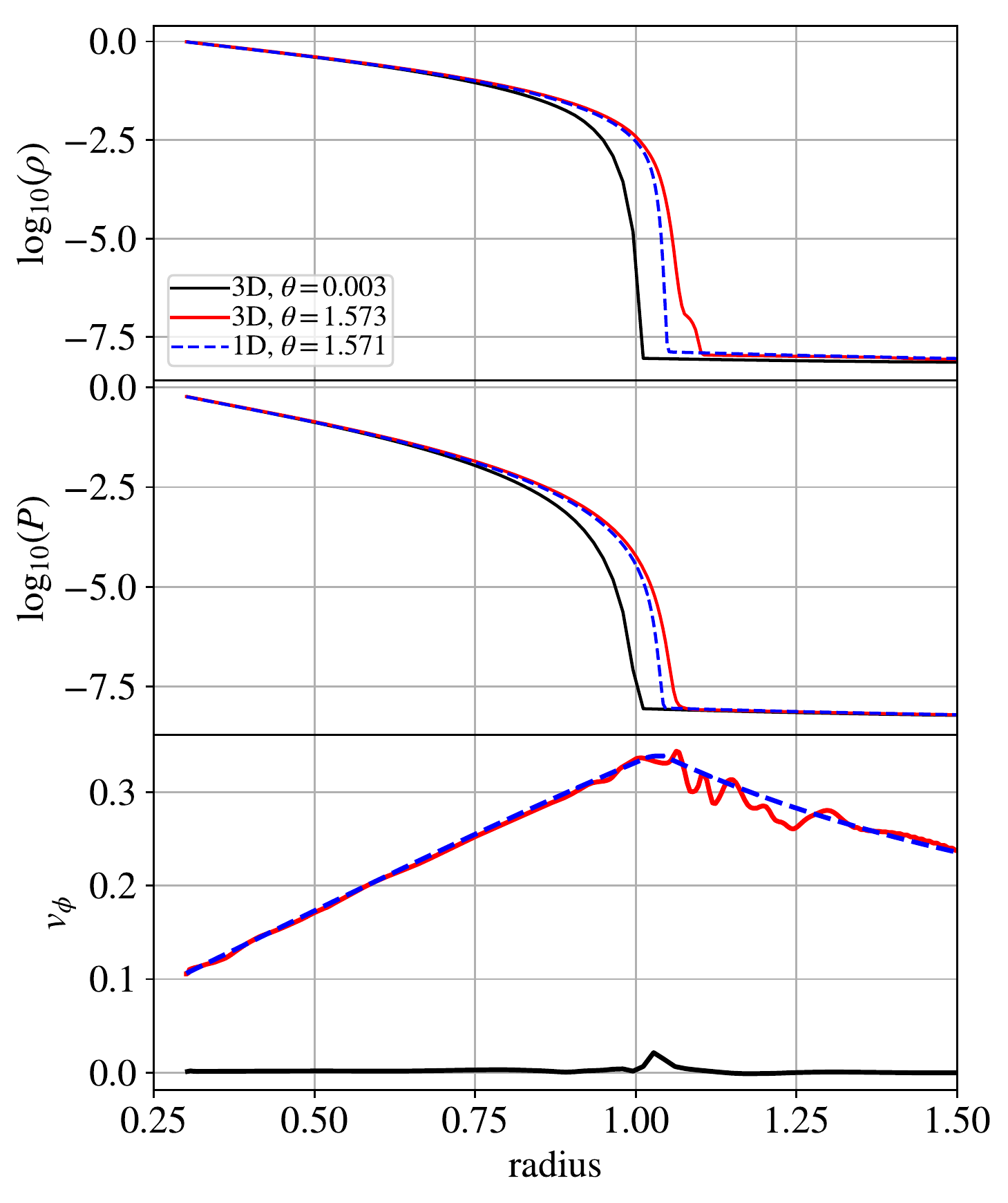}
\caption{ Three-dimensional test of hydrostatic equilibrium at the same resolution as our fiducial simulation. The left panels show polar slices through the simulation domain 100 dynamical times after the simulation is initialized. In the velocity slice, the black contour shows a density of $10^{-5} M_1/R_1^3$.  The right-hand panels compare profiles near the pole to those near the equator with the highest resolution 1D profile.     }
\label{fig:3dhse}
\end{center}
\end{figure}

The ability to preserve a structure in hydrostatic equilibrium with the hydrodynamic algorithm is a key aspect of our simulation method. 
Stars feature very small scale heights of density and pressure near their limbs, which pose significant challenges for any multidimensional approach. 
Approaches with too low resolution near the limb artificially extend the  star's atmosphere, with dramatic effects on the phase of early, low-level mass transfer that we simulate here.  Lagrangian and Lagrangian-Eulerian approaches tend to have very poor resolution near the stellar surface, leading the radius of the numerical star to be larger than the original model by a significant factor  \citep[see, for example, Figures 7-10 of][which show doubling of the radius of low-density material near the stellar limb]{2017A&A...599A...5O}. 
Approaches in Eulerian codes with Cartesian meshes struggle to capture the steepest gradients along non-cardinal directions of the mesh. 

For the above reasons, we have adopted a spherical polar coordinate system centered on the donor star in our simulation. 
Figure \ref{fig:1dhse} shows 1D tests of hydrostatic equilibrium of an envelope with no perturber object. The domain is located along the midplane at $\theta=\pi/2$, and the envelope is initialized with spin frequency equal to that of a test mass at a separation of $2R_1$.  Zones are logarithmically spaced in the $r$ direction such that $dr/r$ is constant (with total extent in $r$ from 0.3 to 30). We vary the total number of radial zones from 96 ($dr/r=0.05$) to 1536 ($dr/r=0.0031$). Profiles are initialized with a relaxation phase of 15 dynamical times, and the simulations run to 100 dynamical times. This test examines the resolution needed to preserve donor hydrostatic equilibrium in the radial direction. 

The left panels of Figure \ref{fig:1dhse} show the full stellar envelope profile; the right panels zoom in near the stellar limb, where differences are most apparent. From top to bottom, panels show logarithm of density, logarithm of pressure, and $\phi$ velocity. Profiles are plotted 100 envelope dynamical times after initialization. The radius of the relaxed stellar profile is larger than $R_1=1$ because of centrifugal forces not accounted for in the initial, spherical model. The relaxation method allows a smooth transition to the hydrostatic rotating solution. All resolutions above the mid-level show reasonable agreement, hence we conclude that the converged hydrostatic solution is attained. 

Next, we test the hydrostatic equilibrium of an identical envelope but in the full 3D domain. We apply our fiducial resolution used in the science runs, with 12 ($r$) x 8 ($\theta$) x 16 ($\phi$) mesh blocks of $24^3$ zones and two levels of nested static mesh in the equatorial plane. This corresponds to $dr/r=0.0167$ at the base level (near the poles) and $dr/r=0.004$ at the maximum level (near the equator). Figure \ref{fig:3dhse} displays the results of this calculation, 100 dynamical times after initialization. Slices are plotted in the vertical axis, such that we see the oblate stellar profile due to rotation. The radial velocity map (with contour showing a density of $10^{-5} M_1/R_1^3$) shows some residual flow inside (and especially outside) the star, where the extra degrees of freedom relative to the 1D simulation result in more fluid motion. We also note that this profile, with $\gamma=\Gamma_{\rm s}=5/3$ is marginally convectively unstable under the Schwarzschild criterion (that states that convective stability is realized when $\gamma > \Gamma_{\rm s}$). 

The right panels of Figure \ref{fig:3dhse} compare the highest resolution 1D calculation to profiles from the 3D run along the pole and equator. The 3D star is somewhat more oblate than its 1D counterpart due to the 3D effect of additional fluid pressing toward the equatorial belt from latitudes above and below the equator.  

The overall result of the tests presented in this section is that, at sufficiently high spatial resolution, the hydrostatic profile of the donor star can be preserved with a high degree of stability. This ensures that any evolution of the binary system is a direct result of the binary's gravitational interaction, not the artificial erosion of the original profile of the stellar model.

\subsection{Numerical Parameters and Spatial Resolution}\label{sec:convergence}

\begin{figure*}[tbp]
\begin{center}
\includegraphics[width=0.75\textwidth]{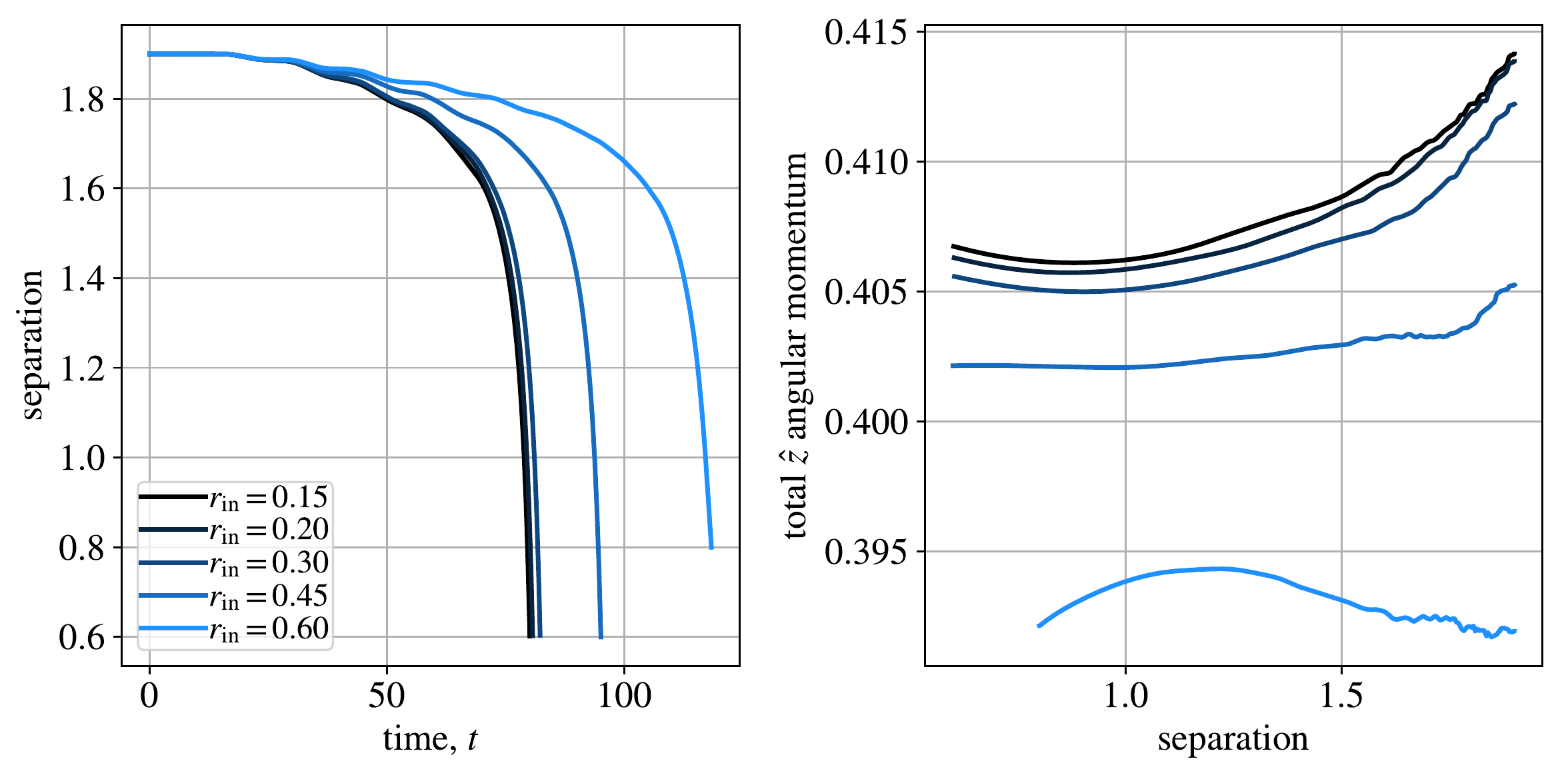}
\caption{Varying the inner boundary condition radius within the donor star.  Left panel shows orbital separation as a function of time (relative to $t_1$, when the separation is $R_1$). The right panel shows total (gas plus particle) system angular momentum as a function of binary separation. Ideally, total angular momentum should be conserved, the amount of departure from conservation is a useful benchmark of the simulations (detailed discussion of reasons for imperfect conservation in the text).  }
\label{fig:conv_rin}
\end{center}
\end{figure*}

\begin{figure*}[tbp]
\begin{center}
\includegraphics[width=0.75\textwidth]{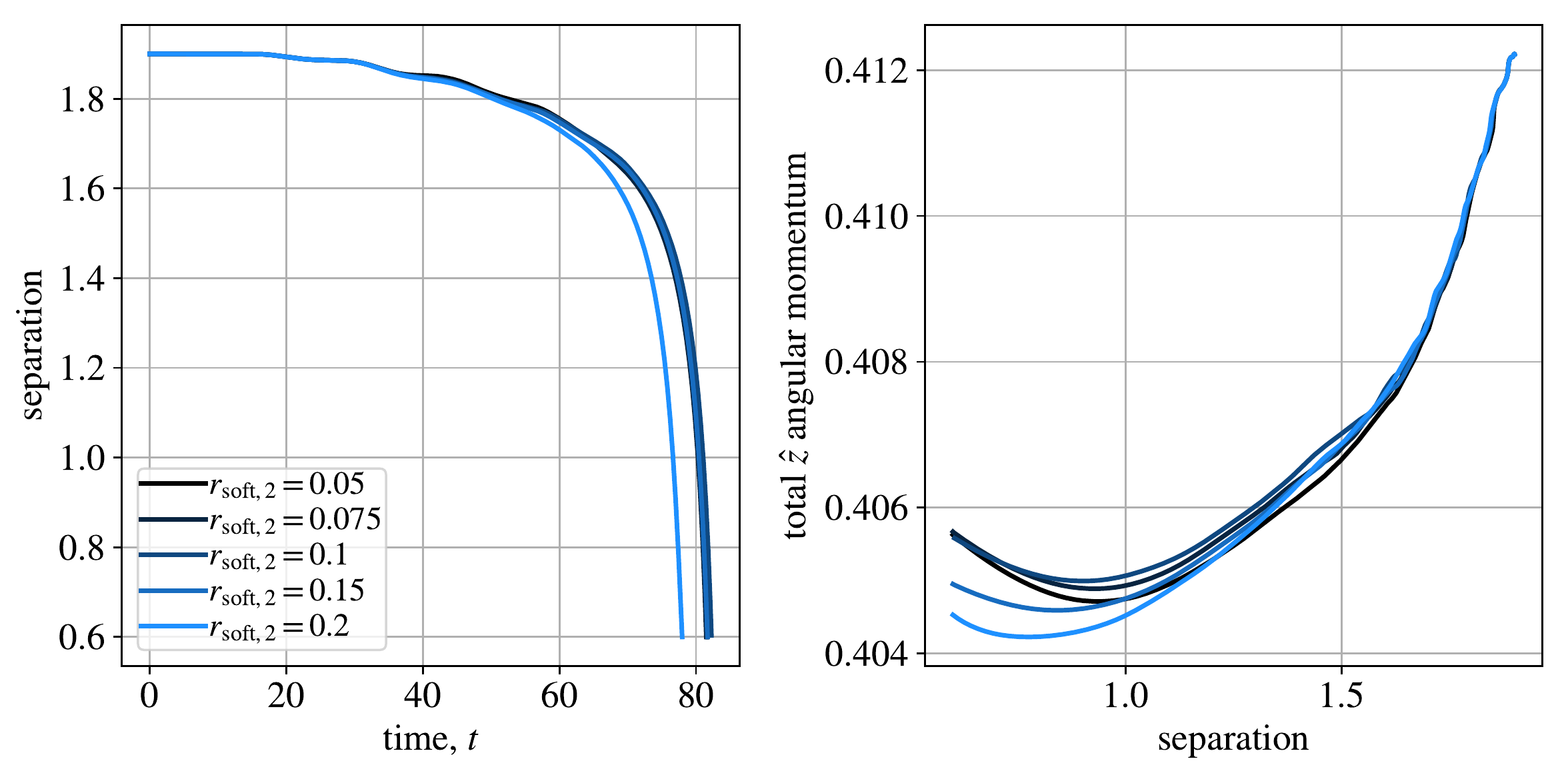}
\caption{ Same as Figure \ref{fig:conv_rin}, but varying the softening radius around the accretor particle.     }
\label{fig:conv_rsoft}
\end{center}
\end{figure*}

\begin{figure*}[tbp]
\begin{center}
\includegraphics[width=0.75\textwidth]{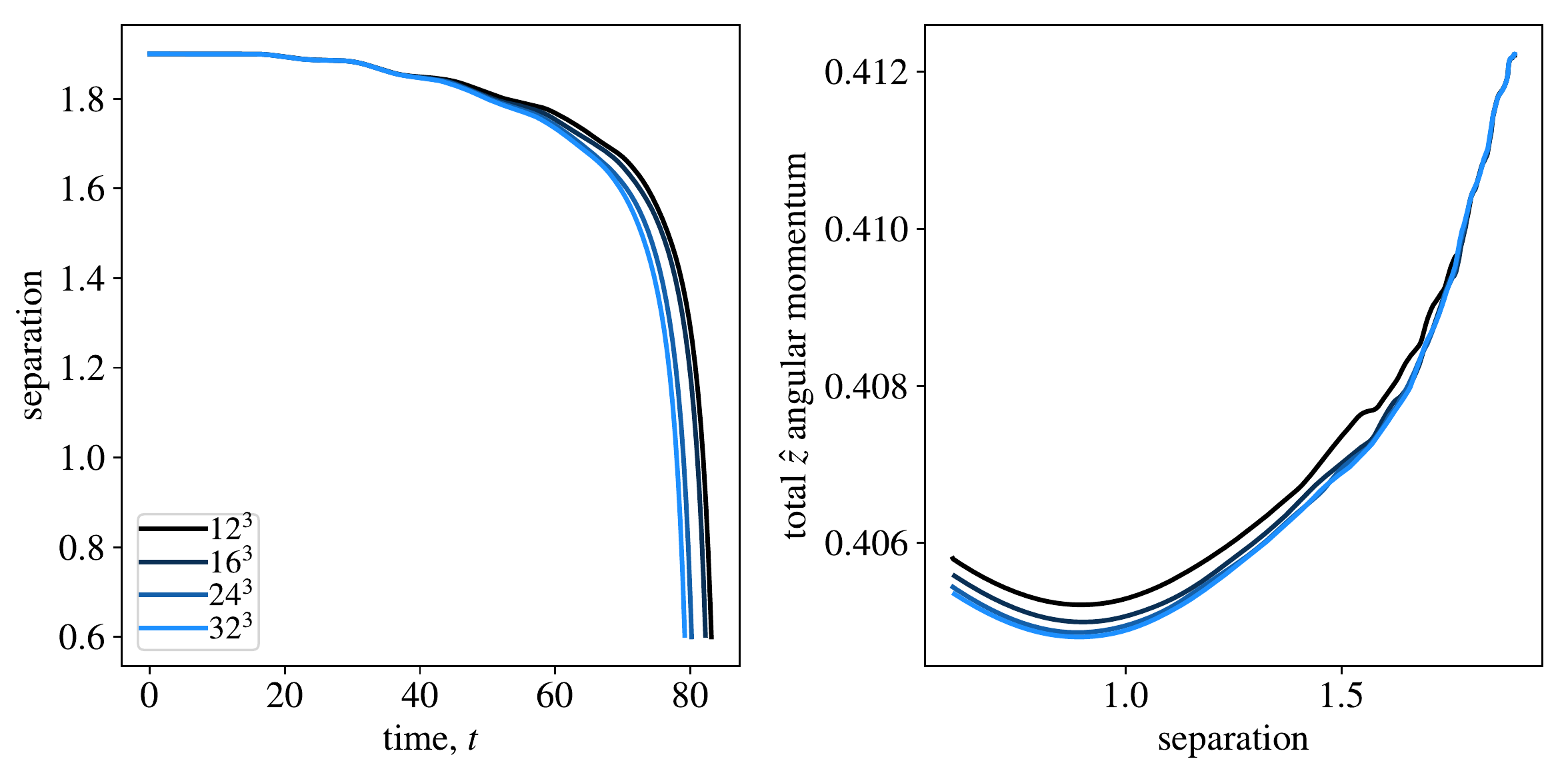}
\caption{ Same as Figure \ref{fig:conv_rin}, but varying spatial resolution. We preserve the same 12 ($r$) x 8 ($\theta$) x 16 ($\phi$) base-level mesh block structure and two levels of nested static mesh refinement, but change the number of zones in each block as labeled in the legend.    }
\label{fig:conv_res}
\end{center}
\end{figure*}

In this section, we test the robustness of our numerical results against several model parameters. These are
\begin{enumerate}
\item the inner boundary radius, controlling how much of the star is excised, $r_{\rm in}$, 
\item the softening radius of the secondary star's gravitational point mass, $r_{\rm soft,2}$, 
\item and the spatial resolution of our mesh.
\end{enumerate}
In what follows, we describe tests against each of these parameters that justify the choices adopted for science runs presented in the main text of this paper. Our fiducial model for this section has an initial separation of $a_0=1.9R_1$, an inner boundary of $r_{\rm in}=0.3 R_1$, a softening radius of $r_{\rm soft,2}=0.1R_1$, and 12 ($r$) x 8 ($\theta$) x 16 ($\phi$) base-level mesh blocks of $16^3$ zones, with additional two levels of static mesh refinement about the orbital plane. 

In Figure \ref{fig:conv_rin} we test the sensitivity of the model to the extent of stellar envelope excised from the center. 
The left panel of Figure \ref{fig:conv_rin} examines the orbital separation evolution, and the right the total angular momentum as a function of separation of the two point masses. In theory, this should be conserved, in practice, it is not perfectly conserved -- for more discussion see Section \ref{sec:Lconservation}.  
We see that when the excision is less than $0.3R_1$, the orbital evolution is converged with respect to smaller excision regions. This suggests that material at those smallest radii within the envelope is not significantly distorted from spherical within the extent of the calculation, so its replacement with a point mass and boundary has little effect on the outcome. 

We test the softening radius of the secondary, accretor, particle, adopting $r_{\rm in} = 0.3R_1$. We test values from $r_{\rm soft,2}=0.05R_1$ to $0.2R_1$. These  softening lengths range from 4 zones across the softening radius at the widest binary separation for the smallest softening length to 16 zones across the largest softening length. Figure \ref{fig:conv_rsoft} demonstrates that values of the softening length less than $0.15R_1$ have little effect on the simulation outcome. 

Finally, we test for convergence of results against spatial resolution of the computational mesh. We do this by fixing the other parameters, and the base mesh and statically-refined mesh block structure, while changing only the number of zones per mesh block. This has the effect of holding everything fixed except the spatial resolution of the calculation. We show a range of resolutions from $12^3$ zones per mesh block to $32^3$ zones per mesh block in Figure \ref{fig:conv_res}. This lowest resolution barely preserves the hydrostatic equilibrium of our chosen stellar profile, while higher resolutions become prohibitively expensive to run. Across this range, we show that neither the runaway orbital behavior nor the overall conservation of angular momentum are sensitive to spatial resolution.

\subsection{Sensitivity to Initial Binary Separation}\label{sec:a0}

\begin{figure*}[tbp]
\begin{center}
\includegraphics[width=0.75\textwidth]{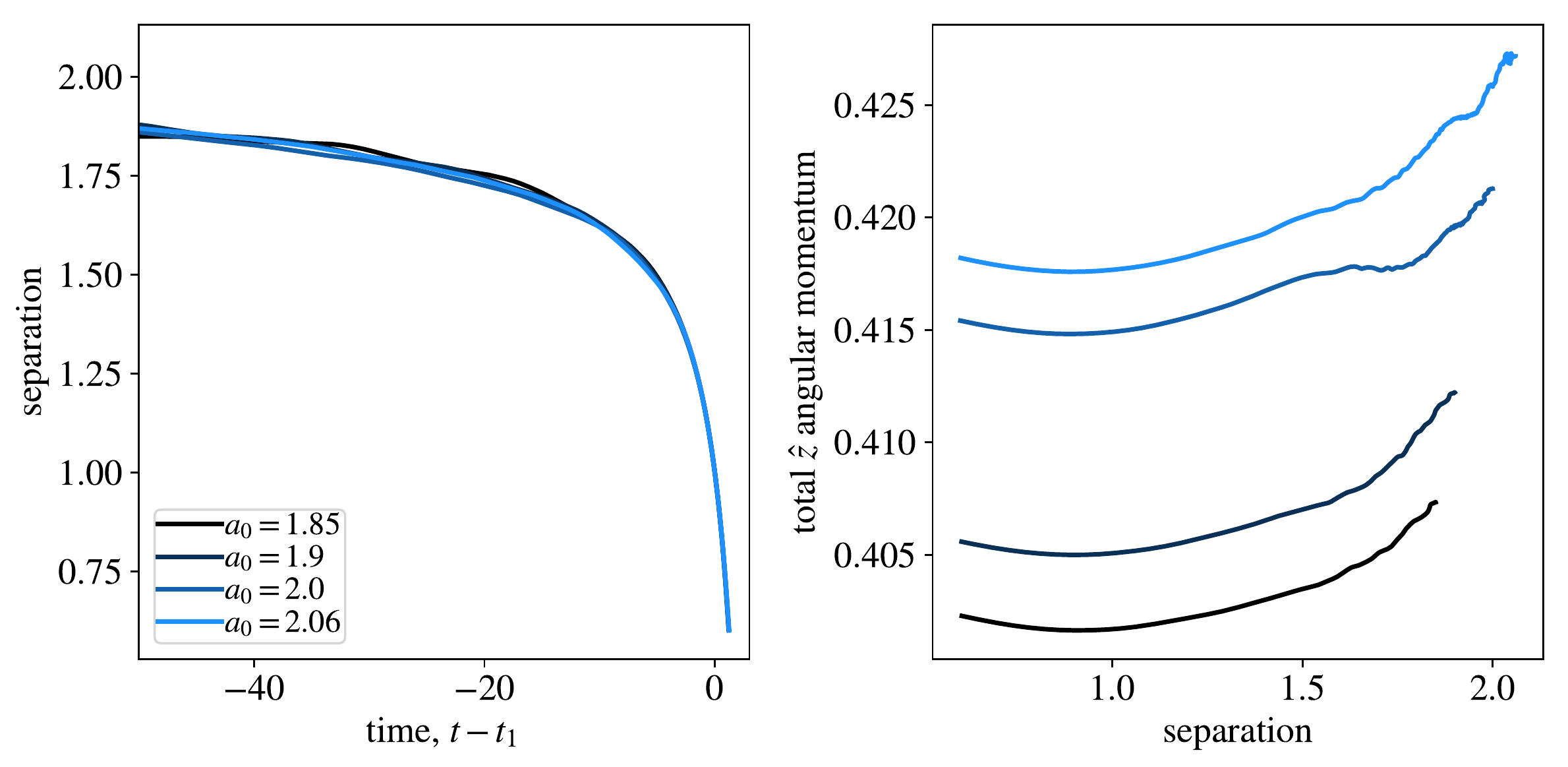}
\caption{ Test of varying initial binary separation, otherwise similar to Figure \ref{fig:conv_rin}.  }
\label{fig:conv_sep}
\end{center}
\end{figure*}

Finally, we consider how the initial binary separation may alter the results of binary coalescence. To test sensitivity to our adopted initial separation ($a_0=1.9R_1$ in the tests above and $a_0=2.06R_1$ in our science runs), we consider a range from $a_0=2.06R_1$ to $a_0 =1.85R_1$. 
The deeper into the phase of runaway mass transfer the simulation is initialized, the less time the simulation needs to run leading into the coalescence of the pair of objects. We run calculations with identical initial stellar spin frequency (equal to the orbital frequency at the analytic Roche limit separation of $2.06R_1$), but with orbital separations of $a_0=2.06$, 2.0, 1.9, 1.85~$R_1$. The left panel of Figure \ref{fig:conv_sep} demonstrates that the overlapping stages of runaway orbital evolution proceed very similarly in each of these cases. We note that the different runs of Figure \ref{fig:conv_sep} have different initial angular momenta because of their differing initial binary separations.

\subsection{Conservation of Total Angular Momentum}\label{sec:Lconservation}

Finally, we comment on the degree of non-conservation of total angular momentum, equation \eqref{Ltot}, observed in the models in Figure \ref{fig:angmom} and Figures \ref{fig:conv_rin}, \ref{fig:conv_rsoft}, \ref{fig:conv_res}, and \ref{fig:conv_sep}. This is due to contributions of several unmodeled physical effects. These effects include boundary interactions and the treatment of self-gravity.  The outer boundary allows outflow of gas (and angular momentum) from the domain, but we keep track of this, as described in equation \eqref{Lgasout}. Unaccounted for effects include the backreaction of the inner, reflecting boundary condition surrounding the donor star core pushing on the gas. One might imagine this adds angular momentum to the system in the case of orbiting particles in gas. Indeed, with the largest central boundary $0.6R_1$ in Figure \ref{fig:conv_rin}, the total angular momentum increases slightly. The other physical effect, which we suggest is most significant, is the unmodeled gas self-gravitational interaction. We treat gas self-gravity statically, by applying the potential of the initial, spherical, undisturbed star, as described by equation \eqref{asg}. This becomes less accurate as mass is removed from the donor and distributes in increasingly complex configurations around the binary.  That said, the overall magnitude of the non-conservation (about 2\% given a factor of similar to 2 change in component momenta) justifies this approximation for the purposes of our analysis. 

\end{document}